\documentclass[twocolumn]{el-author}

\usepackage{soul}
\usepackage{xcolor,colortbl}
\usepackage{romannum} 
\usepackage[superscript]{cite} 
\usepackage{amsmath} 
\usepackage[symbol,perpage]{footmisc} 

\date{August 2022}

\begin{document}

\title{Object Storage, Persistent Memory, and Data Infrastructure for HPC Materials Informatics}

\author{Stephanie R. Taylor}

\abstract{
This report was prepared as a final deliverable from the author's participation in the 2022 GRIPS-Berlin (8-week) summer internship program. For additional details, refer to the Acknowledgements section. For a short, step-by-step sentence summary of the content in this report, refer to the Conclusion section.
}

\maketitle

\section{\textbf{\textcolor{blue}{Introduction}}}

\begin{figure}[h!]
    \centering
    \includegraphics[width=1.05\linewidth]{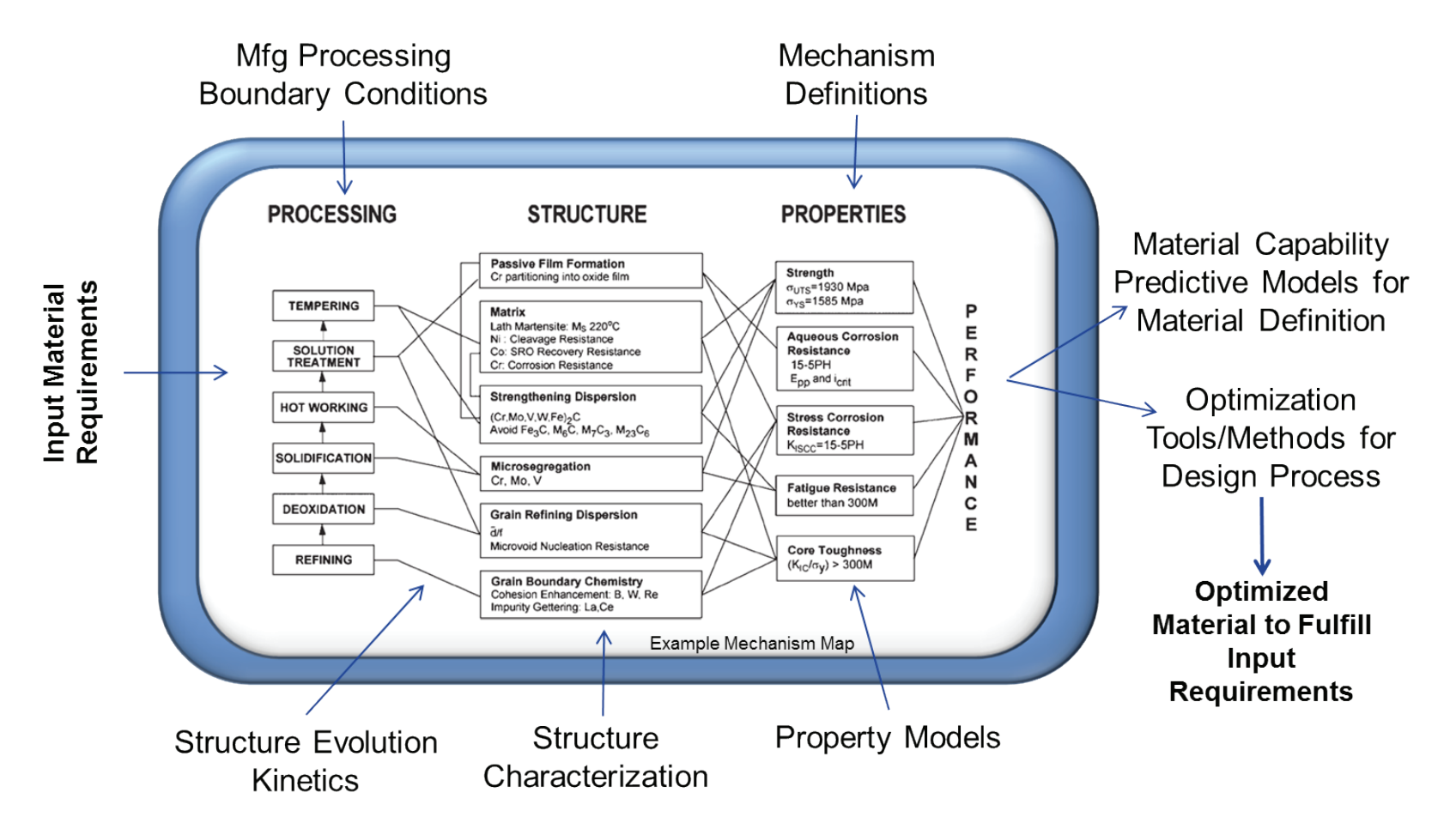}
    \caption{Olson diagram depicting the various streams of data that contribute to the process-structure-property \textbf{(PSP)} relationship.  \cite{NASA_2040_Vision}}
    \label{fig:OlsonDiagram}
\end{figure}

\begin{figure}[h!]
    \centering
    \includegraphics[width=1\linewidth]{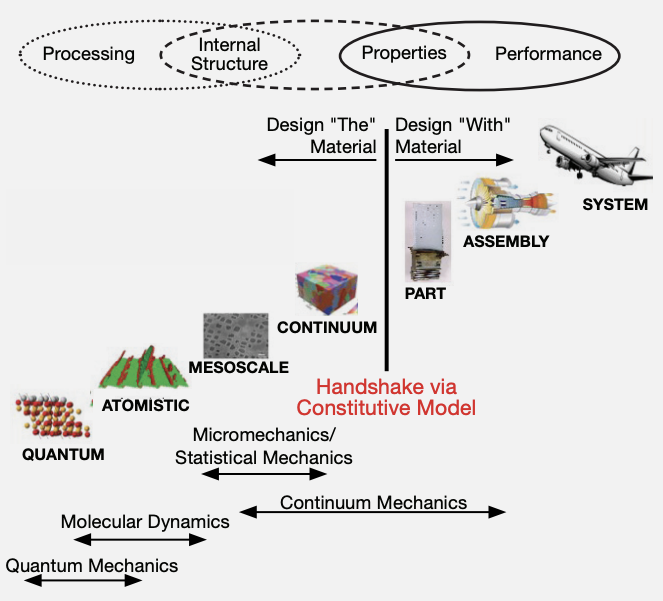}
    \caption{Description of length scale dependence and modeling methods in the context of Integrated Computational Materials Engineering \textbf{(ICME)}. Note that modeling, simulation, and data-driven methods such as artificial intelligence and machine learning may be included in all these stages. \cite{NASA_2040_Vision}}
    \label{fig:ICME}
\end{figure}

	What we can build is ultimately limited by the materials that we have access to. The goal of materials science is to understand and exercise control over the measureable, physical, tangible, macroscale properties of a given material responding to a given environment. Accordingly, materials scientists are concerned with what is known as the ``Process-Structure-Property'' \textbf{(PSP)} relationship. Simply put: processing conditions govern the development of material’s microstructure, which, in turn, governs the properties that the material possesses. 
 An Olson diagram, such as that shown in \textbf{Figure \ref{fig:OlsonDiagram}}, provides one way of visualizing the more general PSP relationships in materials. Because of the wide range of time- and length-scales involved, the questions that materials science strives to answer are inherently complex and multi-scale (see \textbf{Figure \ref{fig:ICME}}).  

	Materials informatics tackles these questions using computational, data-driven methods. The tools of this field are (1) theoretical models, (2) experimental data, and (3) computing power. These three pillars are continually evolving in response to each other. Scientists across history have always strived to leverage theoretical models and experimental data, but it is only since the advent of the modern computer that the third pillar – computing power – was truly unleashed. The Computer Revolution has continued to present day, and it is transforming the fields of science and engineering along with it, materials science included. The prospect of being able to accurately model complex, multi-scale material property phenomena finally appears to be within our grasp. However, it is nascent territory. The data infrastructure required to realize this dream is still being developed, and ascertaining the proper building blocks for this infrastructure is of pertinent concern. High-performance computing \textbf{(HPC)} is one of the primary tools, so developments in the HPC space have a ripple effect within the field of materials informatics.

In this report, speculation is provided on how infrastructure choices fit into the materials data ecosystem.
Special attention is paid to object storage, Intel's DAOS API, storage-class memory \textbf{(SCM)}, and the prospect of non-von Neumann computing. (`Persistent memory' and `SCM' are used interchangeably in this report.) Lastly, the hypothesized implications of data infrastructure choices on a sample materials informatics problem is discussed: computational materials discovery of phase-change materials with properties tailored for phase-change memory \textbf{(PCM)}. The motivation for selecting PCM as a sample materials informatics case study comes from its relevance to emerging SCM hardware.
Overall, this report touches on a variety of subjects, but they are all woven together through the
same underlying story, the plot of which can be summarized as follows.
 \begin{itemize}
     \item  Computational methods can vastly accelerate materials development for `real-world' use in industry, and investments in materials data infrastructure greatly improve the efficacy and reach of such methods.
     \item One of the pillars of this infrastructure is data storage.
     \item Advances in the materials used for data storage hardware have not only improved the capabilities of existing storage paradigms, but have also enabled the development of new ones.
     \item By intelligently incorporating next-generation storage technologies into the materials data infrastructure, materials scientists are able to leverage enhanced computational workflows, which further accelerates materials development.
 \end{itemize}


\section{\textbf{\textcolor{blue}{Historical Context for Materials Data Infrastructure and ICME}}}

The infrastructure building blocks of a successful materials data ecosystem is a topic that has received a significant amount of attention in the materials research community in recent years, especially in the United States. However, stakeholder interest in creating a robust materials data infrastructure \textbf{(MDI)} is nothing new. 

Consider a report from 2004, \textit{Materials Research Cyberscience Enabled by Cyberinfrastructure}\cite{Cyberinfrastructure_2004}, that summarizes the perspectives of the 110 participants (including 44 PIs and co-PIs, 35 graduate students, and 22 post-doctoral fellows) at the National Science Foundation \textbf{(NSF)} Computational Materials Science Review. This report is vague on the actual pillars that ought to comprise an implemented MDI, but clear desire is expressed for the creation of database repositories, such as: ``A central repository for important computational materials results, similar to protein data banks. It should be investigated whether this could be done with the appropriate archival journals. ... A software repository for materials algorithms and for materials software. To be efficient, support for only one such repository should be funded.'' The authors appear to have possessed -- among other things -- a vision of a more organized, centralized, government-funded materials database. Despite the considerable challenges involved, they were optimistic in their tone. They state: ``The possibilities in theory that are dreams today may become realities. ... It is entirely possible that advances in cyberinfrastructure will create revolutionary advances that are not on anyone’s ‘radar screen’ today. That is one of the joys of research.''

Around this time, the concept of Integrated Computational Material Engineering \textbf{(ICME)} was gaining momentum as well. A seminal document from 2008 defined “ICME cyberinfrastructure” as: ``the Internet-based collaborative materials science and engineering research and development environments that support advanced data acquisition, data and model storage, data and model management, data and model mining, data and model visualization, and other computing and information processing services required to develop an integrated computational materials engineering capability''.\cite{ICMEReport_2008} One of the fundamental ideas of ICME is taking output data from one model is and feeding it as input data into another model. This enables multi-scale data flow, as well as the ability for materials data to respond much more dynamically to the product design cycle. Successful case studies of ICME had been demonstrated by companies such as Ford, General Electric, Boeing, QuesTek, etc., piquing interest by others in the materials science \& engineering \textbf{(MSE)} community.\cite{ICMEReport_2008} For instance, Ford Motor Company reported that its ICME-inspired implementation of a Virtual Aluminum Casting process saved them millions of dollars in direct cost savings or cost avoidance.\cite{Ford_2006} The aforementioned 2008 document,  \textit{Integrated Computational Materials Engineering: A Transformational Discipline for Improved Competitiveness and National Security}, prepared for the National Academy of Sciences, also detailed a clear vision for the nascently-emerging field of ICME. Several key conclusions from the report are as follows:
\begin{itemize}
    \item (Conclusion 1) The materials development and optimization cycle cannot operate at the rapid pace required by integrated product development teams, and this potentially threatens U.S. competitiveness in powerhouse industries such as electronics, automotive, and aerospace, in which the synergy among product design, materials, and manufacturing is a competitive advantage. \\
    \item (Conclusion 2) ICME is a technologically sound concept that has demonstrated a positive return on investment and promises to improve the efficient, timely, and robust development and production of new materials and products. \\
    \item (Conclusion 3) While some aspects of ICME have been successfully implemented, \textbf{ICME as a discipline within materials science and engineering does not yet truly exist.} \\
    \item (Conclusion 4) For ICME to succeed, it must be embraced as a discipline by the materials science and engineering community, leading to requisite changes in education, research, and information sharing. ICME will both require and promote a better connection between the science of materials and the engineering of materials. ICME will transform the field of materials science and engineering by integrating more holistically the engineering and scientific endeavors. \\
    \item (Conclusion 5) Industrial acceptance of ICME is hindered by the slow conversion of science-based materials computational tools to engineering tools and by the scarcity of computational materials engineers trained to use them. \\
    \item (Conclusion 6) A coordinated government program to support the development of ICME tools, infrastructure, and education is lacking, yet it is critical for the future of ICME. \\
    \item (Conclusion 7) Although there has been significant progress in the development of physically based models and simulation tools, for many key areas they are inadequate to support the widespread use of ICME. However, in the near term, ICME can be advanced by use of empirical models that fill the theoretical gaps. Thus experimental efforts to calibrate both empirical and theoretical models and to validate the ICME capability are paramount. \\
    \item (Conclusion 8) An ICME cyberinfrastructure will be the enabling framework for ICME. Some of the elements of that cyberinfrastructure are libraries of materials models, experimental data, software tools, including integration tools, and computational hardware. An essential “noncyber” part of the ICME infrastructure will be human expertise. \\
    \item (Conclusion 9) Creation of a widely accepted taxonomy, an informatics technology, and materials databases openly accessible to members of the materials research and development, design, and manufacturing communities is essential for ICME. \\
    \item (Conclusion 10) The development of rapid characterization tools alongside new information technology and materials databases will allow speedy calibration of the empirical models required to fill gaps in theoretical understanding. 
    
\end{itemize}

\noindent However, this report also makes it clear that ICME is not cheap. The authors speculate that ``development of an ICME capability for a given material system to solve a particular foundational engineering problem will require an investment of \$10 million to \$40 million over 3 to 10 years, depending on the completeness and complexity desired.'' Ford's Virtual Aluminum Casting program from 2006, for instance, required about \$15 million in expenditures over 5 years. This resulted in a return-on-investment \textbf{(ROI)} of over 7:1 for Ford, illustrating that such investments can indeed be worthwhile. Even so, they are not trivial undertakings. Furthermore, the multi-year investment typically required for ICME was one of the primary challenges of implementing ICME at this point, since industry typically runs on a 1-year budget cycle. On the other hand, the potential payoffs of ICME were enticing. According to the 2008 report, ``ICME will be transformative for the materials discipline, promising to shorten the materials development cycle from its current 10-20 years to 2 or 3 years in the best scenarios.'' 

\begin{figure}[h!]
    \centering
    \includegraphics[width=0.9\linewidth]{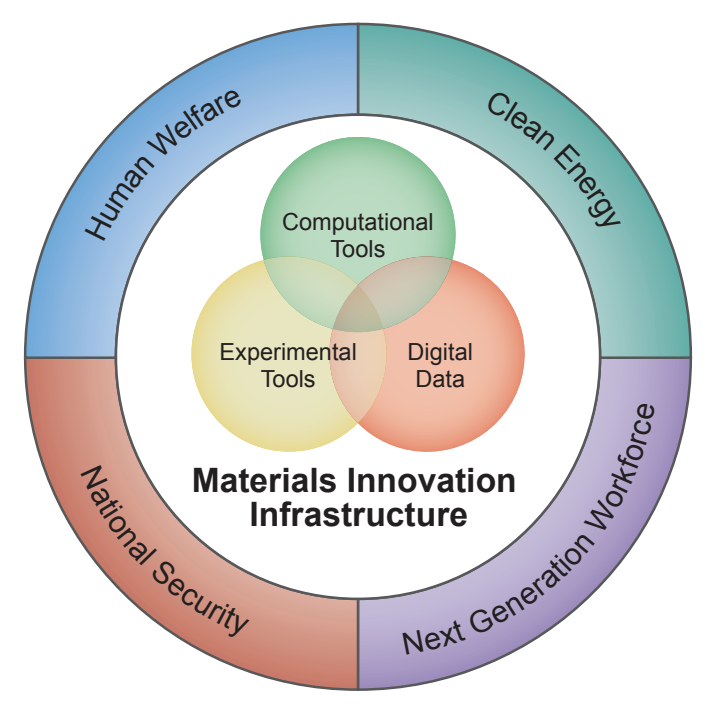}
    \caption{Graphic provided by the 2011 whitepaper to illustrate the goals of the Materials Genome Initiative.\cite{MGI_2011}}
    \label{fig:MGI2011}
\end{figure}

A key milestone for MDI and ICME was the Materials Genome Initiative \textbf{(MGI)}, announced in 2011 by President Barack Obama. The white paper for the MGI states:\cite{MGI_2011} ``This initiative offers a unique opportunity for the United States to discover, develop, manufacture, and deploy advanced materials at least twice as fast as possible today, at a fraction of the cost.'' For fiscal year 2012, the Obama Administration requested \$100M to fund multi-year DOE, NIST, NSF, and DOD programs supporting various components of the MGI. The graphic used to indicate the overarching vision of the MGI is shown in \textbf{Figure \ref{fig:MGI2011}}.

A more detailed overview can be found in the 2014 Strategic Plan for achieving four sets of MGI goals:\cite{MGI_2014} (1) leading a culture shift in materials-science research to encourage and facilitate an integrated team approach; (2) integrating experiment, computation, and theory and equipping the materials community with advanced tools and techniques; (3) making digital data accessible; and (4) creating a world-class materials-science and engineering workforce that is trained for
careers in academia or industry. These efforts have had far-reaching implications on the field of computational materials science. The impact of the MGI has indeed been transformative; its unleashed momentum has continued to present-day. Even so, developing the visioned infrastructure has been slow.

The state of the situation in 2017 was well summarized in the TMS report, \textit{Building a Materials Data Infrastructure}.\cite{BuildingMDI_2017} In this report, the actual term `Materials Data Infrastructure' \textbf{(MDI)} was formally defined as consisting of ``three core digital components -- repositories, tools, and e-collaboration platforms -- as well as the technology, policies, incentives, standards, people, and activities necessary to plan, acquire, process, analyze, store, share, reuse, and dispose of materials data.'' The obstacles to MDI were also identified and grouped into four tiers, depending on the potential impact and probability of successfully overcoming them. These obstacles are reproduced below, and the items with particular relevance to data storage -- specifically object storage (described in the next section) -- are \underline{\textcolor{purple}{\textbf{colored}}}. \\

\noindent \textbf{`Tier I' problems:} \textit{High potential impact, high probability of success}
\begin{itemize}
    \item No unified MSE community approach to its diverse challenges with materials data.
    \item Mismatch between consumers and generators of specific materials data.
    \item Lack of e-collaboration platforms.
    \item \textcolor{purple}{Lack of a qualified knowledge base on data management and analytics in the MSE community.}
    \item \textcolor{purple}{Inadequate awareness of options and best practices for data storage.}
    \item \textcolor{purple}{Limited focus on sustainable long-term data storage strategies and support.}
    \item \textcolor{purple}{Insufficient participation of the computer science community in the MDI.}
    \item Many required elements and solution pathways for the MDI are not defined in enough detail.
    \item \textcolor{purple}{Insufficient mechanisms for crediting data contributors.}
    \item A pathway for industrial participation from large scale manufacturing in the MDI is unclear.
    \item Scarcity of data sharing incentives.
\end{itemize}
\textbf{`Tier II' problems:} \textit{Low potential impact, high probability of success}
\begin{itemize}
    \item \textcolor{purple}{User interfaces for uploading and downloading data can be challenging to design.}
    \item Success stories and proofs of concept are needed to demonstrate the value of data-driven materials science and engineering.
    \item \textcolor{purple}{Need for federated approaches to data sharing and storage.}
    \item \textcolor{purple}{Resources among data infrastructure providers are poorly integrated.}
    \item \textcolor{purple}{Insufficient options for long-term storage of `intermediate' data.}
\end{itemize}
\textbf{`Tier III' problems:} \textit{Low potential impact, low probability of success}
\begin{itemize}
    \item \textcolor{purple}{Lack of robust APIs of connected systems and instrumentation.}
    \item \textcolor{purple}{Complexity and disparate nature of materials data.}
    \item \textcolor{purple}{Inadequate understanding of cost associated with materials data in the short- and long-term.}
    \item Constraints of government technology export regulations.
\end{itemize}
\textbf{`Tier IV' problems:} \textit{High potential impact, low probability of success}
\begin{itemize}
    \item Retraining the existing workforce.
    \item \textcolor{purple}{Limited data repository usage and availability of tools.}
    \item Lack of a clear, unified vision of how the MDI will benefit the community.
    \item Insufficient consensus on uncertainty quantification methods in the MSE community.
    \item Lack of developed, agreed-upon ontologies for materials domains.
    \item \textcolor{purple}{Underdeveloped data management approaches for MSE knowledge.}
    \item Need for standardized components and documented workflows to enable data extraction and reuse.
    \item \textcolor{purple}{Poor interconnectivity of data platforms, which inhibits creation of materials data ecosystems.}
    \item \textcolor{purple}{Lack of mechanisms or use-metrics to indicate when old data is updated.}
    \item Lack of existing long-term sustainable business models for individual elements of the MDI.
    \item \textcolor{purple}{Inadequate IT security and outdated operating systems.}
    \item \textcolor{purple}{Lack of funding and career opportunities for materials data management.}
    \item Ambiguity of federal agency data policies.
    \item Lack of well-defined data sharing norms among publishers and funding agencies.
    \item Inadequate career incentives to share data.
    \item A data sharing culture is hindered by issues such as intellectual property and privacy.
\end{itemize}

\noindent This short-term assessment of MDI needs conveys the conclusion that there remains a great deal of work to be done. Along a similar vein, the 2040 Vision report from NASA (published in 2018) provides a an articulated roadmap of a vision to work towards.\cite{NASA_2040_Vision} A summary of the vision is shown in \textbf{Table \ref{tab:NASA2040Vision}}.
\begin{table}[h!]
    \scriptsize{

    \begin{tabular}{|>{\centering\arraybackslash}p{4cm}|>{\centering\arraybackslash}p{4cm}|}
        \hline
        \textbf{``Today'' (2018)} & \textbf{2040 Vision}\\ \hline
        Design of materials and systems is \textit{disconnected}. & Design of materials and systems is \textit{integrated}. \\ \hline
        Stages of the product development lifecycle are \textit{segmented}. & Stages of the product development lifecycle are \textit{seamlessly joined}. \\ \hline
        Tools, ontologies, and methodologies are \textit{domain-specific}. & Tools, ontologies, and methodologies are \textit{usable across the community}. \\ \hline
        Materials properties are \textit{based on empiricism}. & Materials properties are \textit{virtually determined}. \\ \hline
        Product certification relies heavily on \textit{physical testing}. & Product certification relies heavily on \textit{simulation}. \\ \hline
    \end{tabular} \\ \\
    }
    \caption{NASA 2040 Vision for the future materials data infrastructure.}.
    \label{tab:NASA2040Vision}
\end{table}
Over 450 professionals from industry, government, and academia contributed to the report. It is noted that computational modeling in industry does not usually incorporate materials modeling; simply investing in more `computational modeling' does not necessarily help promote the MDI. Accordingly, the report states: ``While significant progress has been made over the past 20 years within industry to enhance and advance system/structural design and analysis technologies, ... connection of this structural paradigm viewpoint with that of the design of materials paradigm, with all its potential impact, is still lacking and will require more work. Consequently, the authors of this report purposefully focused more attention on the issues surrounding multiscale modeling of materials and the infusion of the associated technologies (e.g., model based definitions) into the systems design/analysis paradigm. As a result, the content of this report is more biased toward computational materials science and engineering than systems/structural design and analysis.'' 
The report then goes on to identify nine key elements necessary for realizing the 2040 Vision and discusses them in depth. The ninth element, computational infrastructure, is of particular interest here, as it includes the needs of the data storage infrastructure. Specifically, Key Element 9 is comprised of the following three topics:
\begin{enumerate}
    \item Computer hardware (storage, CPU, co-processors, memory, backplane), firmware, code base, operating systems, middleware, application software, and the interoperability of these components that enable the numerical simulation of physical phenomena in ways that are useful for engineering purposes.
    \item High-bandwidth networks and software platforms to support simultaneous access to enable global collaborative engineering.
    \item HPC architectures and frameworks that use parallel/distributed, neuromorphic, quantum, cloud, machine learning, etc., processing approaches to solve large scale, computationally and data-intensive analysis and/or optimization problems.
\end{enumerate}

\noindent This report is quite lengthy, and it is not the intention here to summarize the report in its entirety. (However, the  especially interested reader may find the case studies detailed in Appendix A and B to be of particular use.) Rather, the intention is to emphasize the point that the materials data infrastructure remains far from established but that there also currently exists a great deal of momentum in the materials community to pursue its development. Highly experienced and respected members of the community have already collaborated in a variety of instances across academia and industry to produce roadmaps for what they envision the necessary infrastructure to be capable of. There are common themes in all of these. As far back as 2008, the aforementioned ICME report states:\cite{ICMEReport_2008}
\noindent ``The goal of a balanced, well-designed ICME cyberinfrastructure is to give scientists and engineers the means to do a number of things:
\begin{itemize}
    \item Link applications codes—for example, UniGraphics, PrecipiCalc, and ANSYS.
    \item Develop models that accurately predict multiscale material behaviors.
    \item Store and retrieve analytical and experimental data in common databases.
    \item Provide a repository for material models.
    \item Execute computational code anywhere computational resources are available.
    \item Visualize large-scale data.
    \item Enable local or geographically disperse collaborative research.
    \item Measure the uncertainty in a given design and the contributions of individual design parameters or sources.''
\end{itemize}

\noindent These goals of the MDI have not changed much from their inception to present day. What has changed -- in addition to the enabling technologies themselves -- is the sheer size of the community participating in these efforts. Consider, for instance, the Materials Project database, launched in 2011 at Lawrence Berkeley National Laboratory \textbf{(LBNL)} to gather -- and make publicly available --  data on a wide variety of material compounds.\cite{MaterialsProject}

\begin{figure}[h!]
    \centering
    \includegraphics[width=1\linewidth]{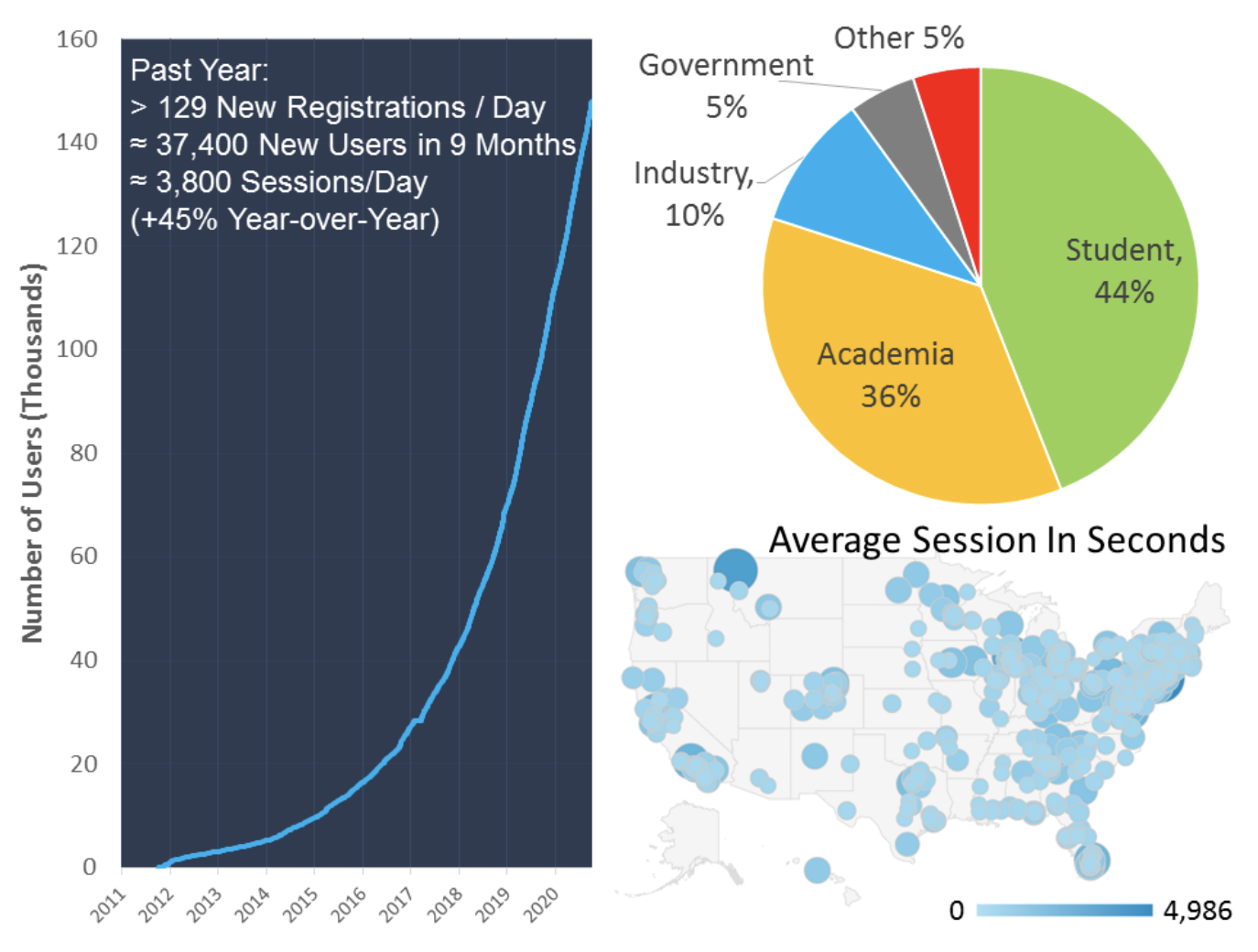}
    \caption{Growth in registered users of the Materials Project by usage and demographic, between 2011-2021.\cite{MGI_2021}}
    \label{fig:MPoverview}
\end{figure}

\begin{figure}[h!]
    \centering
    \includegraphics[width=1\linewidth]{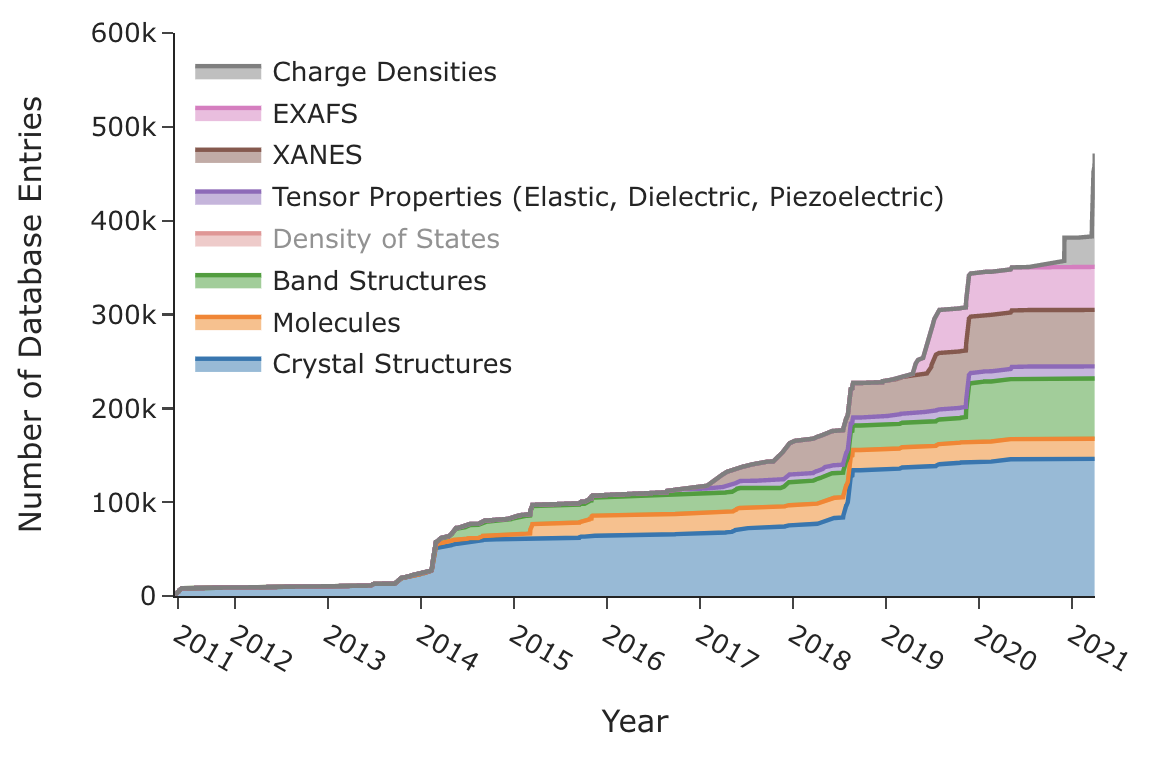}
    \caption{Growth in database entries of the Materials Project, from 2011-2021.\cite{MaterialsProject}}
    \label{fig:MPEntries}
\end{figure}

\noindent \textbf{Figure \ref{fig:MPoverview}} shows the growth in Materials Project users, and \textbf{Figure \ref{fig:MPEntries}} shows the growth in number of database entries on material compounds entered into the Materials Project database. According to the website, the Materials Project currently contains over 131,000 compounds, 49,700 molecules, and 530,000 nanoporous materials. The emergence of databases like this one have propelled the rise of the nascent field of materials informatics. However, the MDI obstacles detailed on the previous page are still salient. A decade after the 2011 launch of the Materials Genome Initiative, present-day perspectives on the state and direction of the MDI are discussed in the 2021 MGI Strategic Plan.\cite{MGI_2021} This report highlights the trends and issues that have emerged. Simply put, the situation remains rather piecemeal; there are areas where computational materials engineering has made its mark, but materials informatics and its data-driven practitioners are still a niche subset of the materials community. One of the consequences of this is, as the 2021 MGI Strategic Plan remarks, ``A sizeable fraction of the materials R\&D community does not yet see the entire value proposition in having a materials R\&D data infrastructure, and the community does not speak with a single voice on the need for it.'' However, a simple fact remains: whether the need for MDI is acknowledged or not, the data is here... and is growing. As the 2021 MGI Strategic Plan states, particularly urgent motivation for investing in MDI comes from ``the growing computational power that will soon reach the exascale (10$^{18}$ floating point operations per second—flops), experiments that are producing massive data sets, and automated synthesis techniques that can systematically produce vast numbers of new materials.'' Accordingly, one of the key strategic objectives -- for the next five years -- discussed in the report is to establish a National Materials Data Network \textbf{(NMDN)}. It will be interesting to see what comes from this effort. Technological advancements are funneling us into a data-driven world, and the competitive edge will go to whoever can leverage this data. 

In many ways, the desired features of an MDI overlap with data infrastructure needs in other scientific fields. What distinguishes the needs of the materials science community is that materials data is extremely heterogeneous. For instance:
\begin{itemize}
    \item Materials data exists across a wide range of time- and length-scales. (Consider the aforementioned PSP relationships.)
    \item Materials data has inherent uncertainty associated with it, which is both difficult to quantify and capable of propagating through higher-level analyses.
    \item Each measurement technique produces data with its own unique nuances, and measurement techniques for obtaining materials data are continually evolving.
    \item The theories are continually evolving as well, and our definitions of what constitutes a property comes from these theories.
    \item Similarly, the labeling schema of materials classes is also in flux. Our ideas of how to categorize materials have changed over time, as theoretical understanding has evolved.
    \item The data itself is inherently incomplete, always a partial picture colored with biases. If a researchers does not think to test for a given property, it will not appear in the data.
    \item Data collected for the same material system can vary wildly depending on the processing history of the material, the environmental conditions, the measurement techniques employed, and even the local region of material assessed.
    \item The phenomena that materials data aim to capture are extremely diverse.

\end{itemize}

\begin{figure}[b]
    \centering
    \includegraphics[width=0.95\linewidth]{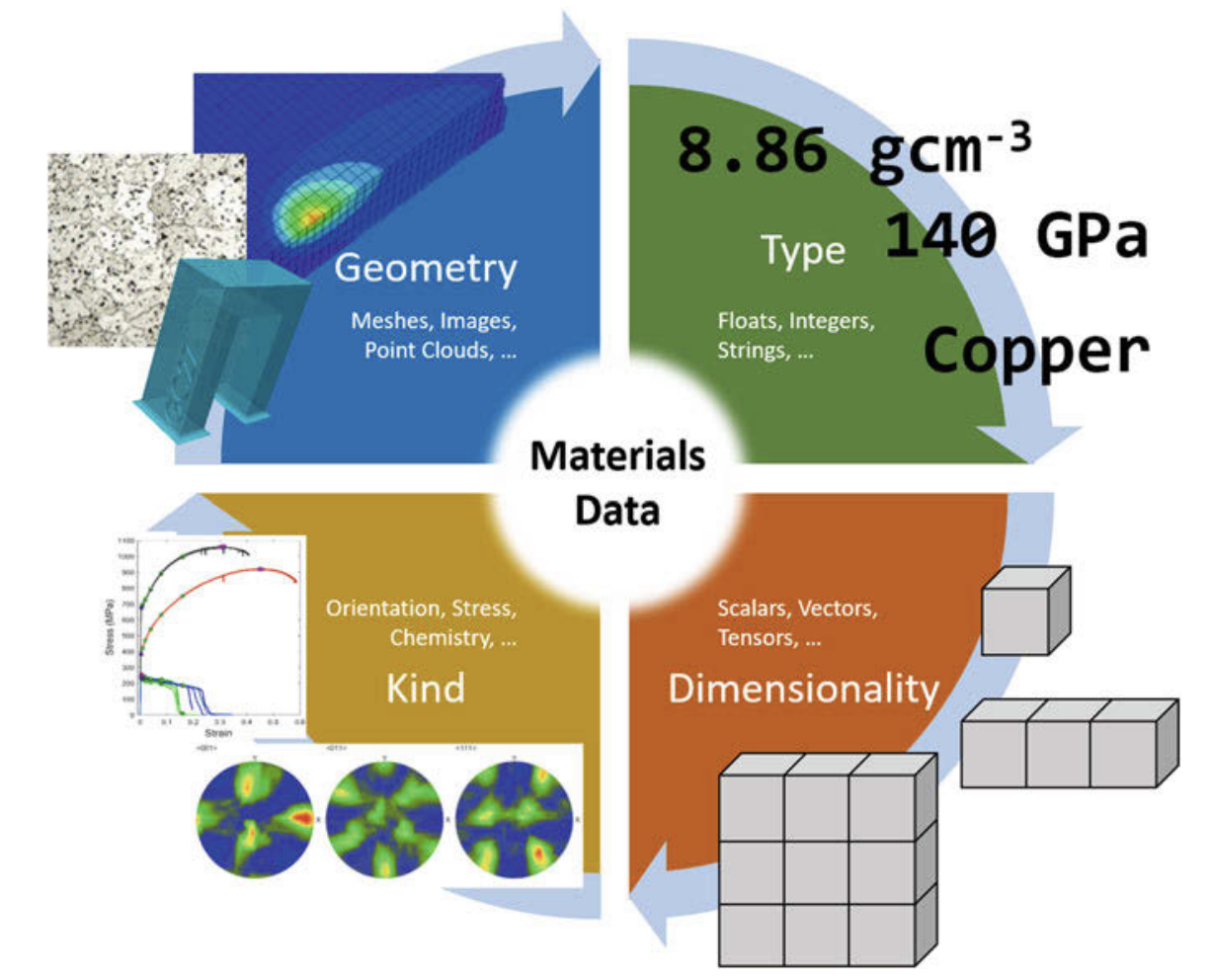}
    \caption{Schematic for four key features the define ICME data. A single dataset often exhibits variation in all four features.\cite{Donegan_2020}}
    \label{fig:MaterialsData}
\end{figure}

\noindent 

\begin{figure}[h!]
    \centering
    \includegraphics[width=0.95\linewidth]{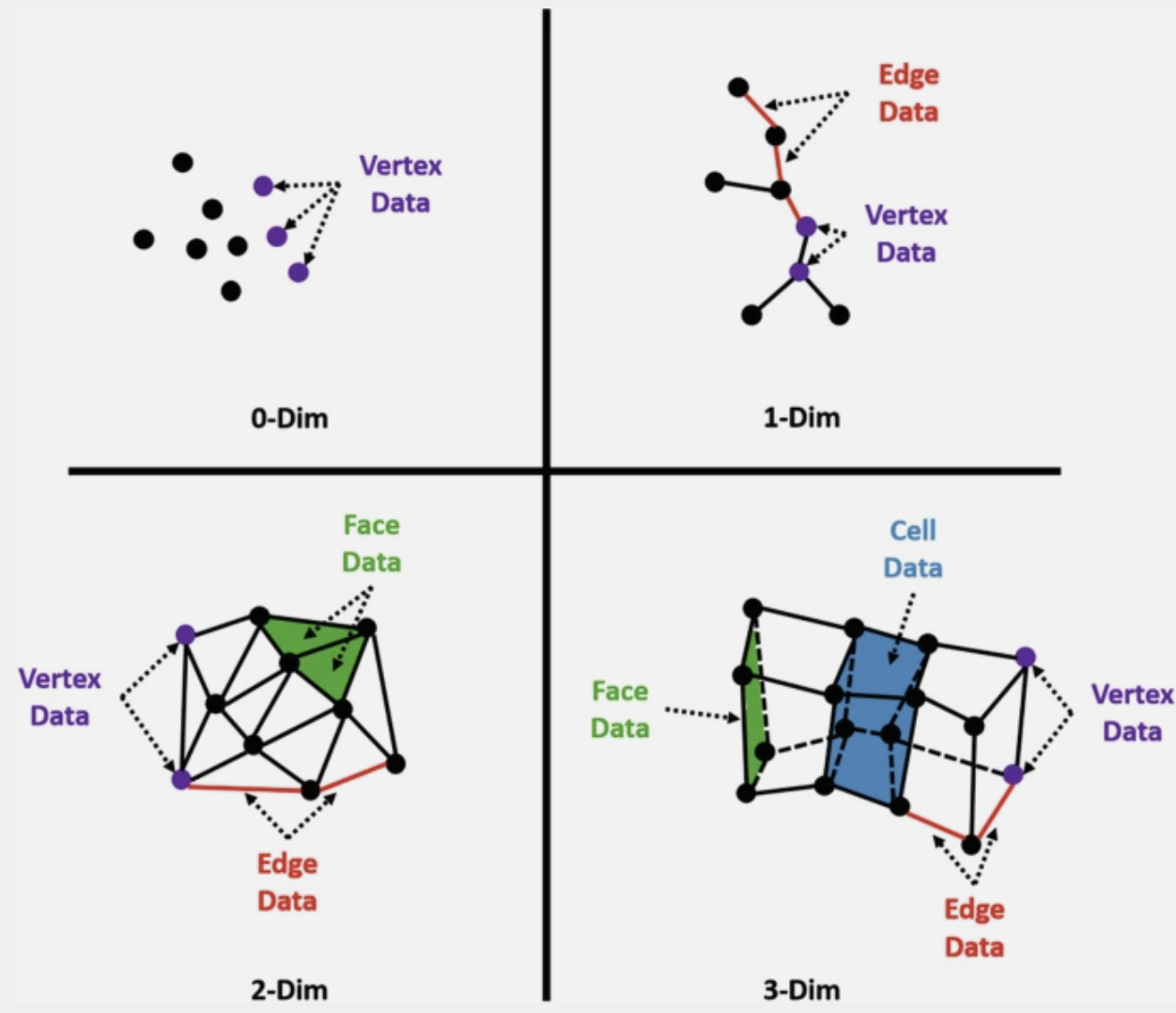}
    \caption{Illustration of how materials data may be stored on any unit element that comprises a given geometry.\cite{Donegan_2020}}
    \label{fig:GeometryData}
\end{figure}

\begin{figure}[h!]
    \centering
    \includegraphics[width=0.95\linewidth]{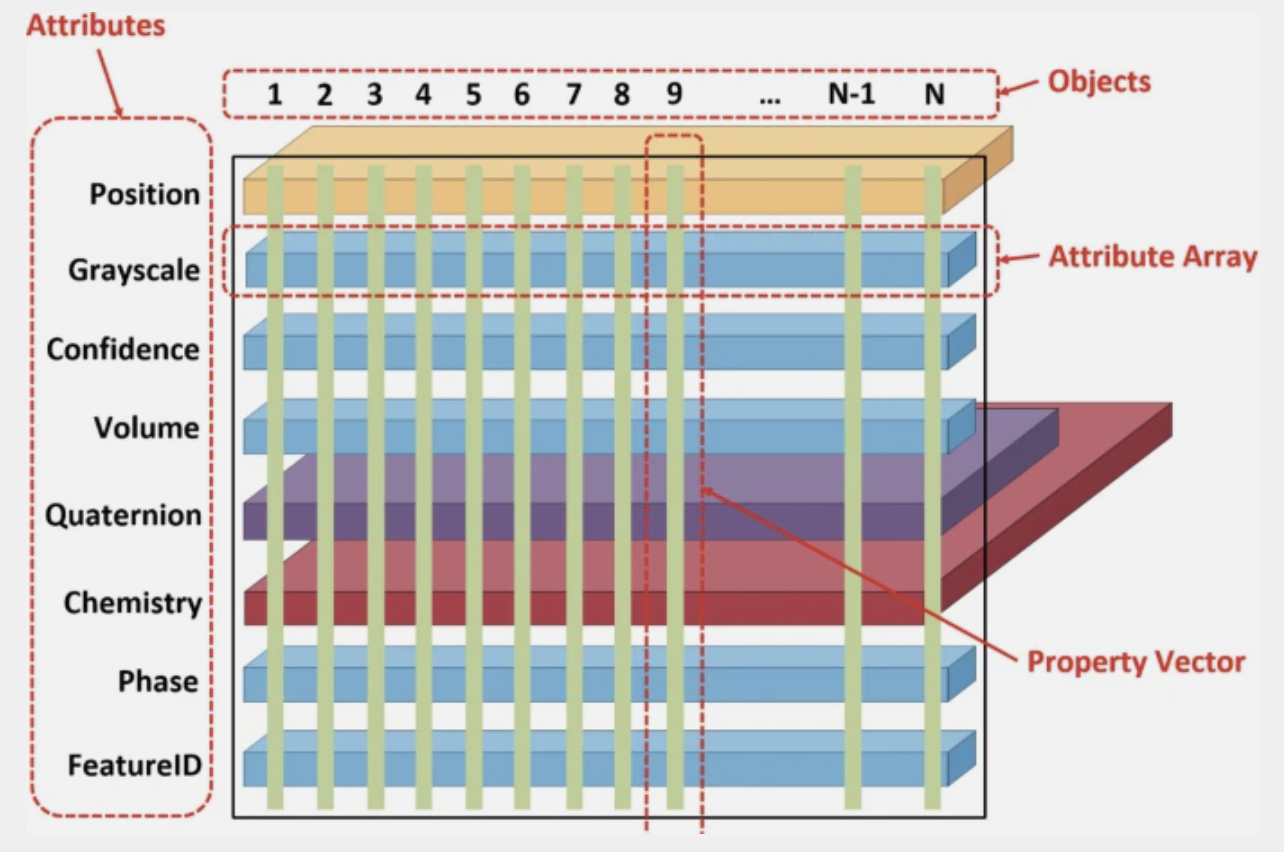}
    \caption{Attribute matrices are a way to organize data that is associated with geometries, features, and ensembles. To quote a paper from 2020: ``Attribute matrices themselves do not store heavy data; instead, they serve to define the type of data being stored and its shape. Dense data are stored in attribute array objects, which are contained within attribute matrices. There are three general types of attribute matrices: \textbf{element}, which store data associated to the unit elements of a geometry; \textbf{feature}, which store data for groups of elements; and \textbf{ensemble}, which store data for groups of features. There are four types of element attribute matrices, corresponding to the four basic unit elements: vertex, edge, face, and cell.''\cite{Donegan_2020}}
    \label{fig:AttributeArrayData}
\end{figure}

Many of the present-day obstacles of materials data are simply reflections of this inherent heterogeneity. Likewise, \textbf{Figure \ref{fig:MaterialsData}} shows an overview of the wide diversity of computational representations for materials data. Additional visualization schemes for materials data are shown in \textbf{Figure \ref{fig:GeometryData}} and \textbf{Figure \ref{fig:AttributeArrayData}}. Trying to standardize materials data is an upwards battle against entropy; even as the large centralized databases grow, so do the number of smaller, more niche-focused ones. To work in this space means using data from many different sources. Standardized file formats would make this easier, but the heterogeneity of materials also presents a hindrance towards broad community acceptance of such a file format. File formats are a form of organization, and different research groups are incentivized to organize their data differently. Regardless, there are efforts underway for standardized file formats, such as the Physical Information File \textbf{(PIF)} hierarchical schema proposed by Citrine Informatics.\cite{PIF_Citrine_2016} Also, by choosing which file formats they support, large-scale databases like the Materials Project can exert a great deal of influence. An additional difficulty with materials data is that the needs of academia (which desires free, open-source data widely shareable across the community, for the purpose of advancing human knowledge) often conflict with the needs of industry (where businesses rely on keeping their materials data proprietary, for the purpose of profit and market competitiveness). Having established the setting for which we find ourselves in today, let us now discuss the building blocks of our computational infrastructure in more detail.

\section{\textbf{\textcolor{blue}{Data Storage Infrastructure and Storage-Class Memory (SCM)}}}

At the simplest conceptual level, all problem-solving workflows consist of:
\begin{itemize}
    \item Accessing data.
    \item Performing operations with that data (creating new data).
    \item Recording data.
\end{itemize}
Accordingly, in the context of high performance computing \textbf{(HPC)} for scientific applications such as materials informatics, the main infrastructural building blocks are:
\begin{enumerate}
    \item The data storage.
    \item The compute nodes.
    \item The intermediate software that facilitate information transfer between the storage and the compute nodes.
\end{enumerate}
The space of implementation choices for these building blocks is innumerably vast, yet the choices have far-reaching implications on the computing capabilities of the end system It is tempting for researchers to become comfortable within a specific framework, taking the problem-solving infrastructure for granted and choosing to expend their focus to the problems themselves. Such disregard -- however understandable that it may be -- is irresponsible, and particularly ill-advised for HPC resources. There are well-documented tangible economic and environmental costs to computing choices, and each researcher should strive to be aware of these as they choose which tools and skills to invest in over the course of their career.

\begin{figure}[h!]
    \centering
    \includegraphics[width=1\linewidth]{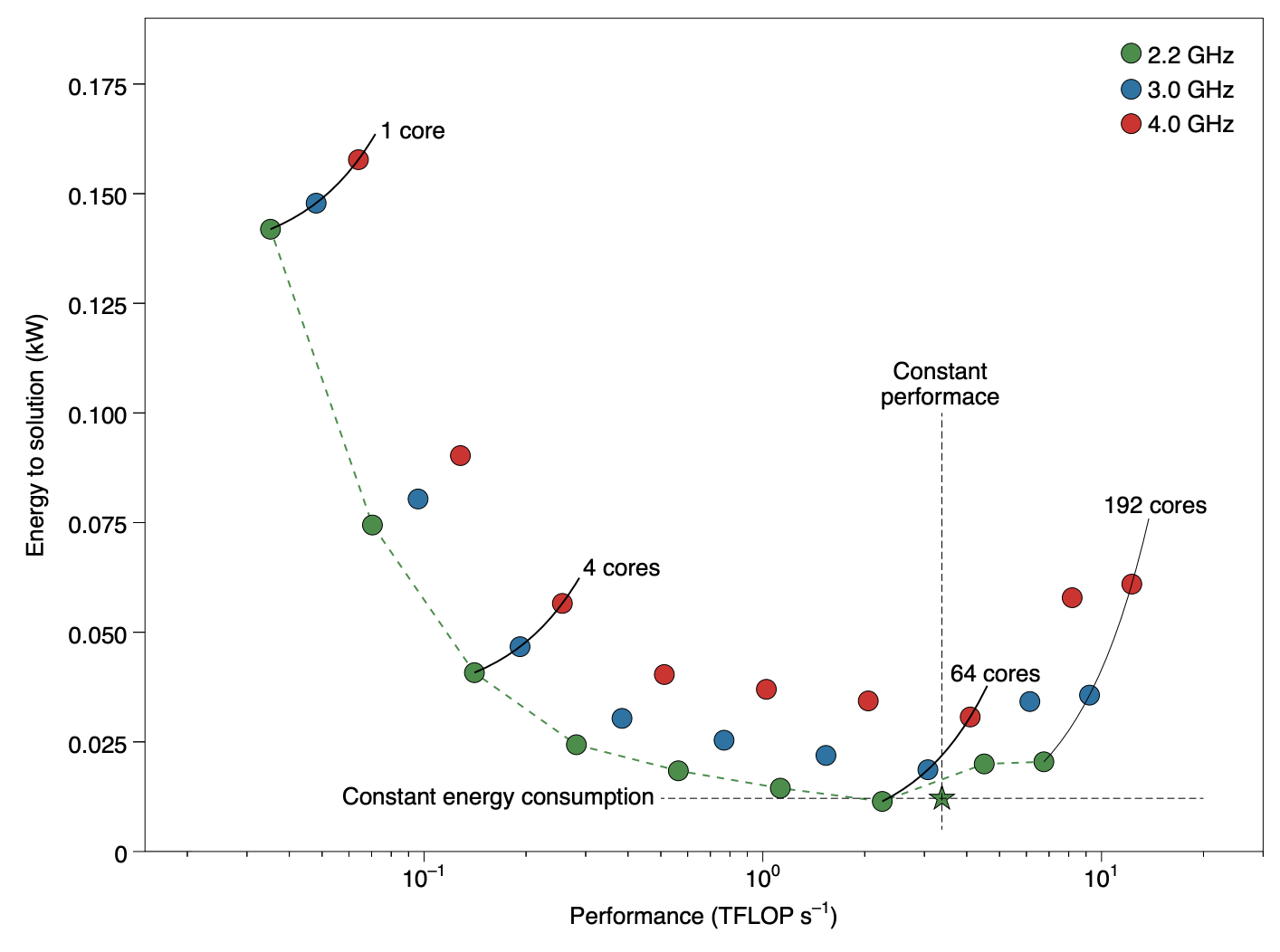}
    \caption{Energy consumption vs. performance measured from using a variety of workstation conditions to solve the N-body problem (a common astronomical calculation). The region indicated near the star is most desirable, because it gives the maximum performance for the minimum energy consumption. \cite{Zwart_2020}}
    \label{fig:CoresCost}
\end{figure}

\begin{figure}[h!]
    \centering
    \includegraphics[width=1\linewidth]{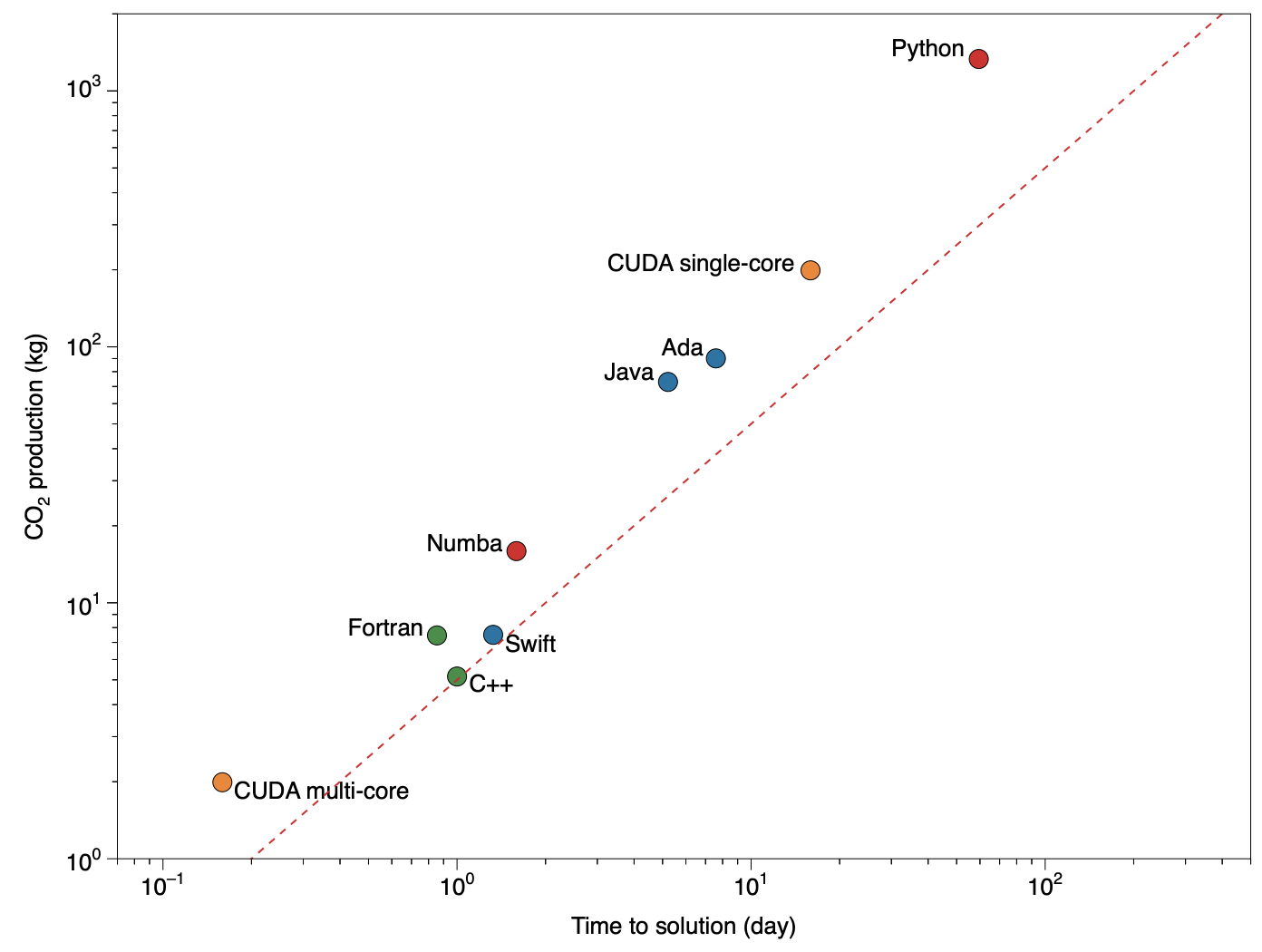}
    \caption{Footprint of various languages used to solve the N-body problem.\cite{Zwart_2020}.}
    \label{fig:ProgrammingLanguageCost}
\end{figure}

Consider a recent 2020 study which examined the ecological impact of high-performance computing in astrophysics \cite{Zwart_2020}. 
\textbf{Figure \ref{fig:CoresCost}} and \textbf{Figure \ref{fig:ProgrammingLanguageCost}} show several takeaways from this study. In \textbf{Figure \ref{fig:CoresCost}}, an N-body solver code is run on a 96-core (192 hyperthreaded) workstation, using a varying number of cores.\footnote{\scriptsize{The specific workstation used was the quad CPU 24-core (48 hyperthreaded) Intel Xeon E7-8890 v4. A `quad 24-core CPU' has 4 $\times$ 24 = 96 cores.}} The green dots show the runs, while the red and blue dots are tests from overclocking the processor. Using one core is the most inefficient scenario. Increasing the number of cores increases the energy efficiency, to a limit. The green star indicates the scenario when all physical cores are occupied (96 physical cores in this case). The workstation can run more cores due to its hyperthreading capability;\footnote{\scriptsize{Hyperthreading is when one physical core is split into multiple virtual cores.}} however, performance continues to improve by adding more cores past this point, but it does so at the cost of increased energy consumption. As a takeaway, the author recommended: ``For optimal operation, run a few ($\sim$1,000) cores on a supercomputer or a GPU-equipped workstation. When running a workstation, use as many physical cores as possible, but leave the virtual cores alone. Over-clocking reduces wall-clock time but at a greater environmental impact.'' \textbf{Figure \ref{fig:ProgrammingLanguageCost}} shows the results of implementing the N-body solver code in a variety of programming languages. Python had the highest carbon footprint.

\begin{figure}[b!]
    \centering
    \includegraphics[width=0.5\linewidth]{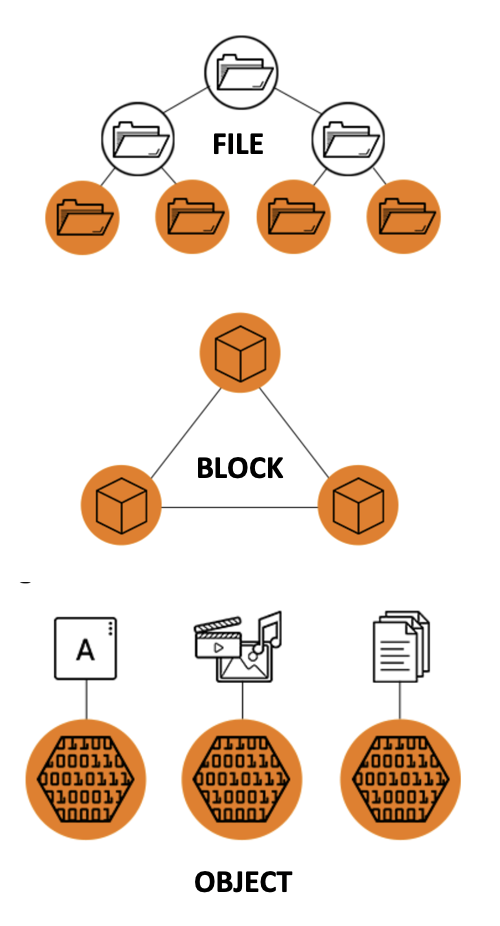}
    \caption{High-level visualization of the three primary storage classes available on the market: file, block, and object. \cite{Redhat_Graphic}}
    \label{fig:FileBlockObject}
\end{figure}

Having established that computational workflow choices have quantifiable environmental impacts and the necessity of due diligence in this space, let us transition into considering different data architecture choices, specifically for data storage. Three organizational schemes for data storage -- file, block, and object storage -- are shown in \textbf{Figure \ref{fig:FileBlockObject}}.

File storage organization is easy to visualize, since the concepts derive from the precursor days of files (as physical sheets of paper) stored in folders and cabinets. 
A key aspect of file storage is that it is hierarchical; file storage systems present a directory hierarchy where files exist within directories,
which can exist within other directories, and so on.
The data for a particular task may be spread across many related files and directories, and modern applications often build specific information layers above these primitives.\cite{AFileIsNotAFile}
The file name includes the path through the folders which one must follow in order to locate the file, as well as a file extension type determined by the application used to create it (such as .txt, .jpg, and so forth). Such a storage architecture works well in many scenarios (particularly when the number of files is not too large and when the location of the file is precisely known), which is one of the reasons that it is so ubiquitous today. The vast majority of personal computers, whether running MacOS, Windows, Linux, CentOS, etc., leverage file storage. The experience of `seeing files on your desktop' comes from filesystem-based storage. Storage allocated to hard drives is usually filesystem-based, and network-attached storage \textbf{(NAS)} devices typically run protocols that presume filesystem formats. 

\begin{figure}[t!]
    \centering
    \includegraphics[width=0.9\linewidth]{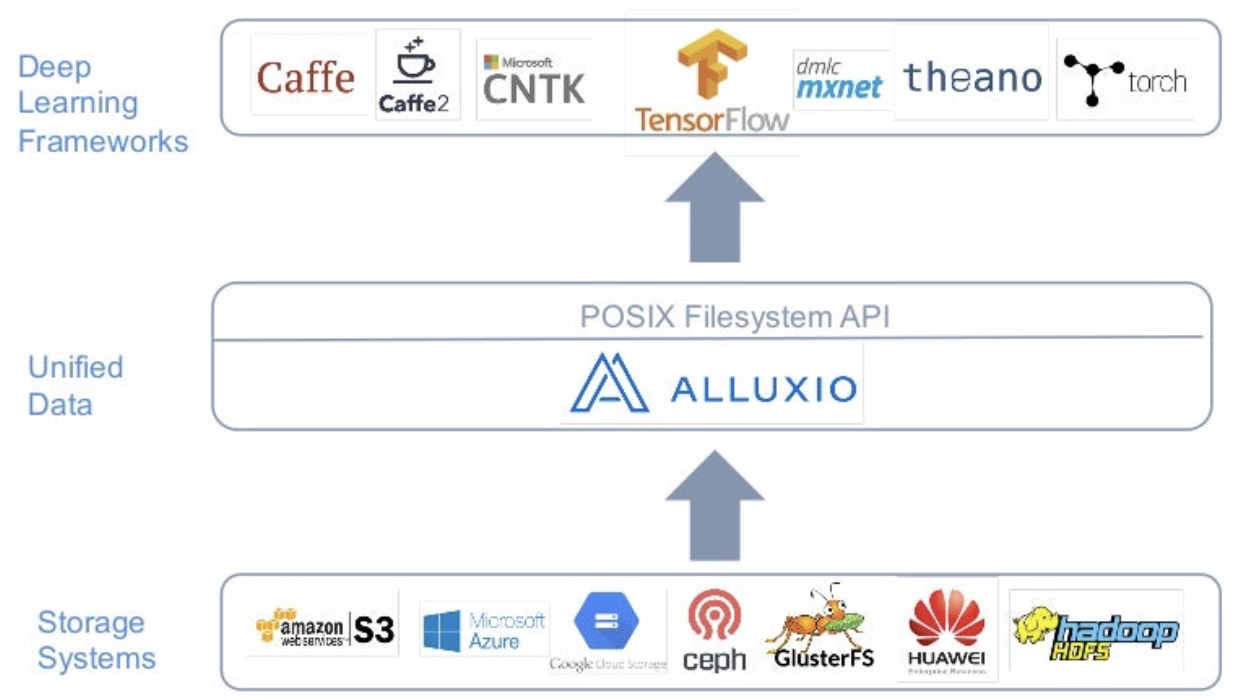}
    \caption{Graphic from Alluxio, indicating how they aim to fit into the deep learning and data storage ecosystem. Note the dependence of the various deep learning frameworks on the POSIX filesystem API.\cite{Alluxio_Graphic}}
    \label{fig:AlluxioGraphic}
\end{figure}

One of the main standards used for filesystem-based storage is POSIX, the Portable Operating System Interface, specified by IEEE; the name POSIX refers to IEEE Std 1003.n family of standards.\cite{IEEE_POSIX1}$^,$\cite{IEEE_POSIX2} POSIX is not static; it is updated regularly by the Austin Common Standards Revision Group \textbf{(CSRG)}. The main reason for adhering to POSIX is portability; if two systems are POSIX-compliant, data transfer between them is presumed to be seamless (at least in theory). Most machine learning applications rely on the POSIX framework; this enables companies like Alluxio to provide the service illustrated in \textbf{Figure \ref{fig:AlluxioGraphic}}. However, adhering to POSIX has limitations associated with it, some of which have led to significant critique.\cite{POSIX_outdated_2016}$^,$\cite{POSIX_IO_bad_2017}
To quote a study from 2018:\cite{Liu_2018} ``POSIX-based parallel filesystems provide strong consistency semantics, which many modern HPC applications do not need and do not want.'' Furthermore, a question arises of whether file storage (POSIX-compliant or otherwise) can scale to handle the enormous quantities of data able to be generated currently and in the future. In an HPC context, high performance is attained via distributed filesystems including, but not limited to: Lustre,\cite{Lustre} HDFS,\cite{HDFS} Spectrum Scale\cite{SpectrumScale}, BeeGFS.\cite{BeeGFS} Yet, as filesystems scale, they become excessively cumbersome to navigate through, and adding more capacity has its limitations. The issues with scaling filesystems have led the data science community to contemplate whether other storage systems may perform better at scale. In moving beyond file storage, it is necessary to ask, `what would files \emph{be} without the filesystem?’

An alternative to file storage is block storage. With block storage, the data is organized into fixed-sized `blocks', each possessing a unique identifier. The convenience of this approach is its flexibility; in a distributed environment, applications need not be aware that blocks are being served from different servers running different operating systems.
For instance, one block of data can be stored in a Windows environment, while another block can be stored in a Linux environment. This allows the freedom of tailoring the data across multiple environments. How the data is broken up, and how it is reassembled when a user requests that the data be retrieved, is determined by the block storage software. Since block storage does not require navigating the metadata associated with directories and file hierarchies, it can exhibit lower latency and greater performance; as such, it is often used for business-critical enterprise applications. Block storage is most commonly deployed as a storage area network \textbf{(SAN)}, which requires a server. RAID volumes, which combine multiple disks through stripping or mirroring, use block storage. A key caveat is that if the data is not highly structured, block storage loses many of its advantages. Furthermore, the capability to handle metadata is extremely limited at best.

\begin{table}[b!]
    \scriptsize{
    \begin{tabular}{|>{\centering\arraybackslash}p{4cm}|>{\centering\arraybackslash}p{4cm}|}
        \hline
        \textbf{Pro} & \textbf{Con}\\ \hline
        \multicolumn{2}{|>{\columncolor[gray]{.9}}c|}{File Storage} \\ \hline

        • Easy to access on a small scale. &  • Becomes expensive at large scales. \\
        • Users can manage their own files. & • Hard to work with unstructured data. \\
        • Allows access rights / file sharing / file locking to be set at user level. & • Challenging to manage and retrieve large numbers of files. \\
        • Familiar to most users.  &  \\ \hline

        \multicolumn{2}{|>{\columncolor[gray]{.9}}c|}{Block Storage} \\ \hline

        • Fast (\textit{high performance with low latency for data retrieval when blocks are stored locally or close together}). &  • Lack of metadata (\textit{block storage does not contain metadata, making it less useful for unstructured data storage}). \\
        • Reliable (\textit{block storage has a low fail rate, because blocks are stored in self-contained units}). &  • Not searchable (\textit{large volumes of block data become unmanageable because of limited search capabilities}). \\
        • Easy to modify  (\textit{changing a block does not require creating a new block, only a new version}). &  • High cost (\textit{purchasing additional block storage is expensive and often cost-prohibitive at a high scale}). \\ \hline
        
        \multicolumn{2}{|>{\columncolor[gray]{.9}}c|}{Object Storage} \\ \hline

        • Handles large amounts of unstructured data (\textit{which is increasingly important for AI/ML and big data analytics}). & • Cannot lock files (\textit{all users with access to the cloud, network, or hardware can access the objects stored there}).  \\ 
        • Affordable consumption model (\textit{instead of paying in advance for a set amount of storage space,  you pay only for the object storage you need}). & • Slower performance than other storage types (\textit{the file format usually requires more processing time than file storage and block storage}). \\ 
        • Uses metadata (\textit{because metadata is stored with the objects, users can quickly gain value from data and more easily retrieve the object they need}). & • Cannot modify a single portion of a file (\textit{once an object is created, you cannot change the object, you can only recreate a new object}).  \\
        • Advanced search capabilities. &   \\
        • Unlimited scalability (\textit{you can add as much additional storage as you need, even petabytes or more}). &  \\ \hline

    \end{tabular} \\ \\
    }
    \caption{Simplified overview of the pros and cons of various types of\\storage, as described by IBM.\cite{IBM_FileBlockObject}}
    \label{tab:FileBlockObjectIBM}
\end{table}

Object storage is another alternative. (A pro/con list from IBM on file, block, and object storage is consolidated below in \textbf{Table \ref{tab:FileBlockObjectIBM}}.) 
With object storage, data and its associated metadata are managed as `objects’ that are allocated unique identifiers in a flat namespace; i.e. without any hierarchy of directories or sub-directories. Objects can contain any type of data, structured or unstructured. This associated metadata provides considerable flexibility for better indexing and management. However, keeping the metadata consistent during updates may be expensive, so many systems discourage or prohibit modifying an object once written. To support updates, applications need to create new objects with adjusted content and delete the old ones.
As such, the ideal use of object storage is in `write once, read many times’ situations, such as accessing video and photo libraries. (Netflix, for instance, relies on Amazon's `S3' object storage to store their content.\cite{Netflix}) Resiliency and scalability are the hallmarks of object storage; the hardware used by the object storage need not be confined to one geographic location, and by adding more hardware devices to the storage pools, object storage can scale without limit. When dealing with Big Data, the failure of individual system components is not an `if’, but rather a question of `when’ and ‘how often’. In addition to the resiliency that arises from the geographic distribution of the data, object storage can offer data-recovery features such as erasure coding, which allows data to be reconstructed in the event of corruption or drive failure.

\begin{figure}[b!]
    \centering
    \includegraphics[width=1\linewidth]{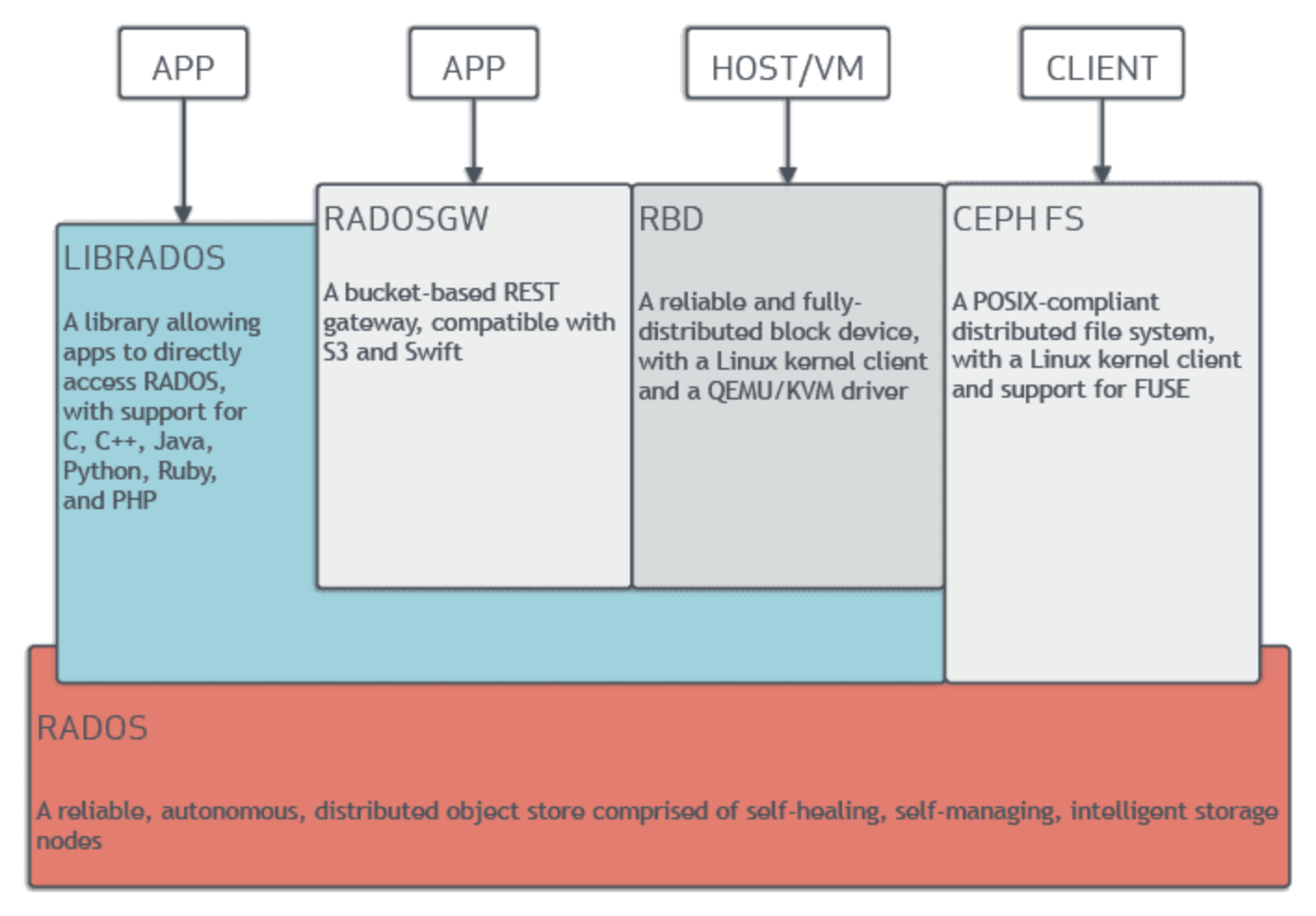}
    \caption{Ceph software stack.\cite{CephSoftwareStack}}
    \label{fig:CephStack}
\end{figure}

The desirable features of resiliency and scalability have led to object storage's rise as the predominant storage type in the cloud computing space. Amazon was one of the earliest commercial object storage providers, debuting its `S3' storage in 2006. \cite{Amazon_CloudStorage}$^,$\cite{Amazon_S3}
Other object storage systems include: Microsoft Azure (`Blob' storage),\cite{MicrosoftAzure} Google Cloud,\cite{GoogleCloud} MinIO,\cite{MinIO} Ceph,\cite{Ceph} OpenStack Swift,\cite{Swift} and StorNext.\cite{StorNext} Object storage can also be made to mimic other types of storage, via use of appropriate interfaces. The Ceph software stack, for instance (shown in \textbf{Figure \ref{fig:CephStack}}), provides interfaces compatible with object, block, and file storage.
While object storage is relatively mainstream in the cloud computing space, it has yet to gain a prominent role for HPC. This is because HPC is generally concerned with performance above all else, and object stores have a reputation of exhibiting higher latencies than file or block storage. 
Considered unsuitable for intensive, continuous I/O, object storage has been relatively confined to the role of archives, back-ups, and long-term storage. However, its prominence in the realm of cloud computing has motivated a push towards improved performance, and in leveraging the technological advances made by present-day storage hardware, object storage has been presented an opportunity to rebrand itself. 

Consider the `Distributed Asynchronous Object Storage' \textbf{(DAOS)} system developed by Intel.\cite{DAOS} DAOS is open-source; the first community release was on June 18, 2020, and `DAOS version 2.0' was released April 20, 2021.\cite{DAOS_2021_YoutubeUpdate} DAOS provides a compelling contradiction to the claim that object storage systems display low performance. The IO500 List is a community ranking of supercomputer performance and storage technologies for submissions from around the world, organized by the Virtual Institute for I/O \textbf{(VI4IO)}.\cite{IO500} DAOS submissions have consistently been among the top contenders on this list since 2020.\cite{DAOS_paper_2020} It also came in first prize for Bandwidth on the `10-node challenge' in 2021, at nearly 400 GIB/s (almost twice as high as the second place contender).\cite{DAOS_2021_YoutubeUpdate} What was the key to their success? The fact that DAOS is the culmination of nearly a decade of work by Intel, focusing specifically on storage for high-performance computing, is part of it.\cite{BriefOverview_2019} The fact that DAOS allows operations to bypass the Linux kernel, saving time, is another part. But the main reason DAOS performs so well is that it was built to leverage next-generation storage hardware, like storage-class memory \textbf{(SCM)}.


\begin{figure}[h!]
    \centering
    \includegraphics[width=1\linewidth]{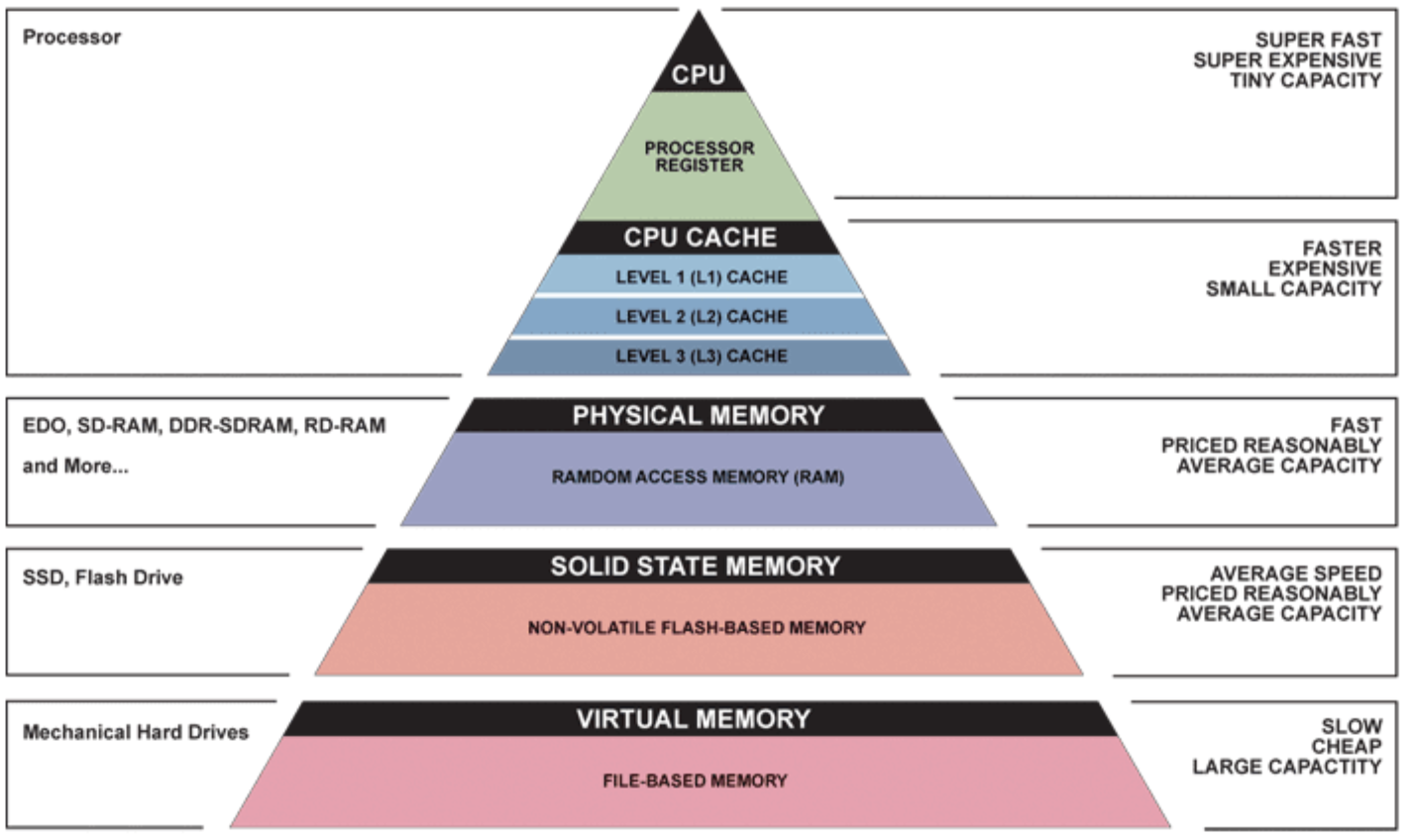}
    \caption{Simplified hierarchy of computer memory.\cite{MemoryHierarchyPic}}
    \label{fig:MemoryHierarchy}
\end{figure}

In discussing next-generation storage technologies, some context on the existing physical hardware can be helpful. In a computer, the central processing unit \textbf{(CPU)} is what performs the computation and data manipulation. One scheme for ranking the tiers of storage, displayed in \textbf{Figure \ref{fig:MemoryHierarchy}}, comes from the distance between the data storage device and the CPU. 
The general trend is that the higher up in the pyramid (at the so-called `hot tiers’ of memory), the faster the data can be accessed, but the trade-off is less data stability and less storage size. The lower down in the pyramid (at the so-called `cold tiers’ of memory), the slower it is to access the data, but more data can be stored and with less likelihood of data loss. Several types of computer memory devices (listed in order of `hot' to `cold') include, but are not limited to:
\begin{itemize}
    \item Registers
    \item Cache
    \item RAM
    \item SCM \textit{(aka persistent memory)}
    \item Flash memory
    \item Hard drive
    \item Tape
\end{itemize}

\begin{figure}[h!]
    \centering
    \includegraphics[width=0.6\linewidth]{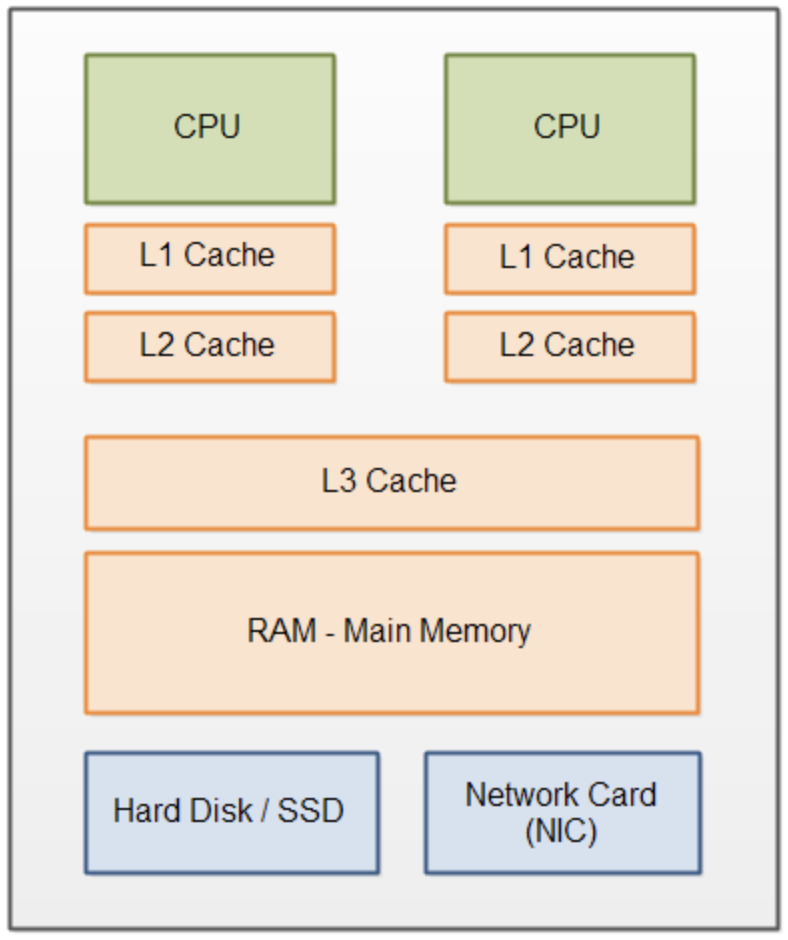}
    \caption{Schematic to illustrate the difference between cache and RAM chips as implemented physically on a computer motherboard.\cite{CachePic}}
    \label{fig:CachePic}
\end{figure}

\noindent Registers are high-speed memory built directly into the CPU chip. However, register memory has a very small capacity and is limited to facilitating low-level instructions. The next tier, cache memory, has a significant impact on the performance speeds of the CPU. Located in close physical proximity to the CPU, the cache memory stores copies of frequently-accessed data from the main memory. Since accessing data stored in the cache is faster for the CPU to do than accessing memory stored in the main memory, the cache memory helps to speed up the computation done by the CPU. Cache memory is designated into three tiers (L1, L2, L3), depending on the capacity and how close the chip is to the CPU. A simplified visualization of these types of memory physically mounted on a computer motherboard is shown in \textbf{Figure \ref{fig:CachePic}}.

\noindent Demonstrating that the tiers of memory classification often overlap, the physical implementation of most cache memory is static RAM \textbf{(SRAM)}. The terms `SRAM' and `cache memory' are often used interchangeably. 

\begin{figure}[h!]
    \centering
    \includegraphics[width=1\linewidth]{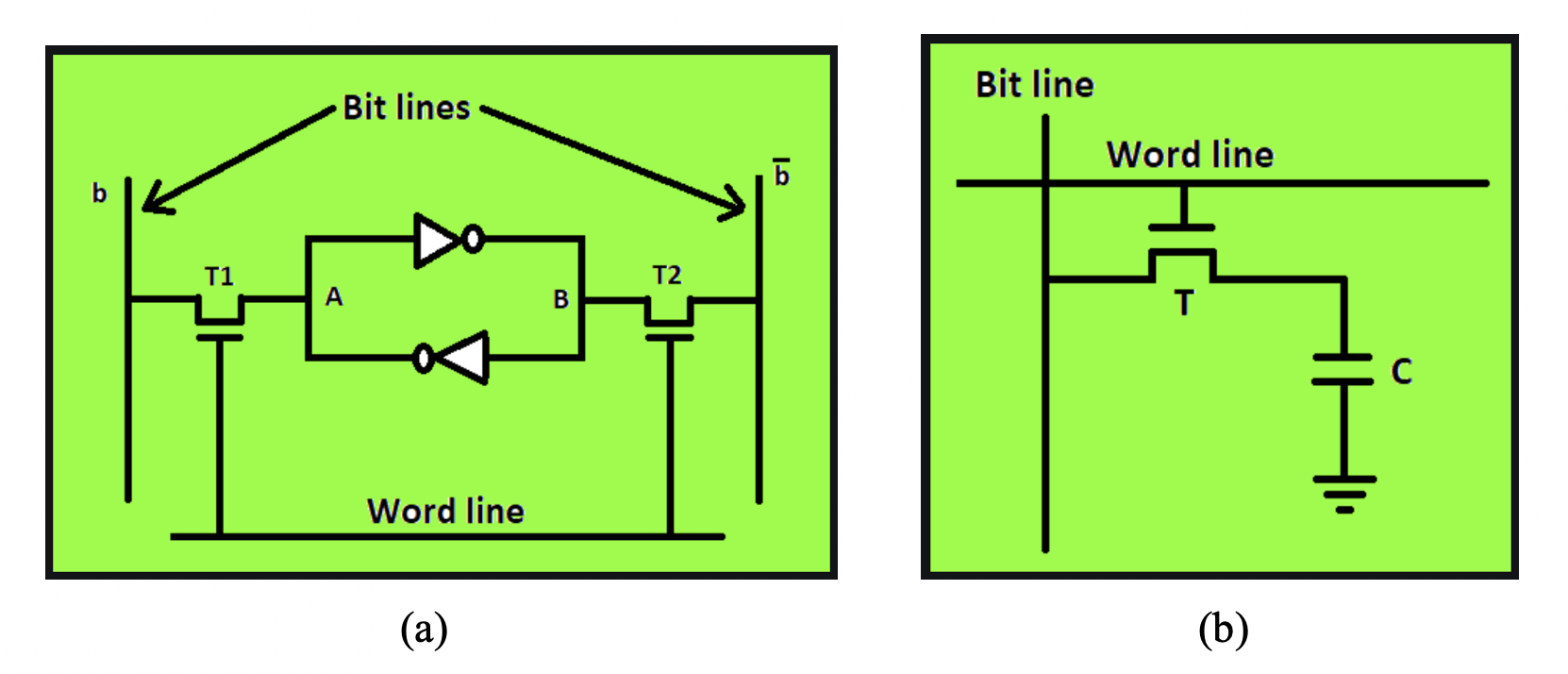}
    \caption{Cell diagrams of (a) SRAM and (b) DRAM. \cite{SRAMvsDRAMpic}}
    \label{fig:SRAMvsDRAM}
\end{figure}

Random access memory \textbf{(RAM)} is a chip on the motherboard that stores the temporary working data used by the CPU, i.e. the real-time data being read and written by programs. The term `main memory' of a computer usually refers to RAM. Inside of a RAM chip, a technology known as a dual in-line memory module \textbf{(DIMM)} is found. These are printed electronic circuits that come in a variety of configurations. The two main types of RAM that a DIMM provides are: static RAM \textbf{(SRAM)} and dynamic RAM \textbf{(DRAM)}. Cell diagrams illustrating the difference in circuit configuration for SRAM and DRAM are shown in \textbf{Figure \ref{fig:SRAMvsDRAM}}.

\noindent As shown in the figure, SRAM cells use multiple transistors, while DRAM cells generally consist of a transistor and a capacitor.\footnote{\scriptsize{The transistor type is usually a Metal-Oxide Semiconductor Field Effect Transistor \textbf{(MOSFET)}, but other types -- such as the Bipolar Junction Transistor \textbf{(BJT)} -- can also be employed.}} The data in SRAM is stored in an orchestrated communication between the transistors, called a `flip flop' gate. The data in DRAM is stored via the charge on the capacitor. Since the capacitor will leak electric charge over time, the charge needs to be periodically refreshed to prevent data loss. An important aspect of SRAM and DRAM is that they both require a power supply to function. When power is turned off, data stored in RAM is lost. This is known as volatile memory.

\begin{figure}[h!]
    \centering
    \includegraphics[width=0.9\linewidth]{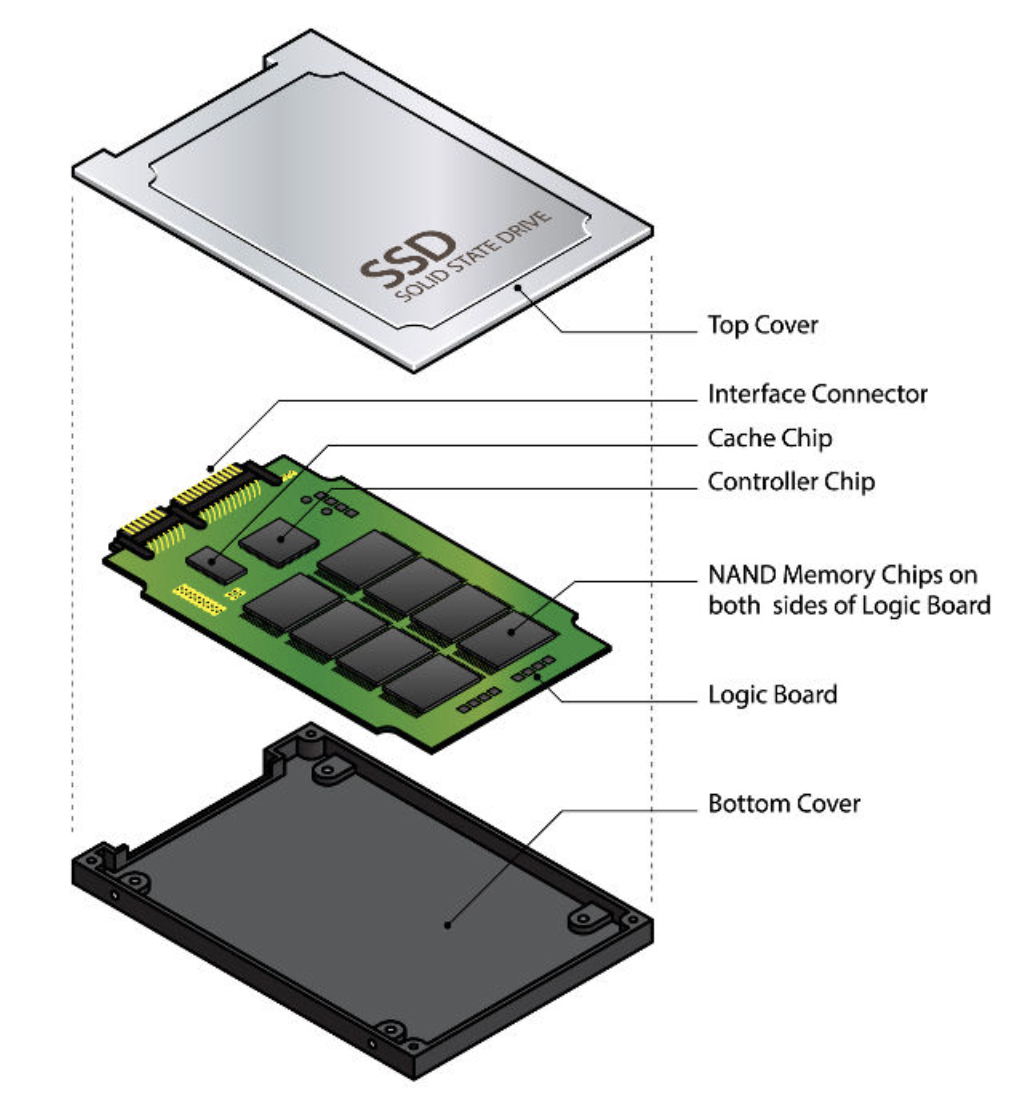}
    \caption{Deconstruction of a generic solid state drive \textbf{(SSD)}, showing its usage of flash memory, in the form of NAND memory chips. \cite{SSDpic}}
    \label{fig:SSDpic}
\end{figure}

Non-volatile memory retains its data even when power is removed. This is not to say that non-volatile memory is invulnerable; the various physical implementations of non-volatile memory have varying strengths and weaknesses that affect their suitability for different use cases. One of the main categories of non-volatile memory devices is flash. Solid state drives (\textbf{SSD}s) usually use flash memory, as demonstrated in \textbf{Figure \ref{fig:SSDpic}}. 

\noindent Aside from SSDs, other devices that use flash memory include: multi-media cards (\textbf{MMC}s), BIOS chips, and USB flash drives. 

\begin{figure}[t!]
    \centering
    \includegraphics[width=1\linewidth]{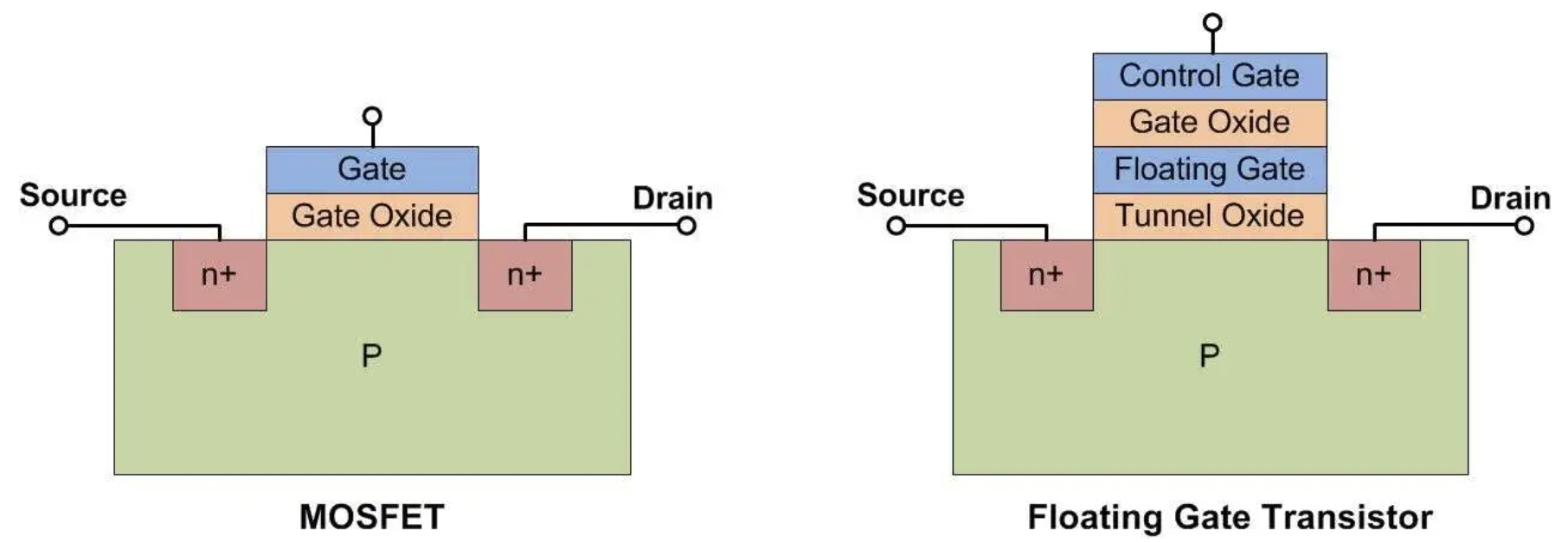}
    \caption{Similarities between the MOSFET and the floating gate transistor. The MOSFET is often used for RAM, while the floating gate transistor is used for flash memory. \cite{FloatingGatePic}}
    \label{fig:floatingMOSFETpic}
\end{figure}

Flash memory relies on a `floating gate' transistor that is essentially a modified version of the MOSFET, as shown in \textbf{Figure \ref{fig:floatingMOSFETpic}}. Because the gate is electrically insulated, the data (which is stored in the form of the electrical charge on the gate) is protected even if an external power source is absent. Placing electrons within the floating gate (charging the gate) is a `write' -- aka `program' -- operation, and removing the electrons (discharging the gate) is an `erase' operation. Two main types of flash memory are NAND and NOR, so named for their implemented architectures within the memory cell -- which resemble NAND (``not AND'') and NOR (``not OR'') logic gates, respectively. Even though it is non-volatile, flash memory can only withstand a finite number of program/erase \textbf{(P/E)} cycles; the erase operation relies on electron tunneling across the metal oxide layer, which has the side effect of degrading the metal oxide over time. The lifetime of a flash storage device is usually expressed in terms of the number of estimated P/E cycles that can be performed before the charged state becomes indistinguishable from the uncharged state. Although, it is noted that the P/E cycle-based aging marker is not always a reliable wear indicator.\cite{NAND_AgingMarker_2021} SSD lifetimes have been quoted at a wide range of values, from over a decade to as little as a year, depending on P/E cycles and operating conditions. \cite{SSD_LifeExpectancy1}$^,$\cite{SSD_LifeExpectancy2}$^,$\cite{SSD_LifeExpectancy3}
The storage temperature also affects the data integrity, as the degradation of the storage device can be drastically accelerated at warmer temperatures.
In extreme worst case scenarios, the data on an SSD might degrade after only seven days.\cite{SSD_7days} Humidity is particularly detrimental as well; a recent study found that humidity significantly affected the degradation of SSDs, even when operated at low temperature.\cite{SSD_Humidity}
For these reasons, SSDs should be stored in climate-controlled environments, especially if implemented for enterprise scale.

At a lower tier of the computer memory hierarchy, below flash storage, lies the hard drive. Inside of a hard drive is a spinning, circular `platter' with a thin magnetic coating. The surface of the platter can be allocated into small, discrete individual regions that are either magnetized (to store a 1) or demagnetized (to store a 0). These individual regions are small, as a single platter can store over a trillion bits of magnetic memory on it. Reading and writing data on a hard drive is also a mechanical process; for the magnetically-stored data to be accessed, the disk needs to rapidly spin, exposing each of the regions to the probe `head' (mounted on an actuator arm). This is visually similar to a needle on a record player. On one hand, hard drives can retain their data at a \emph{much} wider range of temperature and humidity environments than flash memory.
On the other hand, hard drives are more susceptible to physical damage; for instance, dropping an SSD or a flash drive is not likely to compromise the data, but dropping a hard drive can destroy the integrity of the entire disk. According to IBM:\cite{IBM_HDD} ``The longevity under constant use for an internal hard drive is three to five years. The lifespan can be longer if the device is an external hard drive and stored in a controlled space ... HDDs [hard disk drives] are a better long-term storage device. SDDs tend to be less reliable for long-term storage because of data leaks that begin after a year of being unpowered.''
The magnetic memory in a hard drive is quite durable and resilient to data loss; rather, it is the mechanical components that are generally responsible for failures that occur. Another storage medium that relies on magnetic memory is tape. Tape storage consists of a thin plastic ribbon with a magnetic oxide coating (a mixture of powdered iron oxide particles, lubricant, stabilizer, binder, and pigments), wound into spools. Tape is one of the slowest types of storage to access, because the data encoded onto the ribbon must be read in an entirely sequential fashion. However, it has quite good longevity and happens to be fairly inexpensive, making it a popular choice for long-term storage.

\begin{figure}[h!]
    \centering
    \includegraphics[width=1\linewidth]{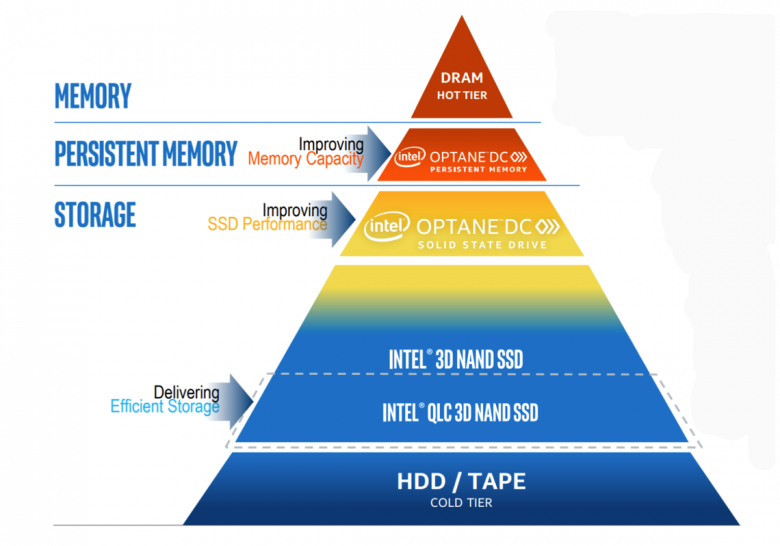}
    \caption{Location of Intel Optane persistent memory in the memory hierarchy. Note the conceptual illustration of `hot memory' and `cold memory'. \cite{HotColdOptanePic}}
    \label{fig:HotColdTiers}
\end{figure}

Having provided a brief overview of several of the main tiers in the memory hierarchy, let us now consider a relatively nascent tier that has been interjected between DRAM (volatile) and flash memory (non-volatile). This nascent tier is called storage class memory \textbf{(SCM)}, but it also goes by the name of `persistent memory' \textbf{(PMem)}. SCM has the dual advantages of being non-volatile -- like flash -- and relatively quick to access -- like DRAM. The archetypal enabling physical medium for this technology is a hardware design called `3D XPoint' (pronounced "3D cross-point"). 3D XPoint emerged after a decade-long collaboration between the companies Micron and Intel; they began development on it in 2006 and officially released it to market in 2016, under the name of `Intel Optane'.\cite{OptaneRelease} \textbf{Figure \ref{fig:HotColdTiers}} illustrates where Intel Optane was envisioned to dwell within the memory hierarchy; it can be used for both working memory (improving the performance of the CPU) and storage (improving the overall capacity of the system). 

\begin{figure}[h!]
    \centering
    \includegraphics[width=0.45\linewidth]{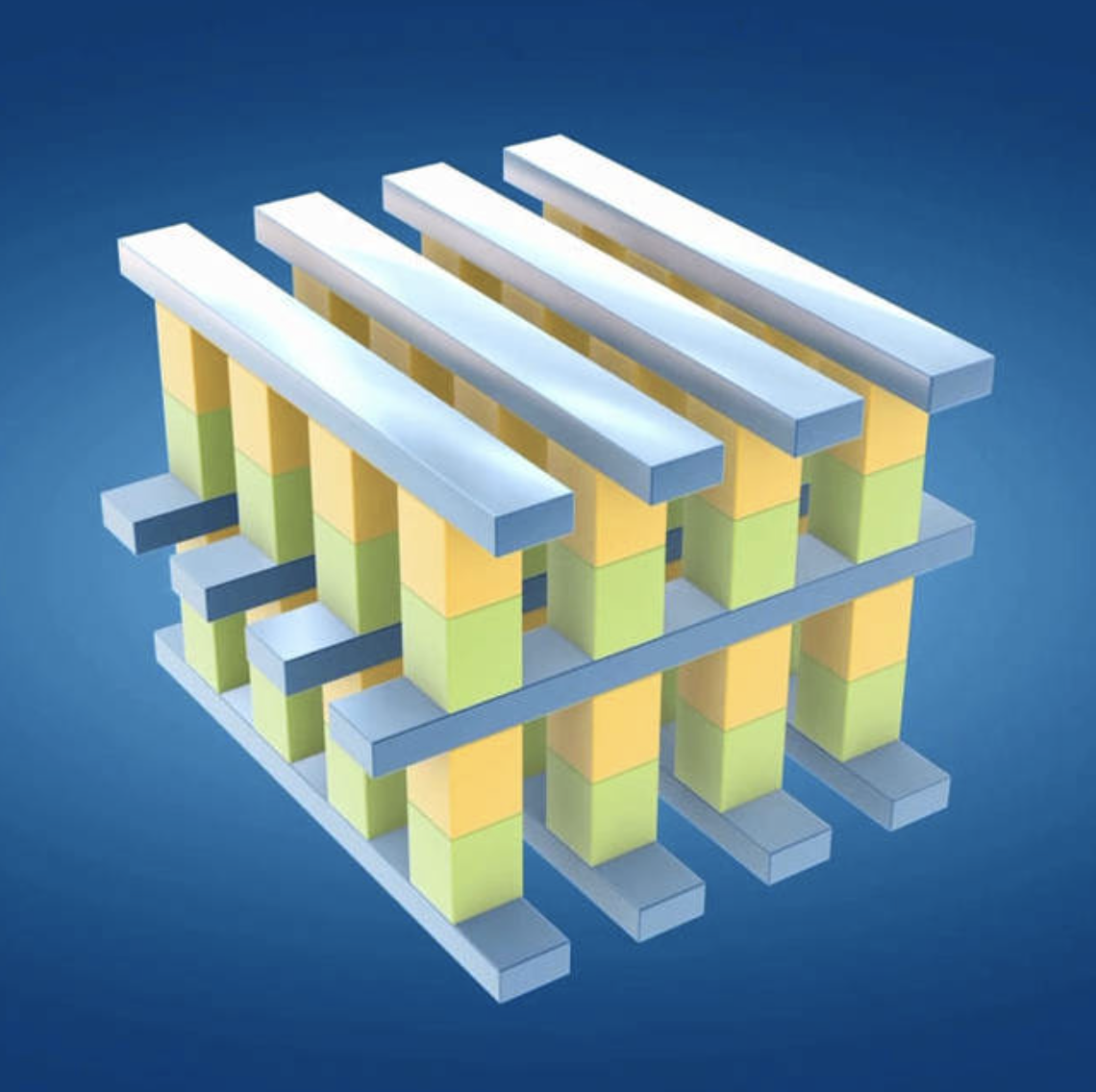}
    \caption{The 3-dimensional structure leveraged by 3D XPoint. \cite{OptaneGenericPic}}
    \label{fig:xpointGeneric}
\end{figure}

\noindent An interesting aspect of Intel Optane is that it can be implemented in a variety of `modes': memory mode, app-direct mode, and mixed mode.\cite{Optane3Modes}$^,$ \cite{Optane3Modes2} When used in memory mode, the Optane PMem device acts as RAM, and (used in conjunction with actual RAM) the actual RAM can act as cache. This has the benefit of greatly increasing the working memory available. However, somewhat misleadingly, Optane in memory mode is not persistent; it behaves as volatile memory. This is not because the storage medium itself is physically volatile. Rather, it uses a cryptographic key that gets flushed if the power turns off. Optane used in app-direct mode, in contrast, does behave as non-volatile PMem. In app-direct mode, the Optane PMem device does not act as RAM; it acts as storage. As storage, it can have a namespace or a file system laid atop it. The benefit of using Optane PMem here, in app-direct mode, is that it provides an ultra-fast connection between the storage and the processor (which is ideal for big data workloads, in-memory databases, and so forth). Mixed mode is a combination of memory mode and app-direct mode. Since Optane features can vary significantly depending on how the hardware is implemented, the marketing around it has been exceedingly vague, exceedingly broad, and -- at the same time -- highly technical. It should be noted that Intel and Micron have been quite secretive on what the physical medium that enables this technology actually is. A generic graphic usually shown to illustrate 3D XPoint is shown in \textbf{Figure \ref{fig:xpointGeneric}}.

\begin{figure}[t!]
    \centering
    \includegraphics[width=0.9\linewidth]{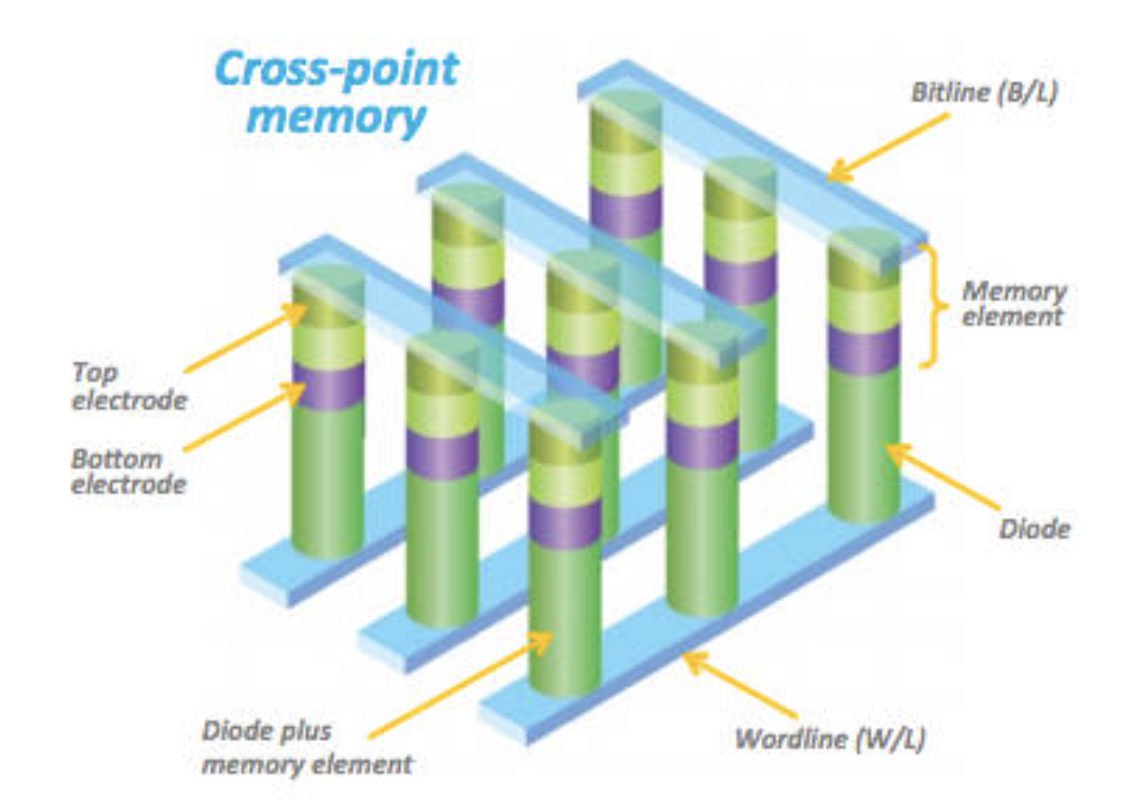}
    \caption{A labeled, simplified graphic of the 3-dimensional structure used by Micron for their 3D XPoint technology.\cite{OptaneConferencePic}}
    \label{fig:XpointConferenceSlide}
\end{figure}

\noindent A labeled graphic presented at a conference (back in 2015) by one of Micron's executives is shown in \textbf{Figure \ref{fig:XpointConferenceSlide}}. 
 Several features of 3D XPoint are described as follows:\cite{OptaneDescription}
\begin{itemize}
    \item Perpendicular wires connect submicroscopic columns; an individual memory cell can be addressed by selecting its top and bottom wire.
    \item The amount of voltage sent to each 3D XPoint `selector' enables its memory cell to be written to or read without requiring a transistor.
    \item 3D XPoint is not significantly impacted by the number of write cycles it can endure, making it more durable.
\end{itemize}

\begin{figure}[b]
    \centering
    \includegraphics[width=1\linewidth]{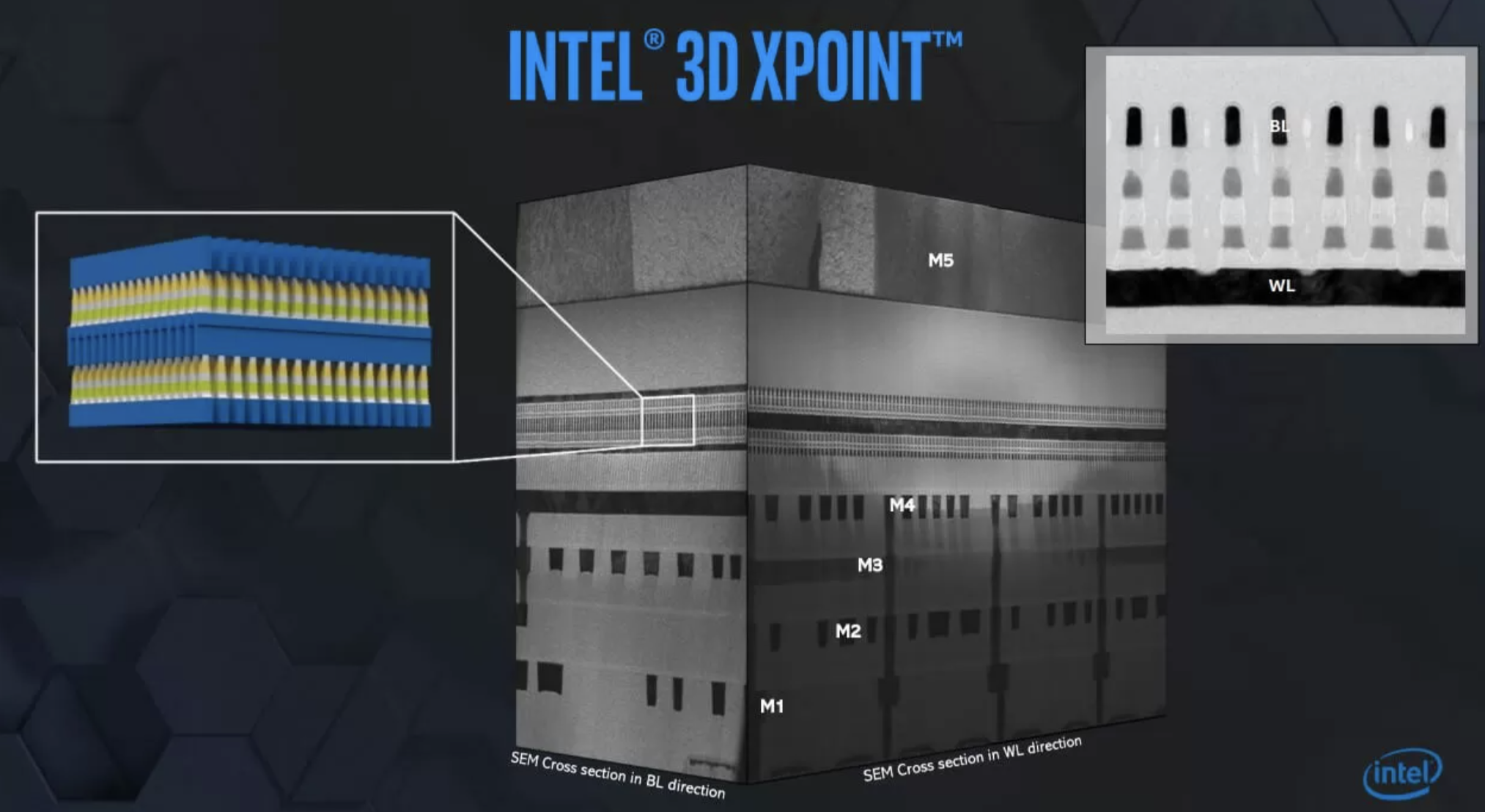}
    \caption{SEM images of 3D XPoint hardware, viewed along two directions. \cite{XpointSEMpic2}}
    \label{fig:xpointSEM2}
\end{figure}

\begin{figure}[b!]
    \centering
    \includegraphics[width=1\linewidth]{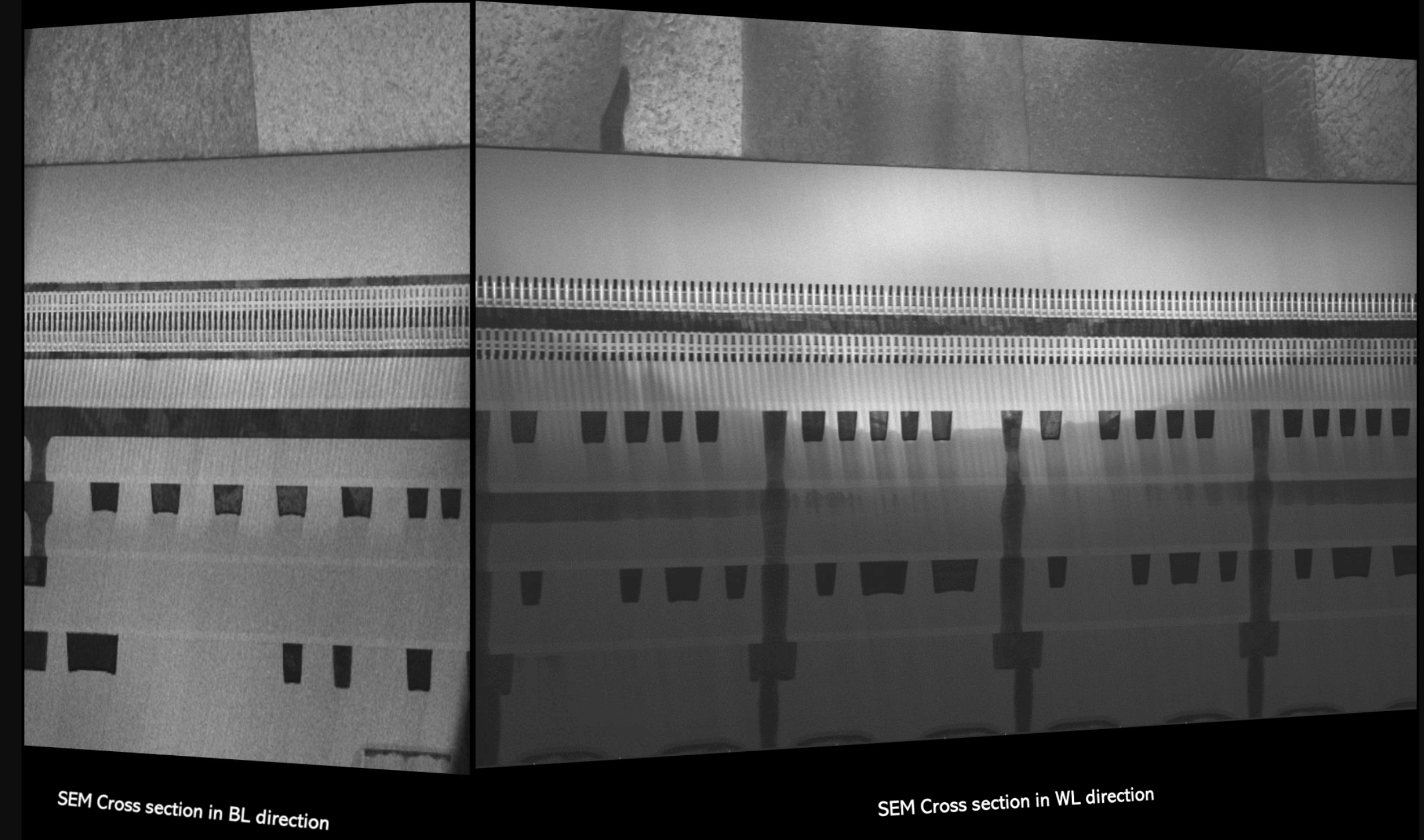}
    \caption{Close-up of the SEM images of 3D XPoint hardware. \cite{XpointSEMpic}}
    \label{fig:xpointSEM}
\end{figure}

\noindent Scanning electron microscopy \textbf{(SEM)} images of the 3D XPoint structure are shown in \textbf{Figure \ref{fig:xpointSEM2}} and \textbf{Figure \ref{fig:xpointSEM}}. Because Intel and Micron are not particularly forthcoming with information surrounding the material system or fabrication process, a lovely bit of controversy has arisen over what terminology ought to be used to describe 3D XPoint; in the years since Optane's debut, a rather broad consensus has emerged that 3D XPoint is a `phase-change memory' \textbf{(PCM)} variant on resistive random-access memory \textbf{(ReRAM)}, although neither Intel or Micron will confirm this. An article from 2017 states: ``An Intel Optane SSD [has been reverse-engineered] to cross-section the XPoint cells within, so we have confirmation that the devices use chalcogenide glasses for both the switching layer and the selector diode. That the latter is labeled `OTS' (for Ovonic Threshold Switch) explains the confusion over the last year as to whether this device is a Phase-Change Memory \textbf{(PCM)} or Resistive Random Access Memory \textbf{(ReRAM)}... it seems to be the special variant of ReRAM using PCM material that has been branded Ovonic Unified Memory or `OUM'... [XPoint Optane SSDs] are cross-bar architecture ReRAM arrays of PCM materials, and had the term not been ruined by 17-years of over-promising and under-delivering they would likely have been called OUM (ovonic unified memory) chips.'' As will be made apparent from the discussion in the next section, those terms are worth deconstructing briefly.

The concept of phase-change memory \textbf{(PCM)} has been around for a while. As far back as 1968, it was observed that a chalcogenide-based\footnote{\scriptsize{A chalcogenide is an ionic compound that contains at least one Group 16 element (chalcogen) anion and at least one electropositive element (like a metal). While oxygen is a chalcogen, the term `chalcogenide' usually does not refer to oxides.}} glass (Si$_{12}$Te$_{48}$As$_{30}$Ge$_{10}$) exhibited a rapid and reversible phase-change between a conductive state and a highly resistive state, and the implications for memory were speculated upon.\cite{Ovskinsky_1968} A detailed description of how to use amorphous semiconductors (including chalcogenide glasses) as devices for computer memory is even provided in a paper from 1973.\cite{Neale_1973} Essentially, PCM can be summarized as follows. Certain types of semiconducting glasses can undergo reversible transitions between ordered (crystalline) and disordered (amorphous) states, upon application of a voltage. This reversible transition is a useful feature, because it is also accompanied by strong optoelectronic changes; the amorphous phase has a high resistivity and low optical reflectivity, and the crystalline phase has a low resistivity and a high optical reflectivity. The optical properties of PCM materials are already widely exploited in CD, DVD, Blu-ray technologies. In the context of electronic data storage, it is the electrical resistance changes that are useful.\footnote{\scriptsize{Actually, in the overwhelming majority of solids, changing between amorphous and crystalline states is accompanied with a pronounced change in electrical resistance. What makes PCM materials unique is that the reversible recrystallization process can occur at lower power. A review article from 2007 states:\cite{Wuttig_2007} ``The two key properties that characterize phase-change materials are their remarkable crystallization kinetics and their contrast between the amorphous and crystalline phases.''}} To quote a review article from 2020: \cite{Gallo_2020} ``The stored data can be retrieved by measuring the electrical resistance of the PCM device. An appealing attribute of PCM is that the stored data is retained for a very long time (typically 10 years at room temperature), but is written in only a few nanoseconds.'' These appealing features of PCM have led to decades of research: going from concept, to prototype, to commercialization. 

Along the way, a variety of insights have been gained. For instance, in 2010, when a team of professors at UC San Diego unveiled their `Moneta' prototype PCM storage system, they note:\cite{Moneta_2010} ``Achieving high performance in Moneta requires simultaneously optimizing its hardware and software components.'' In a press release, Steven Swanson, one of the professors involved in its development, is quoted as follows:\cite{SwansonQuote_2011} ``We’ve found that you can build a much faster storage device, but in order to really make use of it, you have to change the software that manages it as well. Storage systems have evolved over the last 40 years to cater to disks, and disks are very, very slow. Designing storage systems that can fully leverage technologies like PCM requires rethinking almost every aspect of how a computer system’s software manages and accesses storage.'' Swanson's group at UCSD continues working on software for persistent memory to this day.\footnote{\scriptsize{The group website for the Non-Volatile Systems Lab states:\cite{SwansonGroup} ``We built NOVA, the world’s fastest persistent memory file system, Orion, the most capable distributed file system for persistent memory, and Onyx, the world’s first phase change memory SSD. Most recently, we described easy techniques to adapting existing programs to use PMEM, provided the first independent characterization of Optane persistent memory performance, and crafted simple rules for how programmers can maximize performance.''}}$^,$\footnote{\scriptsize{A variety of aspects about Intel Optane persistent memory are also described in depth in a 2019 \emph{arXiv} whitepaper from his group.\cite{SwansonOptaneWhitepaper}}} 

While PCM has been around for a while, there have historically been a variety of factors that limited its widespread adoption. An article from 2020 provides a review of fifty years of PCM development.\cite{Fantini_2020} Several insights of the article are summarized as follows:
\begin{itemize}
    \item The first chalcogenide materials used to build PCM devices were not appealing for commercialization, because they had slow transition speeds and the cells required high programming currents.
    \item The first two decades of PCM had little success; the density of the solid-state memory was not very high. However, when Energy Conversion Devices commercialized the Ge-Sb-Te \textbf{(GST)} phase-change alloy and demonstrated its use for rewritable optical disks, renewed interest arose around PCM.
    \item Significant effort was put into commercializing PCM to replace NOR in cellular phones. Yet this might not have been the best business move; ``With the advent of `smart phones', the cellular phone architecture started to rapidly move from a NOR based `execute in place' architecture to a NAND based `store and download' architecture. ... The NOR Flash market in cellular phones has decreased dramatically over the past 12 years, and thus the PCM opportunity in this segment.''
    \item PCM-based SSDs were tested extensively but they ended up evading production, because even though they did offer improvements, the density was still not high enough to justify the cost of switching to their adoption. 
    \item In order for PCM to be successfully commercialized on the market, it needs to be in an area with ``overwhelming compelling value.'' Smaller performance improvements have not historically been enough.
    \item PCM can show ``overwhelming compelling value'' by targeting novel memory paradigms, where it does not have to overcome the significant difficulties of market competition with incumbent well-established technology. For instance, it is considered to be particularly well-suited for storage-class memory \textbf{(SCM)} and neuromorphic computing applications. ``The SCM arena represents the ideal realm for the PCM technology.'' ``It has been shown that PCM devices can really reproduce the synapse plasticity, fitting the biological behavior of synapses.''
\end{itemize}

\begin{figure}[t!]
    \centering
    \includegraphics[width=1\linewidth]{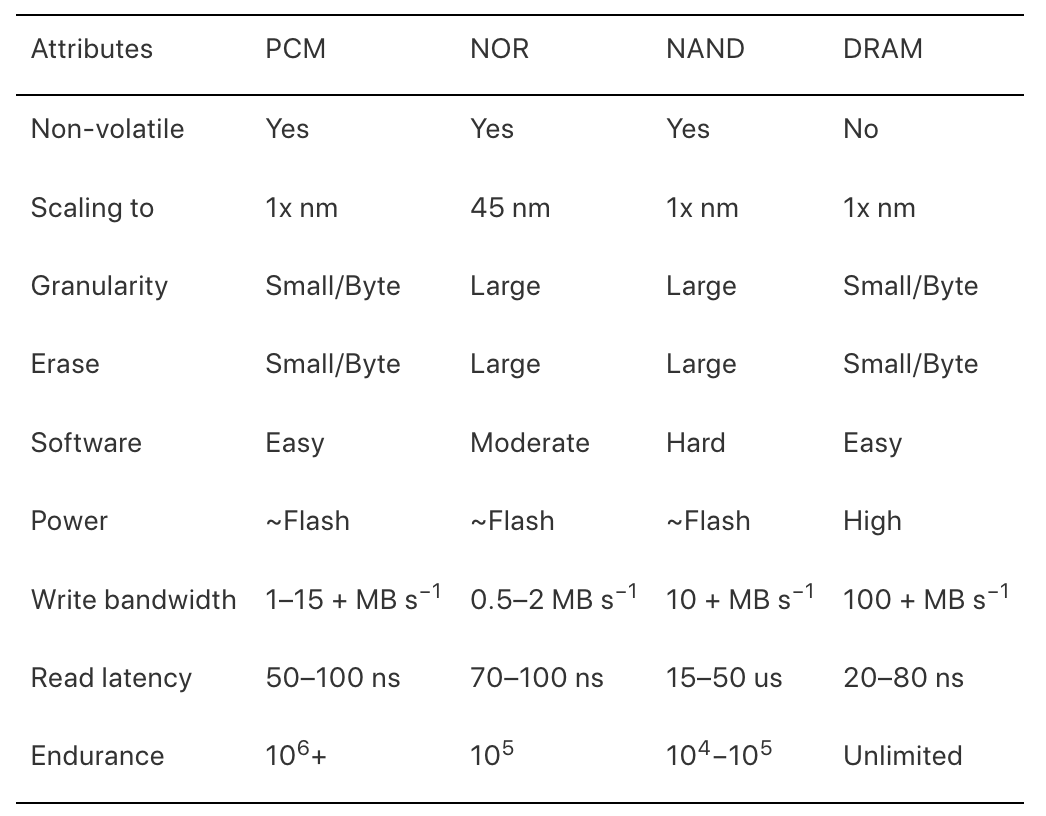}
    \caption{Comparison of various memory technologies, from a 2020 review. \cite{Fantini_2020}}
    \label{fig:PCMTable}
\end{figure}

\noindent An interesting note regarding PCM is that it has been entangled with a vision of `universal' memory, i.e. combining the best of DRAM with the best of NAND. However, as the 2020 review states:\cite{Fantini_2020} ``Instead of becoming `universal' in the sense of bundling all the advantages of DRAM and NAND simultaneously, PCM turned out to have speed and endurance worse than DRAM and a cost structure (density) not as favourable as NAND.'' Thus, instead of overhauling the system and replacing the RAM (DRAM) and Flash (NAND) memory tiers, PCM ends up settling between them at the nascent persistent memory tier.

The other terms mentioned in the context of Intel Optane's 3D XPoint were `ovonic unified memory' \textbf{(OUM)} and resistive random-access memory \textbf{(ReRAM)}. These terms can overlap with PCM. PCM refers to data storage that relies on a phase-change material. ReRAM refers to data storage where the data is stored in a resistor. Thus, many of the current PCM efforts are a subset of ReRAM. These terms should not be used interchangeably, however, because ReRAM cells usually use different materials. The most common ReRAM `memristor' cell is a metal-insulator-metal \textbf{(MIM)} device, in which a transition metal oxide (such as TiO$_2$) is sandwiched between two conductive metal electrodes. The other term - OUM - can be understood as a variant of ReRAM that uses PCM. Use of `ovonic' in this context can be traced back as far back as 1970, to a paper titled \textit{Chalcogenide Glass Bistable Resistivity (Ovonic) Memories.} \cite{Pohm_1970}

Other emerging persistent memory technologies exist as well, such as magnetoresistive RAM \textbf{(MRAM)} and ferroelectric RAM \textbf{(FeRAM)}. With MRAM, the data is stored in a magnetic tunnel junction \textbf{(MTJ)} instead of electrical charges; electric signals are still important though, because they are used to detect and control the magnetic orientation. FeRAM cells are similar to DRAM in that they consist of one transistor and one capacitor; unlike DRAM, however, the capacitor is a ferroelectric capacitor. Since ferroelectric capacitors can retain their polarity without the need for power, this means that -- unlike DRAM -- FeRAM is non-volatile. A theme that emerges when reviewing PCM, ReRAM, MRAM, FeRAM, and so forth, is that using these technologies in the SCM niche can enable the design of entirely new storage systems.

In discussing Optane and 3D XPoint, it would be remiss to overlook a recent business development; over the summer, Intel executives recently confirmed that they were dropping their Optane business.\cite{IntelOptaneDrop2022} Micron had already sold off its assets the previous year, in summer 2021, stating that ``it wanted to shift resources from 3D XPoint to CXL-enabled memory products\footnote{\scriptsize{CXL (Compute Express Link) is a CPU-to-memory interconnect, that is built on top of PCIe; it overlays the caching and memory protocol on top of the existing PCIe protocol. One nice aspect is that it works with the same slots as PCIe; you can plug in a PCIe chip and the slot will function as PCIe, or you can plug in a CXL chip and the slot will function as CXL. CXL focuses on targeting the memory and heterogeneous computing needs that have recently been driving the evolution of the compute landscape. It facilitates extremely high bandwidth, low latencies, and pooled memory. (It also can work in tandem with SCM hardware.) CXL is a very trendy topic at the moment, with a simply overwhelming wave of industry support. \emph{All} of the leading companies that work in this space are involved with it. But this report is long enough as it is, so readers that are interested in CXL are referred to the following citations.\cite{CXL1}$^,$ \cite{CXL2}$^,$ \cite{CXL3}$^,$ \cite{CXL4}}} ... [and was planning] to apply the knowledge [Micron had] gained from the breakthroughs achieved through its 3D XPoint initiative, as well as related engineering expertise and resources, to new types of memory-centric products that target the memory-storage hierarchy.'' \cite{MicronLeaves2021} There was speculation that Intel would be next; a Forbes article in February 2022 asked, ``Is Intel Going To Drop Optane?''\cite{IntelOptaneDropSpeculation2022} This speculation in the article came from the observation that Intel had been selling off parts of the business and had not made any new Optane announcements or talked about Optane during its investor calls for several quarters. This is another instance of PCM having a difficult time on the market; the technology itself is well-received, but profitability is a fluctuating dance with market forces. Earlier successes -- like being awarded a \$500 million contract to provide one of the United States' first exascale supercomputers to Argonne National Laboratory, incorporating Optane memory\cite{AuroraPressRelease2019} -- were unfortunately not sufficient. Optane might still receive some attention; there is still an estimated 2-year supply of available Optane chips that have already been produced.\cite{OptaneSupply} However, for now, the dance with persistent memory must be continued at other companies. Other companies that have announced comparable SCM products include: Kioxia\cite{KioxiaSCM}$^,$\cite{KioxiaSCM2} , Everspin\cite{EverspinSCM}$^,$\cite{EverspinSCM2} , SK Hynix.\cite{SKHynixSCM}$^,$\cite{SKHynixSCM2}

Intel Optane `failed' on the market, but it does not mean that these storage technologies -- or their implications -- are going away anytime soon. Data infrastructure needs to be scalable. The looming issue of the currently implemented storage/memory paradigms is that they may not be sufficient to handle the needs of future data workloads.


\section{\textbf{\textcolor{blue}{DAOS and Object Storage APIs}}}

\begin{figure}[h!]
    \centering
    \includegraphics[width=1.05\linewidth]{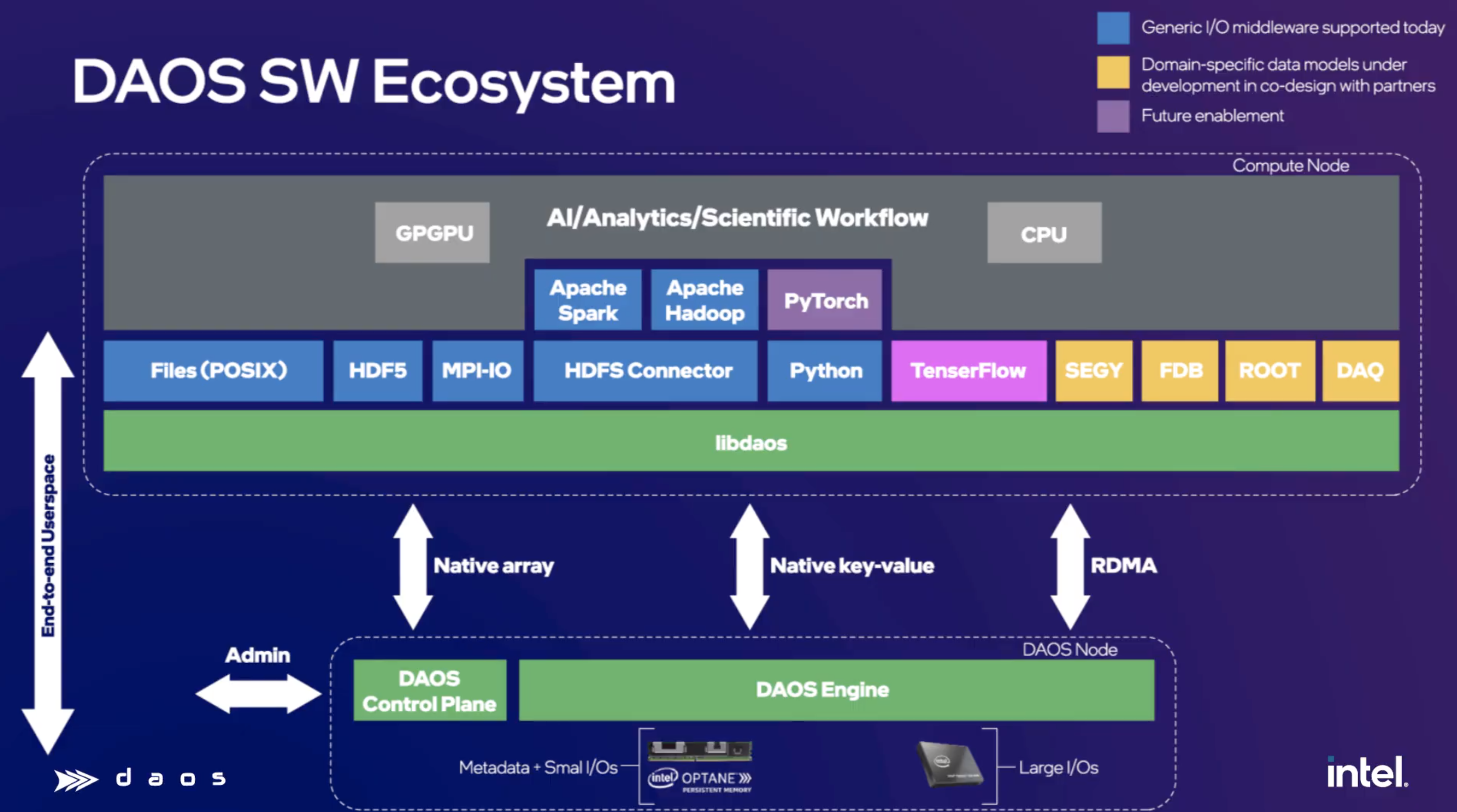}
    \caption{The DAOS ecosystem previously envisioned by Intel. \cite{DAOSEcosystemPic}}
    \label{fig:DAOSecosystem}
\end{figure}

Another consequence of Intel's Optane efforts is DAOS -- the Distributed Asynchronous Object Storage system -- (as mentioned briefly on page 7). Intel's motivation in developing DAOS was to create a `DAOS software ecosystem', as shown in \textbf{Figure \ref{fig:DAOSecosystem}}.

To truly leverage SCM hardware and to capitalize on the performance gains that it can enable, the storage system and software architecture need to be adapted, too. The idea of the DAOS ecosystem was that it would enable the customers of Optane persistent memory to leverage the high-performance potential of their devices, but it would also have the dual benefit of expanding the customer base. By making DAOS open-source, Intel was essentially providing a `free sample' to a nice object storage system. DAOS users might then be persuaded to invest in more Optane devices, in order to improve their performance further. The recent developments over the summer mean that Optane might be receding from the market, but DAOS -- on the other hand -- will still be around. The developers have stated that DAOS ``continues to be a strategic part of the Intel software portfolio.'' They are quoted as follows:\cite{DAOSfutureDeveloperComments} \\

\noindent``We remain committed to supporting our customers and the DAOS community. In parallel, we are accelerating efforts that have already been under way for DAOS to utilize other storage technologies to store metadata on SSDs through NVMe and CXL interfaces. ... The DAOS architecture won’t fundamentally change and the plan is to become more flexible in the configurations we support.  We will continue to store metadata and data on different devices and use direct load/store for the metadata. The DAOS metadata will be stored on either persistent (e.g. apache/barlow/crow pass, battery-backed DRAM or future SSD products supporting CXL.mem) or volatile (e.g. DRAM or CXL.mem) devices. The persistent option is what DAOS supports today. As for the volatile one, there will be an extra step on write operations to keep a copy of the metadata in sync on CLX.io/NVMe SSDs. This work was already underway with community partners and is going to be accelerated. ... Once done, this change should allow DAOS to run on a wider range of hardware while maintaining our performance leadership.'' 
\\

\begin{figure}[h!]
    \centering
    \includegraphics[width=1.1\linewidth]{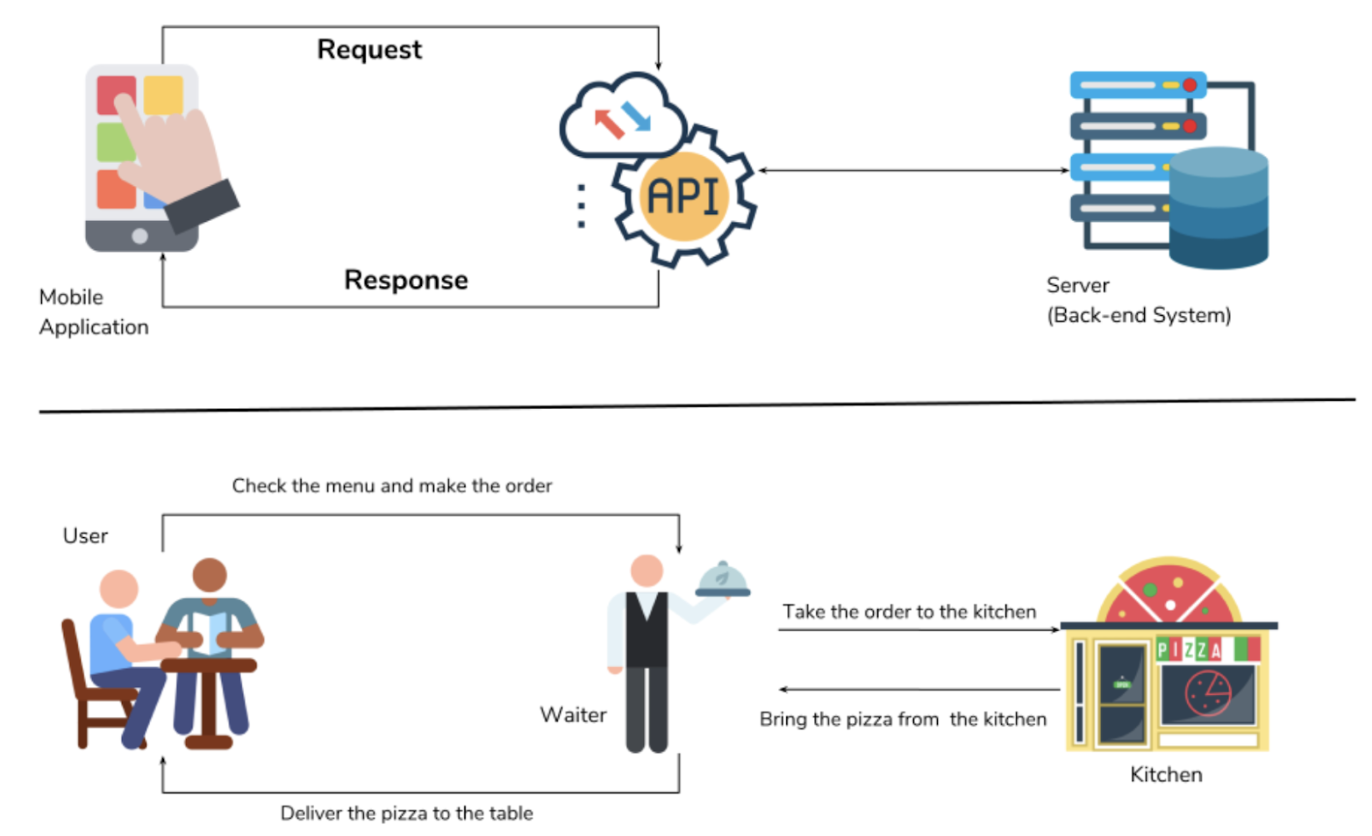}
    \caption{A simple restaurant analogy to explain the basic role of an API.\cite{APIComicPic} (Bottom:) The customers request pizza. The kitchen has the pizza. The waiter is the intermediary, passing pizza between the customers and the kitchen. (Top:) The client requests data. The server has the data. The API is the intermediary, passing data between the client and the server.}
    \label{fig:APIcomic}
\end{figure}

Administrators can interact with DAOS (setting up pools, allocating resources, etc.) via command line, but the user experience is primarily through the API. As such, a brief introduction here to APIs is merited. To access data using file storage, the user simply needs to specify the file name and location. 
Object storage is a little more complicated, in that a client application
using an object storage scheme will access data through use of a URL-based application programming interface \textbf{(API)}.
An API is a 
software interface that facilitates the transfer of data between a client (which requests the data) and a server (which provides the data). A simplified analogy is shown in \textbf{Figure \ref{fig:APIcomic}}. 

\noindent Stated otherwise, an API is a communication bridge. It connects the requests from the \emph{front-end} client to the \emph{back-end} server. 

There is a wide assortment of APIs that can be implemented. The term itself is quite broad. To quote a review article from 2021:\cite{Lamothe_2021} ``Software libraries, software frameworks, and Web services -- either RESTful or SOAP -- have all been interchangeably referred to as APIs, because they all allow pieces of software to communicate, albeit in different ways.'' 
In materials science, researchers are often exposed to (URL-based) APIs in the context of the large public data repositories that have been expanding along with the nascent materials data infrastructure. Several example databases are shown in \textbf{Table \ref{tab:MaterialsDatabasesAPIs}}, along with the relevant type of API that is provided for programmatic access.\footnote{\scriptsize{Standardization efforts with the goal of enabling these materials database APIs to communicate with each other are also underway. 
For instance, in August 2021, the Open Databases Integration for Materials Design \textbf{(OPTIMADE)} consortium outlined a ``universal API to make materials databases accessible and interoperable... OPTIMADE defines a RESTful API that is queried with URLs, with responses adhering to the JSON:API specification.''\cite{OPTIMADE_2021}}}

\begin{table}[h!]
    \scriptsize{
    \begin{tabular}{|>{\centering\arraybackslash}p{3.2cm}|>{\centering\arraybackslash}p{2.8cm}|>{\centering\arraybackslash}p{1.55cm}|}
        \hline
          \textbf{Database Name} & \textbf{Website} & \textbf{Type of API} \\ \hline

         Materials Project & materialsproject.org & RESTful\cite{MaterialsProjectAPI}$^,$\cite{MaterialsProjectAPI2} \\ \hline
         AFLOW Materials Property Repository & aflowlib.org & RESTful\cite{AFLOWAPI}$^,$\cite{AFLOWAPI2} \\ \hline
         Materials Cloud & materialscloud.org & REST\cite{MaterialsCloudAPI}$^,$\cite{AiiDA_API}\\ \hline
         NOMAD (Novel Materials Discovery) Archive & nomad-lab.eu & RESTful\cite{NOMAD_API} \\ \hline
         Open Quantum Materials Database & oqmd.org & RESTful\cite{OpenQuantumMaterialsDatabaseAPI}$^,$\cite{OpenQuantumMaterialsDatabase} \\ \hline
         Catalysis-Hub & www.catalysis-hub.org (suncat.stanford.edu) & GraphQL\cite{CatalysisHubAPI}$^,$\cite{CatalysisHub} \\ \hline
         Open Materials Database & openmaterialsdb.se & REST\cite{OpenMaterialsDatabaseAPI}$^,$\cite{OpenMaterialsDatabaseAPI2}\\ \hline
         Theoretical Crystallography Open Database & www.crystallography.net/tcod & RESTful\cite{TCOD_API} \\ \hline
         Citrination Platform & citrination.com & HTTP\cite{CitrinationAPI} \\ \hline
         Joint Automated Repository for Various Integrated Simulations \textbf{(JARVIS)} & jarvis.nist.gov & REST\cite{JARVIS_API} \\ \hline
         FIZ/NIST Inorganic Crystal Structure Database & icsd.fiz-karlsruhe.de & RESTful\cite{InorganicCrystalStructureDatabaseAPI}$^,$\cite{InorganicCrystalStructureDatabase} (paid) \\ \hline
         PAULING FILE & paulingfile.com & various\cite{PaulingFile}$^,$\cite{PaulingFile2} \\ \hline
         PubChem & pubchem.ncbi.nlm.nih.gov & SOAP, REST \cite{PubChemAPI} \\ \hline
         Organic Materials Database \textbf{(OMDB)} & omdb.mathub.io & web interface \cite{OrganicMaterialsDatabase} \\ \hline
         NREL MatDB & materials.nrel.gov & web interface \cite{NRELMatDB} \\ \hline

    \end{tabular} \\ \\
    }
    \caption{There is a dazzling array of materials databases on the internet.\\ This is a small sampling of some of them.}
    \label{tab:MaterialsDatabasesAPIs}
\end{table}

\noindent In \textbf{Table \ref{tab:MaterialsDatabasesAPIs}}, the `REST' and `RESTful' APIs appear quite predominantly. Since REST APIs are also the predominant API used for object storage, it is worth explaining what this means. A source of confusion arises, because the terms `REST' and `RESTful' are used interchangeably in many cases. They are different, though. A REST-based API \textbf{(REST API)} follows \emph{some} of the REST architectural principles, while a RESTful API follows \emph{all} of them, fully. The term `RESTful' comes from a 2007  book.\cite{RESTfulOrigin2007} The term `REST' (short for `Representative State Transfer') was introduced in a 2000 PhD thesis, and was presented again to the world in a 2002 literature article (same author).\cite{RESTorigin2000}$^,$\cite{RESTorigin2002} Several quotes from the 2002 article are as follows:\cite{RESTorigin2002} 
\begin{itemize}
    \item REST was originally referred to as the `HTTP object model,' but that name often led to its misinterpretation as the implementation model of an HTTP server.
    \item The name `Representational State Transfer' is intended to evoke an image of how a well-designed Web application behaves: a network of Web pages forms a virtual state machine, allowing a user to progress through the application by selecting a link or submitting a short data-entry form, with each action resulting in a transition to the next state of the application by transferring a representation of that state to the user.
    \item The modern Web is one instance of a REST-style architecture.
    \item The central feature that distinguishes the REST architectural style from other network-based styles is its emphasis on a uniform interface between components. By applying the software engineering principle of generality to the component interface, the overall system architecture is simplified and the visibility of interactions is improved. ... The tradeoff, though, is that a uniform interface degrades efficiency, since information is transferred in a standardized form rather than one which is specific to an application’s needs. The REST interface is designed to be efficient for largegrain hypermedia data transfer, optimizing for the common case of the Web, but resulting in an interface that is not optimal for other forms of architectural interaction.
    \item REST is defined by four interface constraints: identification of resources; manipulation of resources through representations; self-descriptive messages; and, hypermedia as the engine of application state.
    \item The key abstraction of information in REST is a resource. Any information that can be named can be a resource.
    \item REST uses a resource identifier to identify the particular resource involved in an interaction between components.
    \item REST components perform actions on a resource by using a representation to capture the current or intended state of that resource and transferring that representation between components. A representation is a sequence of bytes, plus representation metadata to describe those bytes.
    \item A representation consists of data, metadata describing the data, and, on occasion, metadata to describe the metadata (usually for verifying message integrity). Metadata is in the form of name-value pairs.
    \item All REST interactions are stateless. That is, each request contains all of the information necessary for a connector to understand the request, independent of any requests that may have preceded it. 
    \item The primary connector types are client and server. The essential difference between the two is that a client initiates communication by making a request whereas a server listens for connections and responds to requests in order to supply access to its services.
    \item The stateless nature of REST allows each interaction to be independent of the others, removing the need for an awareness of the overall component topology, an impossible task for an Internetscale architecture, and allowing components to act as either destinations or intermediaries, determined dynamically by the target of each request. Connectors need only be aware of each other’s existence during the scope of their communication.
    \item The Hypertext Transfer Protocol \textbf{(HTTP)} has a special role in the Web architecture as both the primary application-level protocol for communication between Web components and the only protocol designed specifically for the transfer of resource representations. ... REST has been used to limit the scope of standardized HTTP extensions to those that fit within the architectural model.
\end{itemize}

\noindent In simpler words, a 2019 book provides the following summary:\cite{RESTfulBook2019} ``The fundamental principle of REST is to use the HTTP protocol for data communication (between distributed hypermedia systems), and it revolves around the concept of resources where each and every component considered as a resource, and those resources are accessed by the common interfaces using HTTP methods.'' These HTTP methods include simple commands like `POST' (to create a record), `PUT' (to update a record), `GET' (to retrieve a record), and `DELETE' (to remove a record). With object storage, `POST' and `PUT' are essentially analogous, because objects cannot be modified; editing an object means overwriting it. The REST architecture aligns well with the object storage framework, which is one of the reasons for its ubiquity in this use case.

In the context of object storage APIs for high performance computing \textbf{(HPC)}, the \emph{back-end} server is the object store database, and the \emph{front-end} client is the compute node on the HPC system; the API passes data between them. \textbf{Table \ref{tab:UsingObjectStorage}} provides a higher-level summary regarding the experience of incorporating object storage -- and an object storage API -- into one's workflow.

\begin{table}[h!]
    \scriptsize{
    \begin{tabular}{|>{\centering\arraybackslash}p{1.3cm}|>{\centering\arraybackslash}p{3.1cm}|>{\centering\arraybackslash}p{3.1cm}|}
        \hline
          & \textbf{HPC Networked Filesystems} & \textbf{HPC Object Storage} \\ \hline

         \textit{Locating data} & Files exist within a hierarchy of directories and are accessed by path name. & Objects are stored in a flat namespace and are accessed by an unique identifier. \\ \hline

         \textit{Reading and writing data} & Uses standard Linux commands (cat, grep, vi, emacs, etc.) to read and write data. Programs can access files by name. & Objects can only be accessed by programs that call specialized functions. \\ \hline

         \textit{Data storage and resiliency} & Data is stored on a more tightly coupled group of disks within the same data center. The system is engineered such that failures of individual disk drives or other components will not cause data loss. & Data is stored on many, loosely coupled servers that may be geographically distributed to ensure very high reliability. The system is engineered such that multiple servers can fail and data will not be lost. \\ \hline

         \textit{Performance} & Highly optimized for a parallel computing workload. Uses fast, high-quality components. & Favors a more distributed workload. Components are high-quality, but performance is limited by network bandwidth between storage devices. \\ \hline
    
    \end{tabular} \\ \\
    }
    \caption{Broad description of how an HPC user's workflow differs between \\ networked filesystems and object storage, as deployed on the NIH Biowulf \\ supercomputer cluster. The NIH object store uses an API that is identical \\ to Amazon's Simple Storage Service (\textbf{S3)} API. \cite{HPCFilesystemsVsObjectStorage}}
    \label{tab:UsingObjectStorage}
\end{table}

DAOS has been marketed for a variety of use cases. Accordingly, the developers have made it accessible through multiple APIs. Over the summer, during the internship,
the author had the opportunity to set up an emulated version of DAOS (on a CentOS system) and interact with it, first as an administrator and then as a user.\footnote{\scriptsize{A brief comment about this experience: perhaps reflective of who DAOS has been marketed to so far, the online documentation for the administrative functionalities was \emph{much} more understandable and easy to follow than the online documentation for the downstream user functionalities; i.e. actually creating objects and accessing their contents was a surprisingly nontrivial effort. Some sample scripts are cited here.\cite{TestScript1}$^,$ \cite{TestScript2}$^,$ \cite{TestScript3}  }} 
For interacting with DAOS in the role of a user, the native DAOS API was selected.
In contrast to the higher-level object storage APIs discussed on the previous page (which access data over a TCP/IP network), the native DAOS API is presented as C library called ‘libdaos’ (visible in \textbf{Figure \ref{fig:DAOSecosystem}}).\cite{NativeObjectInterface} Several of the available lower-level functions that the native DAOS API provides for a key-value object were:\cite{keyvalueobjectDAOS}

\begin{itemize}
    \item daos\_kv\_close()
    \item daos\_kv\_destroy()
    \item daos\_kv\_get()
    \item daos\_kv\_list()
    \item daos\_kv\_open()
    \item daos\_kv\_put()
    \item daos\_kv\_remove()
\end{itemize}

\begin{figure}[h!]
    \centering
    \includegraphics[width=0.9\linewidth]{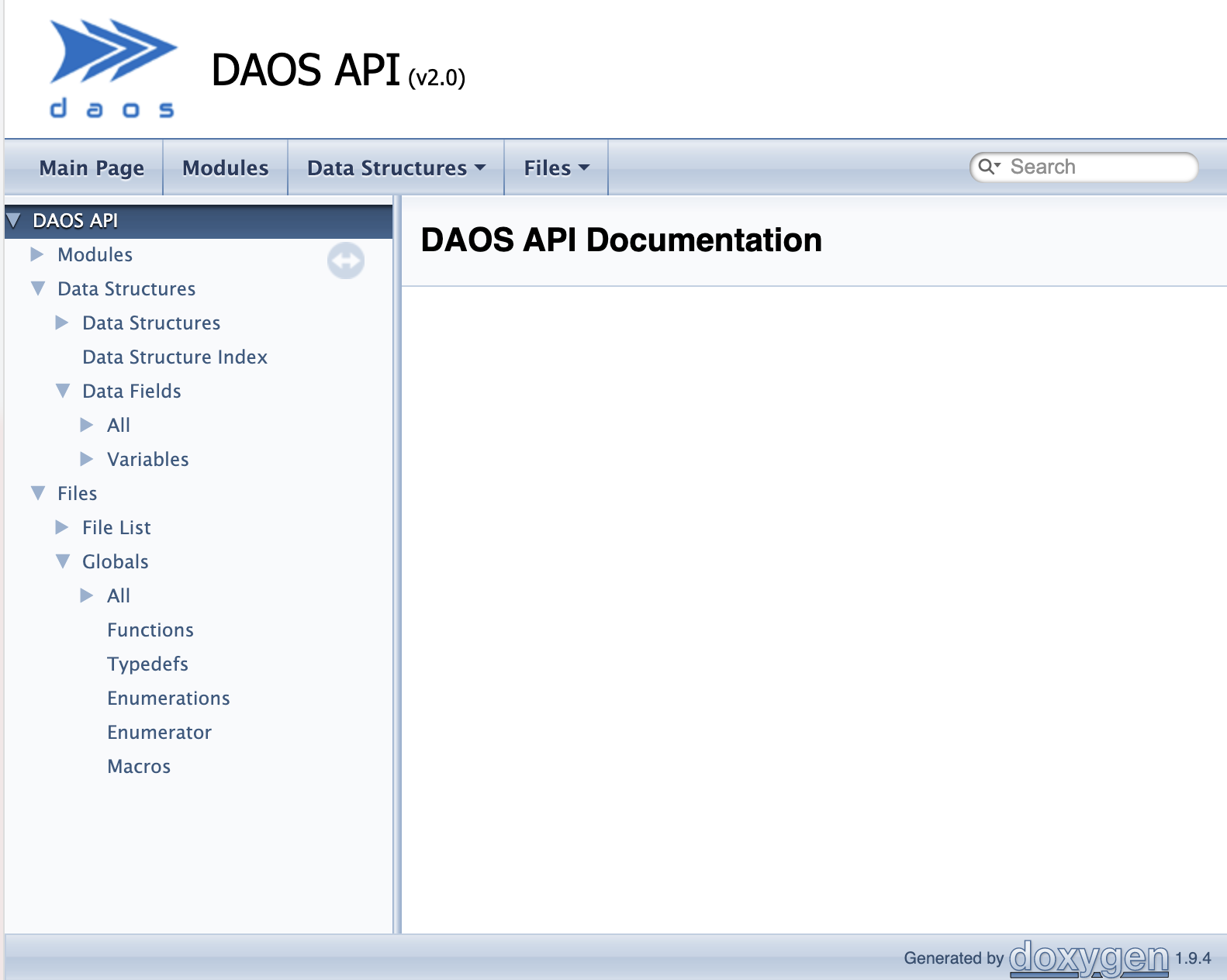}
    \caption{Homepage of the online documentation for the DAOS API. \cite{doxygen} }
    \label{fig:doxygen}
\end{figure}

\begin{figure}[t!]
    \centering
    \includegraphics[width=1\linewidth]{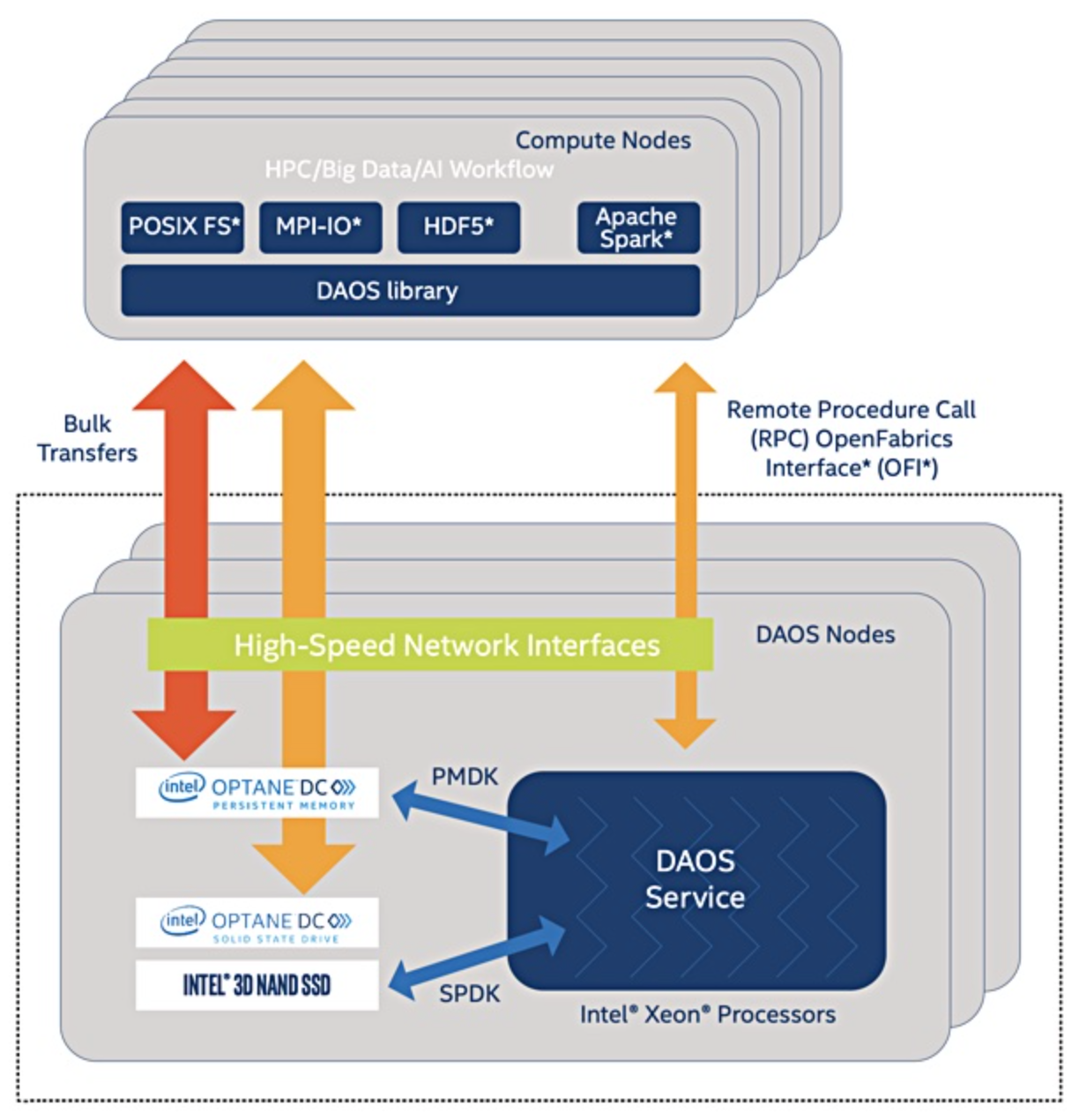}
    \caption{Higher-level overview of the DAOS software stack.\cite{BriefOverview_2019}}
    \label{fig:DAOSsoftwarestack}
\end{figure}

\noindent In addition to the key-value object type, the DAOS API also provides functions for a key-array object type. A 2022 paper has the following comments:\cite{Soumagne_2022}
``A DAOS object can be accessed through different APIs but the multi-level key-array API, which is the DAOS native object interface with locality feature, is of special interest. The key is split into a distribution key \textbf{(dkey)} and an attribute key \textbf{(akey)}. Both the dkey and akey can be of variable length and type (i.e., a string, an integer or even a complex data structure). All entries under the same dkey are guaranteed to be colocated on the same storage target—this is a very important point for the rest of the discussion. By being able to control locality, keys for a given object can not only be accessed with minimum latency but they can also be spread over multiple storage targets, increasing overall bandwidth. The value associated with an akey can either be a single variable-length value that cannot be partially overwritten or an array of fixed-length values. Both the akeys and dkeys support enumeration.'' A screenshot of the online documentation for `libdaos' is shown in \textbf{Figure \ref{fig:doxygen}}. DAOS and its APIs continue to be under development; the online developer forum is cited thusly.\cite{DAOSgroup} DAOS has many compelling features, but in the context of this report, it receives particular interest because of its ability to leverage the persistent memory tier. There are a variety of graphics describing the DAOS architecture; one of them is shown in \textbf{Figure \ref{fig:DAOSsoftwarestack}}. The developers state:\cite{DAOS_paper_2020} ``When the system is up and running, DAOS can \emph{directly access} persistent memory in user space by memory instructions like load and store, instead of going through a thick storage stack.'' For this reason, and since it is open-source, DAOS is a promising testbed for exploring the novel storage paradigms that persistent memory and object storage may enable.

\section{\textbf{\textcolor{blue}{Novel Computing Enabled by SCM}}}

Whenever computing limitations are identified, efforts to circumvent them soon follow. SCM provides a two-fold route to circumvent storage/memory limitations.

On one hand, SCM enables enhancements to existing HPC paradigms. Consider data movement; it is expensive. As much as 25\% of the power consumption of a HPC supercomputing cluster comes from the data movement itself.\cite{Kestor_2014} A report from 2016 states:\cite{Cicotti_2016} 
``[dependent on the application,] 18--40\% of the energy of running an application is spent in moving data up and down the memory hierarchy and that 19--36\% of the total energy is wasted waiting for data and control flow dependencies to be solved (stalled cycles). The energy cost of moving data already account for a large part of the total energy consumption and, when combined with energy wasted while waiting, almost outweighs the energy spend in computation.
\textbf{On average, only about 50\% of the total energy cost is actually spent performing computation.}''
These estimates are somewhat dated, but scaling trends indicate that these values are getting worse, not better, over time. 
More recent figures quote these estimates even higher; a 2021 review stated that over 50\% of power consumption is on data movement.\cite{Si_2021} The report also raised the concern:\cite{Cicotti_2016}  ``The cost of executing a floating point operation has been decreasing for decades at a much higher rate than that of moving data. Bandwidth and latency, two key metrics that determine the cost of moving data, have degraded significantly relative to processor cycle time and execution rate. Despite the limitation of sub-micron processor technology and the end of Dennard scaling, this trend will continue in the short-term making data movement a performance-limiting factor and an energy/power efficiency concern. Even more so in the context of large-scale and data-intensive systems and workloads.'' Since SCM fits between RAM and Flash on the storage/memory hierarchy, it can be leveraged to decrease data movement and improve computing efficiencies.\cite{Cheng_2019_3D}
As a 2020 review article points out:\cite{Fantini_2020} ``Looking at the application requirements, the so large gap between DRAM and NAND (several orders of magnitude!) represents a big opportunity for any emerging memory ... with properties in between DRAM and NAND, both in term of latency (access time < 10 msec) and cost (cost < 1\$/GB).'' Of course, if SCM was able to demonstrate improvements over both DRAM and NAND, it could enable the coalescing off both tiers together into `universal' memory. However, in practice, this is not going to be feasible unless additional breakthroughs occur. For now, SCM can simply be interjected into the memory hierarchy and -- if leveraged properly -- enables reduced data movement.

\begin{figure}[t!]
    \centering
    \includegraphics[width=0.9\linewidth]{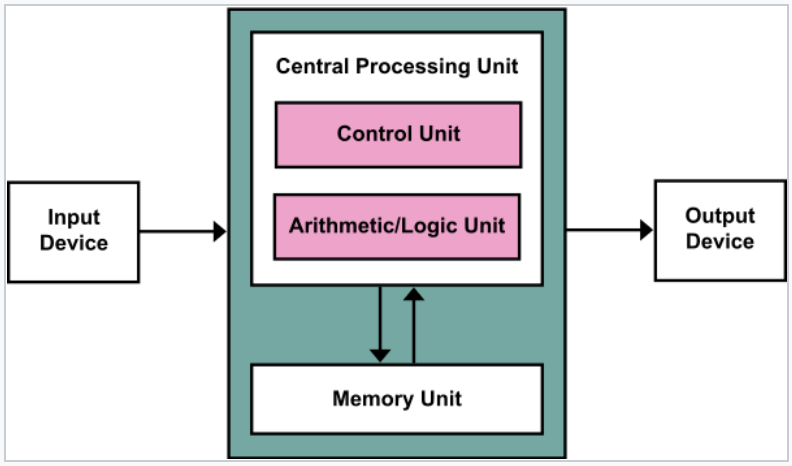}
    \caption{In the von Neumann architecture, a clear distinction is made between the CPU and the memory/storage.\cite{vonNeumannPicture}}
    \label{fig:vonNeumann}
\end{figure}

On the other hand, rather than simply augmenting existing workflows, SCM also enables the implementation of novel capabilities. Much of the existing computational infrastructure leverages what is known as `von Neumann' architecture, shown in \textbf{Figure \ref{fig:vonNeumann}}. As widely implemented as this architecture is, it has faced criticism. Two notable areas of critique are the `von Neumann bottleneck' and cybersecurity. The latter issue is described thusly:\cite{Rosenburg_2017} ``In 1945, Von Neumann and others described a simple but powerful processor architecture with a single internal memory. This architecture continues to dominate the architecture of processors in billions of devices today. This single memory where instructions, data, pointers, and all the data structures needed by an application are stored with no way to tell what is what. It is this memory sometimes called `raw, seething bits' that prevents the processor from cooperating with the program to enforce security. ... As our processors became much more powerful and sophisticated, people began to trust them to protect increasingly valuable things. As processor power made more things possible, people’s expectations also grew and so software size has grown exponentially to keep up. All the while programming languages like C and C++ continued to provide direct access to the `raw, seething bits' of memory without manifest identity, types, boundaries, or permissions to help explain what each word in memory was for. ... When programming in C/C++/Java, the default is unsafe. ... A buffer when allocated does not protect itself from being overwritten; the programmer always has to do extra work (write more code) to make it safe. This is not so much the fault of the programming language as it is of things below the level of the language the programmer is writing in. It is this undefined behavior of the language, where most of the problems (which become cyber vulnerabilities) arise.'' The von Neumann architecture is inherently designed without security or access permissions in mind, and as a result, it takes extra work to incorporate security features at higher levels. 

The other oft-discussed issue, the `von Neumann bottleneck' refers to the limitations that the data transfer between CPU and memory impose, also referred to as the `memory wall'. According to the 2020 review article, the von Neumann bottleneck not only limits the bandwidth, but also leads to a 40\% power waste.\cite{Fantini_2020} In particular, the three primary factors that limit this data transfer are:\cite{Si_2021}
\begin{enumerate}
    \item High energy consumption.
    \item Low bandwidth, where the bandwidth inside the memory chip is 100 times greater than the bus between CPU and memory unit such as DRAM and SSD.
    \item High latency, where memory access to storage in the SSD or HDD is much slower than that of SRAM or DRAM.
\end{enumerate}

\begin{figure}[h!]
    \centering
    \includegraphics[width=0.7\linewidth]{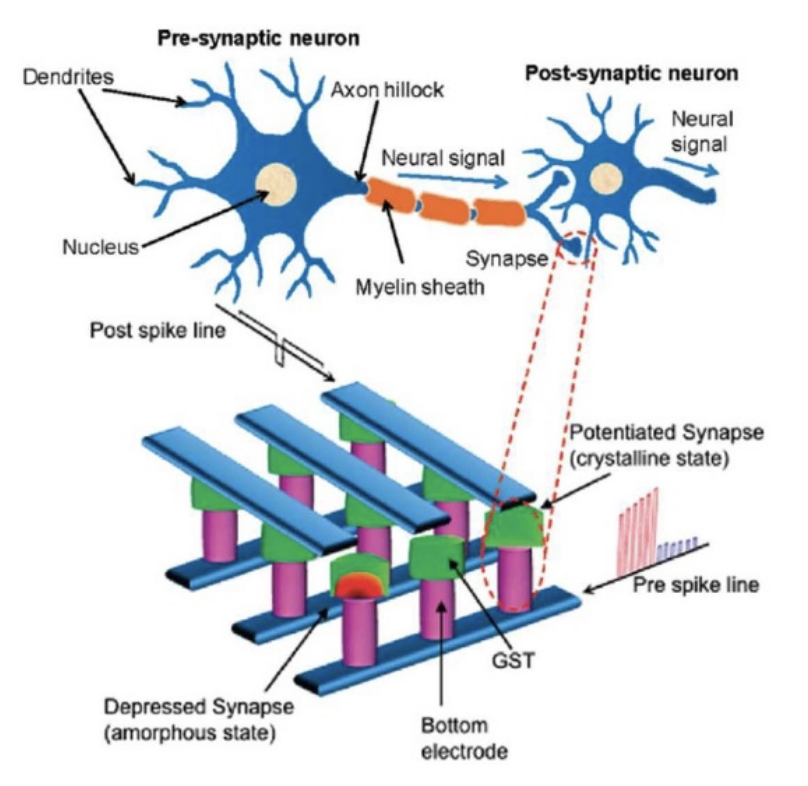}
    \caption{Bio-inspired cross-point phase-change memory array, intended to mimic neuron behavior for neuromorphic computing.\cite{Fantini_2020}}
    \label{fig:bioinspired}
\end{figure}

\noindent These issues have been known for quite some time, and attention on addressing them has existed for over a decade. A significant amount of the hype that has been generated over the years for SCM comes from its suitability to implement various non-von Neumann architectures. Neuromorphic computing, driven in part by the observation that the human brain operates at incredible energy efficiency (with power of about 10-15 W)\cite{Burr_2016}, is one such idea; see \textbf{Figure \ref{fig:bioinspired}}. As the 2020 article states:\cite{Fantini_2020} ``A unique feature of neuromorphic systems at the basis of the brain learning is the process to modulate the connection strength of the synapse linking two neurons (referred to as synaptic weight). In particular, the spike-time-dependent-plasticity \textbf{(STDP)} is a form of synaptic plasticity that updates the synaptic weight on the relative timing between pre-synaptic and post-synaptic spikes in biological systems. It has been shown that [phase-change memory] devices can really reproduce the synapse plasticity, fitting the biological behavior of synapses.'' An assortment of papers reflecting the prototyped devices and review articles in this research area are referenced.\cite{Burr_2016}$^,$\cite{Shi_2021_book}$^,$\cite{Markovic_2020}$^,$\cite{Zhu_2020_review}

\begin{figure}[b!]
    \centering
    \includegraphics[width=0.6\linewidth]{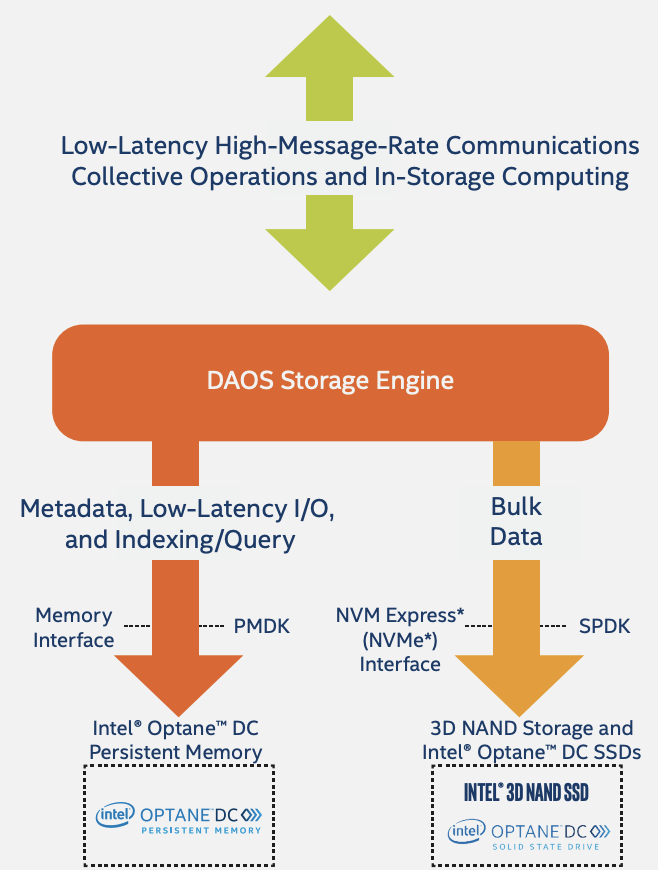}
    \caption{Intel marketing graphic for showcasing the DAOS architecture.\cite{DAOSmarketingGraphic}}
    \label{fig:DAOSmarketing}
\end{figure}

An alternative route for SCM to bypass the von Neumann bottleneck is via in-memory computing or in-storage computing. The enabling factor for this paradigm shift is SCM's ability to behave in a nonvolatile fashion. Nonvolatile memory with a large capacity can be -- for practical purposes -- essentially indistinguishable from storage. However, in distributed computing (and especially in most object storage use cases), there is still the matter that the physical hardware that comprises the computing resources can be geographically dispersed. Here, SCM can be implemented to extend client/server capabilities. While in-storage computing has yet to be achieved in an HPC environment (to the best of the author's knowledge), three potential routes, using DAOS, are proposed as follows: (1) Shipping code to the DAOS server (similar to stored procedures); (2) Activating a software container deployed on the DAOS server; (3) Lazy objects within DAOS. Each of these three routes would rely on alterations to the DAOS API such that a user process can trigger data processing on the DAOS server. One desired outcome of the summer internship was to explore these routes in more depth; the progress was ultimately limited due to time constraints, but the idea is interesting regardless. In the context of HPC object storage, this `in-storage computation' refers to data processing performed locally at the back-end server where the data is stored, rather than on the compute nodes (the front-end clients). The implication is that some data processing is offloaded, freeing up the resources of compute nodes and allowing them to focus on additional, more intensive tasks. Not all computation is created equal, so it would be an interesting question to explore the extent of in-storage computing's capabilities. It is noted that this term has already been exposed to the world; observe in \textbf{Figure \ref{fig:DAOSmarketing}}, a graphic from Intel marketing their DAOS architecture, that ``in-storage computing'' is present, although they do not expand upon it further.

\begin{figure}[t!]
    \centering
    \includegraphics[width=1\linewidth]{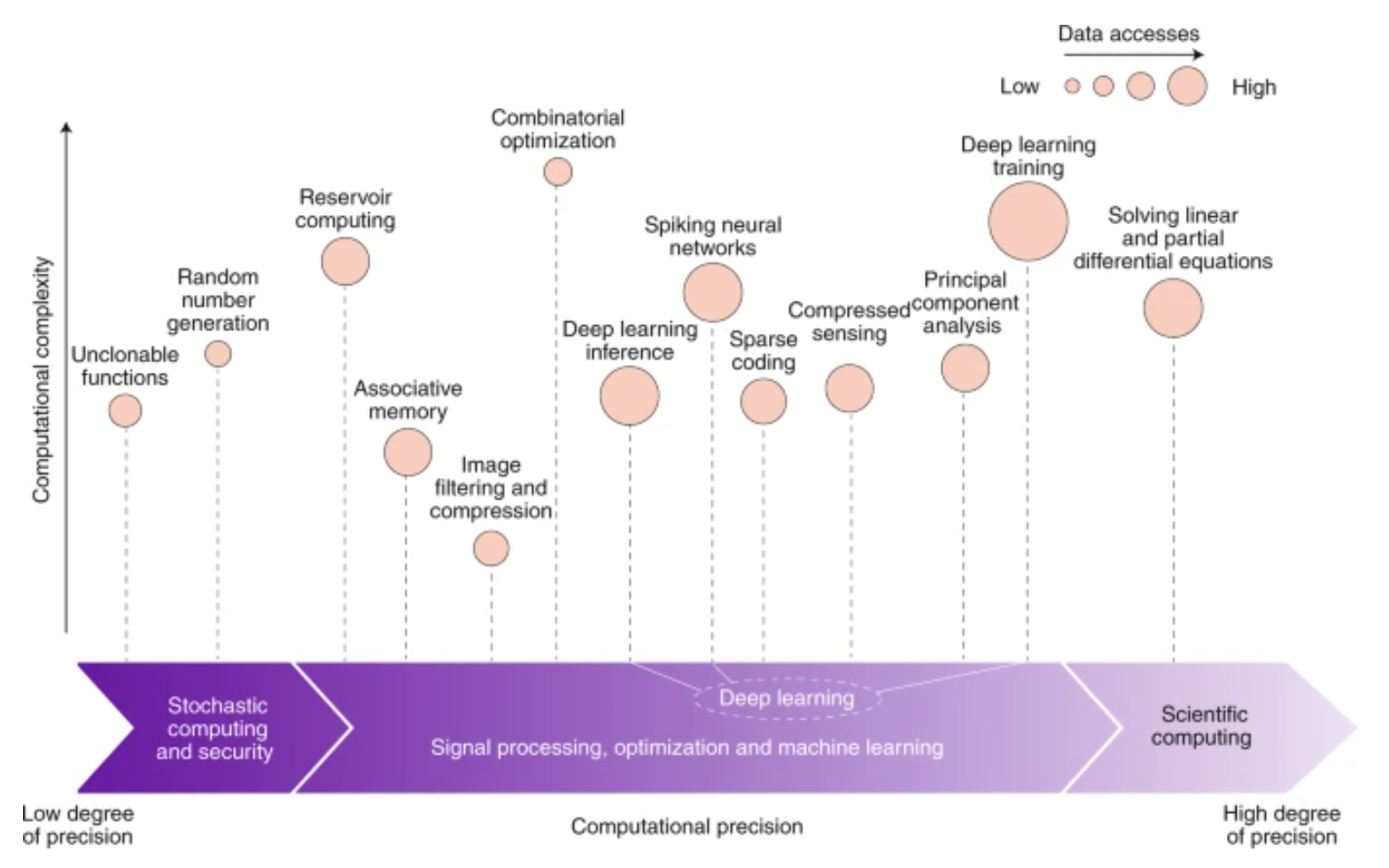}
    \caption{Landscape of applications for in-memory computing, grouped into categories.\cite{Sebastian_2020}}
    \label{fig:memorylandscape}
\end{figure}

\begin{figure}[b!]
    \centering
    \includegraphics[width=1\linewidth]{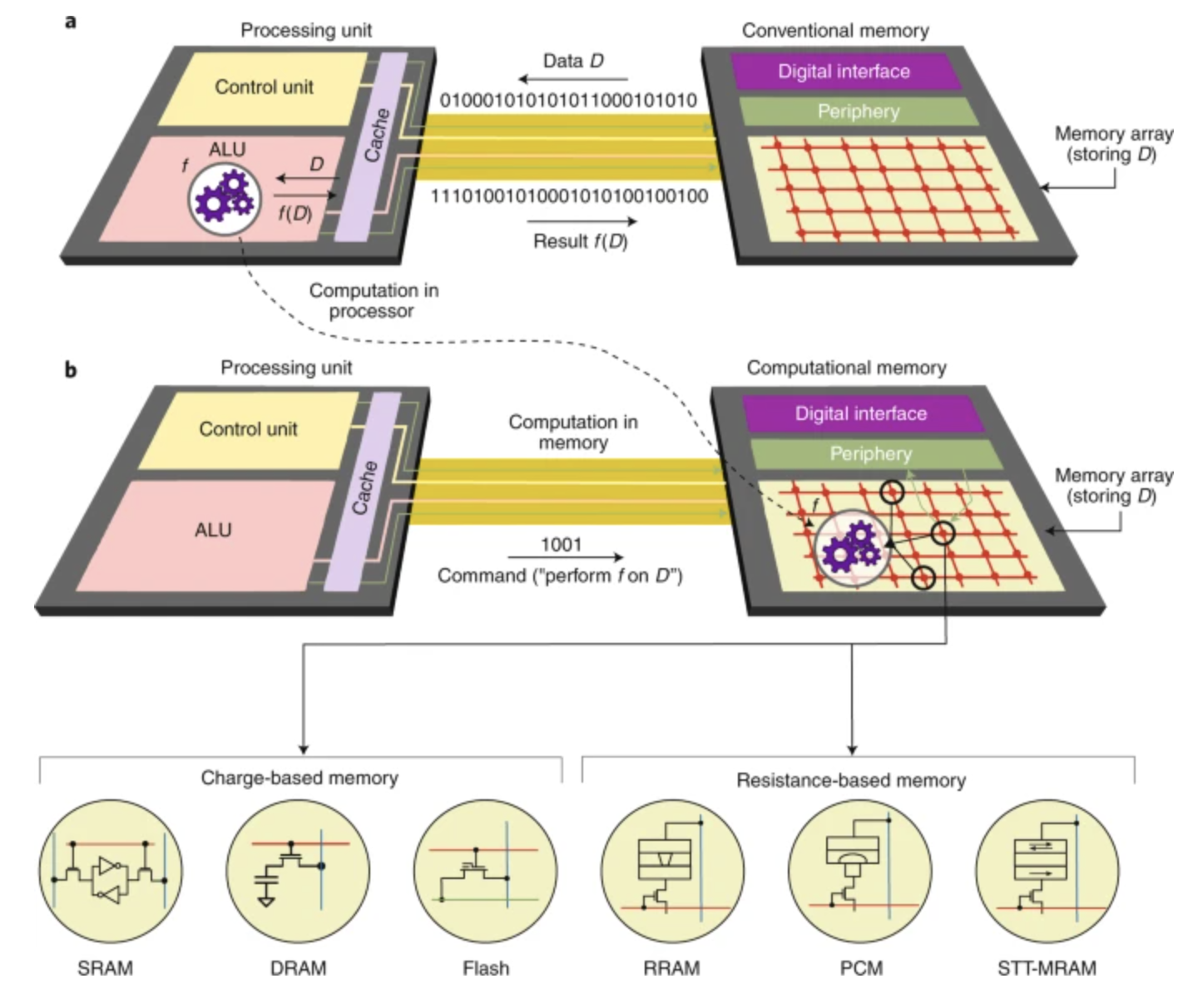}
    \caption{(a) An operation \emph{f} is performed in the context of a conventional von Neumann computing system. (b) Implementation of in-memory computing alleviates the need to move data, \emph{D}, to the processing unit, instead allowing the operation \emph{f} to be performed within the computational memory unit.\cite{Sebastian_2020}}
    \label{fig:inmemorycomputation}
\end{figure}

\begin{figure}[h!]
    \centering
    \includegraphics[width=1\linewidth]{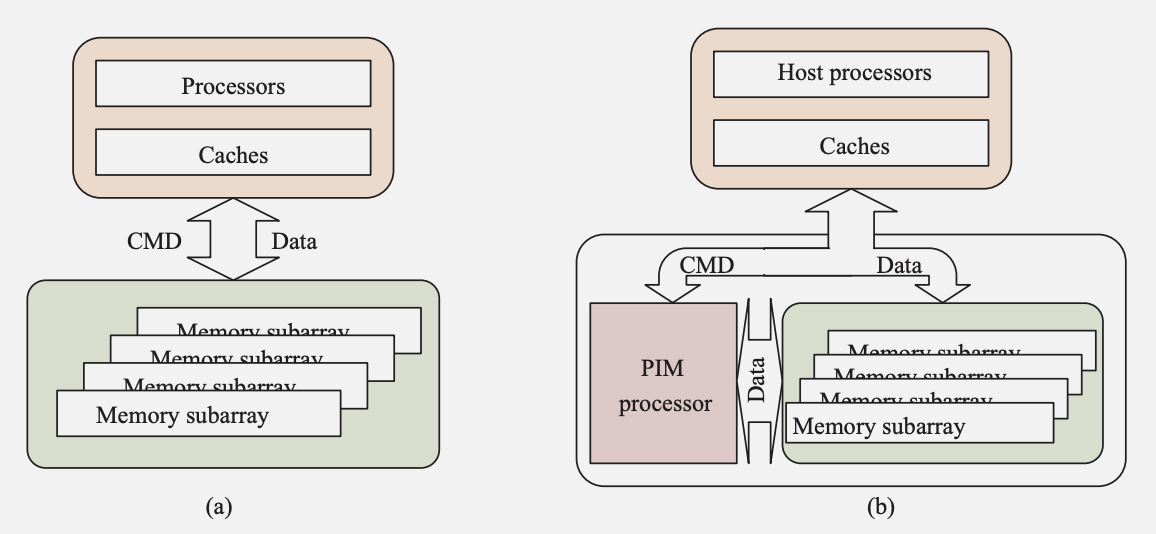}
    \caption{(a) Sample conventional (von Neumann) computing architecture. (b) Sample Processing-In-Memory (in-memory computing) architecture. \cite{Zou_2021_bottleneck}}
    \label{fig:PIM}
\end{figure}

The concept of SCM-enabled `in-memory computing' is still an active area of research as well. Several notable application areas that have been demonstrated are shown in \textbf{Figure \ref{fig:memorylandscape}}, grouped qualitatively by the computational complexity, computational precision, and number of data accesses involved in the computation required for that use case. As a 2020 review article summarizes:\cite{Sebastian_2020} ``By blurring the boundary between processing and memory units (an attribute that is also shared with the highly energy-efficient mammalian brain where memory and processing are deeply intertwined), we gain significant improvements in computational efficiency. However, this is at the expense of the generality afforded by the conventional approach where memory and processing units are functionally distinct from each other.'' An additional graphic for visualization is shown in \textbf{Figure \ref{fig:inmemorycomputation}}. As the bottom row of the figure illustrates, charge-based memory can -- in fact -- be implemented for this as well. In-memory computing using non-volatile Flash memory has been demonstrated too.\cite{MerrikhBayat_2018} However, as mentioned earlier when discussing PCM, incumbent technologies are difficult to compete with; to be commercially successful, emerging memory technologies need to demonstrate ``overwhelming compelling value''. By targeting less-established niche areas like in-memory computing, SCM can attempt to develop in tandem with these application fields and grow its influence. It can also market itself as a good longer-term investment, since SCM technologies have demonstrated scalability to nanometer dimensions. (PCM, for instance, gets more energy efficient the smaller it is scaled, for reasons that will be touched on in the following section.) A comparision of off-memory learning vs. in-memory learning for neural networks is described as follows:\cite{Xi_2021} ``Off-memory learning is a relatively simple way to perform the learning of neural networks on the computer. Each device is programmed to the target value so that the entire neural network has the correct function for inference. After writing, the synaptic weights are almost never changed. Therefore, off-memory learning is currently limited to tasks that do not need [to] update the synaptic weights after programming. This is a serious drawback that limits its applications. In-memory learning involves training the entire neural network on its hardware. The configuration of the entire network changes during learning with new incoming samples. ... For in-memory learning, there are two aspects need to be considered: how to efficiently calculate weight update value, and how to efficiently tune the conductance in the presence of device nonideal effects.'' Furthermore, regarding the use of nonvolatile SCM to the neural network training:\cite{Xi_2021} ``There is no doubt that longer retention time would help both the synaptic weight update and the inference after learning and avoid adverse effects caused by serious conductance drift.'' It is noted, however, that in-memory computing does have additional complications associated with it. Consider \textbf{Figure \ref{fig:PIM}}, which shows a conventional computing architecture compared with a sample processing-in-memory \textbf{(PIM)} architecture. In the PIM system, data is handled by both the host processor and the PIM processor; if this feature is not properly accounted for, it can lead to data coherence issues. \textbf{Figure \ref{fig:PIM}} is certainly not representative of \emph{all} in-memory computing architectures, but it does make plain the need to properly coordinate the data streams.
Further articles on in-memory computing (also referred to as `computational storage', `in-storage processing', `processing-in-memory', etc.) are referenced.\cite{Sebastian_2020}$^,$\cite{Xi_2021}$^,$\cite{Torabzadehkashi_2019}$^,$\cite{Lukken_2021}$^,$\cite{Aljameh_2022}$^,$\cite{HeydariGorji_2022}

As an additional note for a truly unique computing approach, enabled by PCM, consider the concept of hardware that could actively switch between 1-bit and 2-bit (or even 4-bit) modes, depending on system needs. The standard description of PCM is as a 1-bit single-level cell \textbf{(SLC)}. However, since the resistance difference between the crystalline and amorphous states of the PCM is several orders of magnitude, and since intermediate resistance values are able to be obtained depending on the degree of crystallization, intermediate resistance levels can be introduced to store additional data. A PCM multi-level cell \textbf{(MLC)} can be designed for 2-bit, 3-bit, even 4-bit memory. Although MLC does come with materials science challenges of its own (the device endurance is significantly lessened), these unique aspects of the device physics have some neat ramifications, as a 2015 article describes:\cite{Qiu_2015} 
``The memory capacity of a traditional computing system is usually stronger than the requirements of applications, in order to avoid performance loss caused by memory missing. However, this scheme leads to a memory waste since a large portion of the memory is not used in most periods of running time. The SLC/MLC PCM memory architecture can greatly improve the efficiency of the main memory by switching the mode of PCM cells between the SLC and MLC modes. These existing SLC/MLC memory methods can adjust the configuration based on the statistics information acquired during the runtime.'' Overall, it is clear that SCM (and PCM) present opportunities for novel computing possibilities. These technologies are often described as paradigm-shifting, but whether they will experience commercial success is dependent on cost above all else. As illustration: DAOS, combined with Intel Optane, has received glowing reviews, \cite{Fridman_2021}$^,$\cite{LopezGomez_2021}$^,$\cite{Manubens_2022}
 yet Optane is being discontinued due to lack of profitability. Still, it is helpful to be informed on the developments in this space; should these visions come to fruition, they would have a substantial impact on the computing landscape.

\begin{figure}[h!]
    \centering
    \includegraphics[width=1\linewidth]{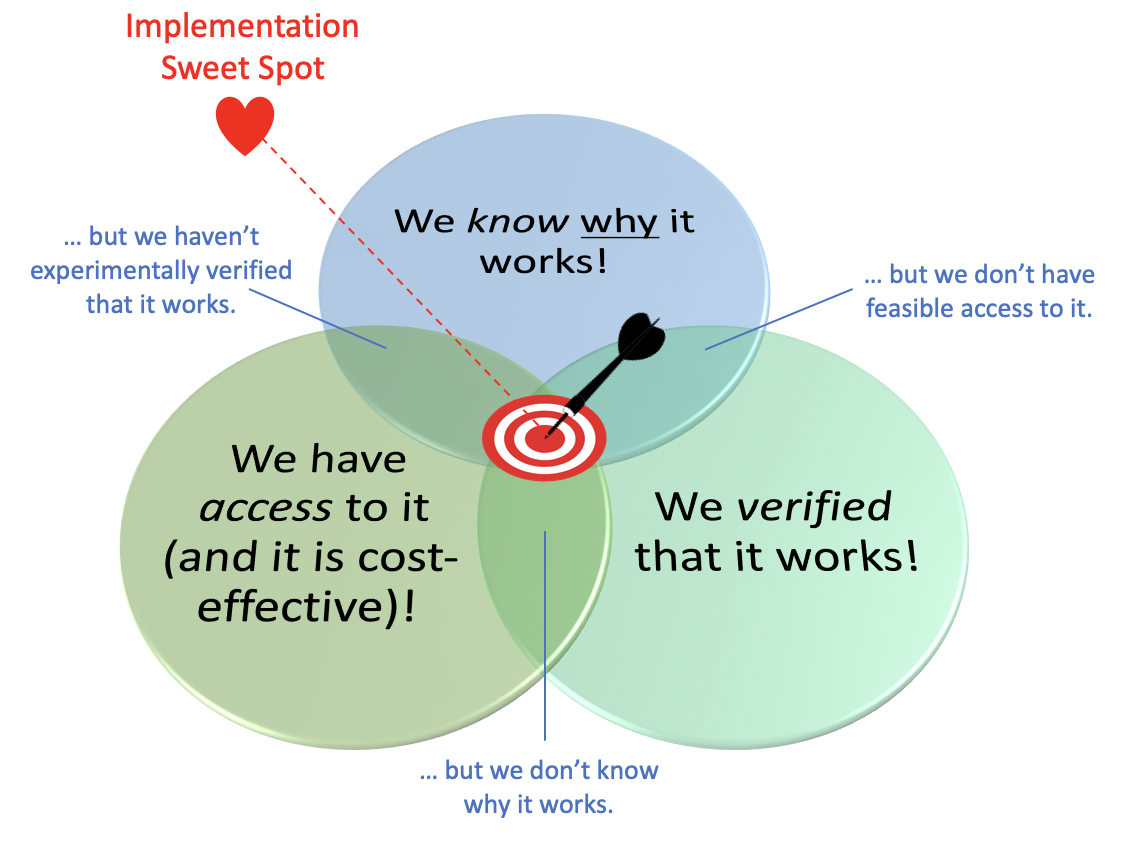}
    \caption{Generalized visualization of the `sweet spot' that a given solution will possess when it meets the desired criteria for implementation. Oftentimes, in advocating for novel solutions, arguments are made that -- despite being quite compelling -- may neglect simultaneous consideration of these areas. Of course, any decision will carry some inherent degree of risk, but recognizing how great that risk is and where it may come from is important for successful decision-making.}
    \label{fig:GeneralVenn}
\end{figure}

\begin{figure}[h!]
    \centering
    \includegraphics[width=1\linewidth]{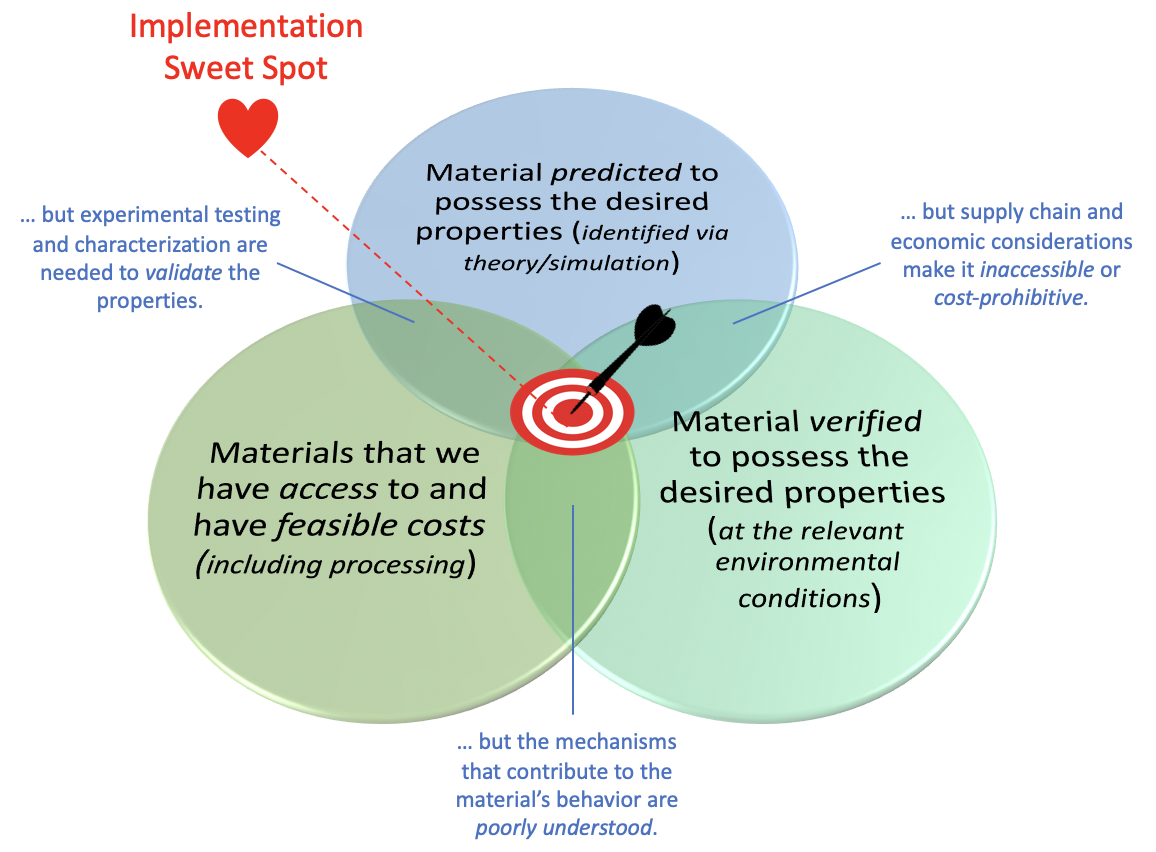}
    \caption{Visualization of the `sweet spot' that materials scientists are targeting when guiding the implementation of a given material for a given application.}
    \label{fig:VennMaterials}
\end{figure}

\begin{figure}[h!]
    \centering
    \includegraphics[width=1\linewidth]{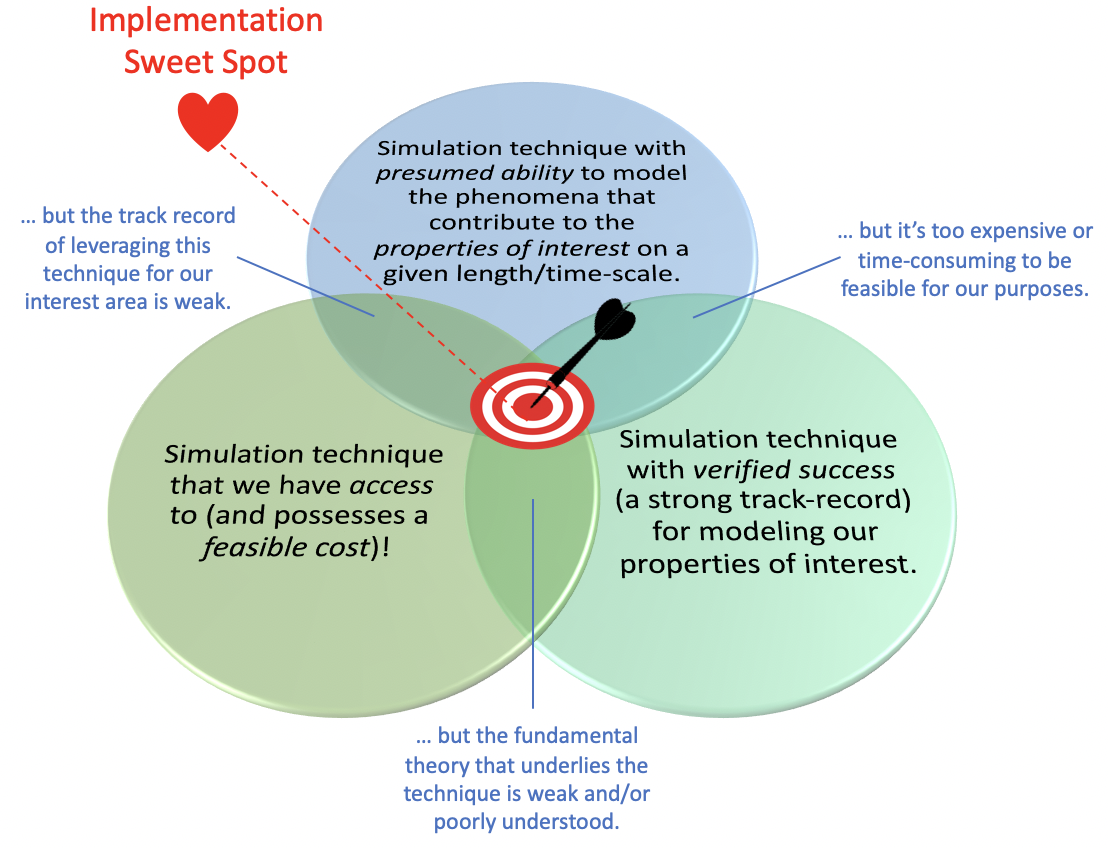}
    \caption{Visualization of the `sweet spot' that computational materials scientists target when deciding the modeling route to implement when studying a specific property/phenomenon.}
    \label{fig:VennSimulation}
\end{figure}

\section{\textbf{\textcolor{blue}{Sample Materials Engineering Problem}}}

Since the point of leveraging materials data infrastructure is to accelerate the pace at which tangible, real-world problems are solved, let us consider an engineering problem relevant to our previous discussion of PCM and SCM: phase-change materials. How could we accelerate the design of a better phase-change material? The questions to consider are: what methods could we use, what properties would we need to target, and how could we implement our strategies effectively to lessen the time and energy consumed by the computation? These are not hypothetical questions; there are already stakeholders working in this space.
The first question to address is that of the properties. Since there is yet to exist a comprehensive theory that links properties across all length- and time-scales, we must identify the niche that captures our needs most adequately, and select the computational approach(es) that are most suited to this niche. Each computational approach has its strengths and weaknesses; it is up to us to navigate these choices. 
Having a clear grasp of the desirable properties for this application is the first step towards efficacy in identifying (and designing) suitable candidate materials.

A simplified framework for navigating this territory is shown in its most generalized form in \textbf{Figure \ref{fig:GeneralVenn}}. Applying this framework to materials selection yields the Venn diagram shown in \textbf{Figure \ref{fig:VennMaterials}}. The most compelling argument for implementing a material comes when the material is situated at the center of the Venn diagram. Likewise, applying this framework to selection of a computational method for modeling the material yields the Venn diagram shown in \textbf{Figure \ref{fig:VennSimulation}}. Simplified though these diagrams are, they do provide a map for identifying the bottlenecks that limit successful implementation of a given solution.

To avoid spending unnecessary time and energy (and money) in obtaining a solution for the wrong problem, it is necessary to clearly define the objectives: what does a ``better'' phase-change material mean? This is certainly going to be a device-specific question, but even \emph{without} optimizing for a specific application area (in which case materials selection would be optimized for factors such as operating temperature, lifetime requirements, intensity of the computational workload, etc.), the first response to this question should be: what are the limitations of the materials that we have? Perhaps a material that will suit our needs is already available. So, is the material system itself one of the bottlenecks that has prevented PCM from achieving more widespread commercial success thus far? To some degree, yes. Consider the engineering criteria of PCM devices:
\begin{itemize}
    \item In order for PCM to compete with DRAM, displaying extremely low \textbf{latencies} is crucial.
    \item  In order for PCM to compete with Flash memory, the \textbf{density} of information able to be stored on the device needs to be extremely high.
    \item In both these competitions, \textbf{endurance} remains important; a device that performs well, only to degrade after a short amount of use, is not desirable for the mainstream market.
    \item \textbf{Power consumption} also needs to be minimized; the performance metrics need to be met without an excessive carbon footprint.
    \item Furthermore, many of the novel computing paradigms that have people interested in PCM rely on its ability to store \textbf{multiple bits} in one cell; this multi-level cell \textbf{(MLC)} architecture requires stable intermediate states, which in turn requires sufficiently large differences between the resistance of each state.
    \item Additionally, since PCM is marketed as \textbf{nonvolatile}, it needs to live up to that image by maintaining its data integrity.
\end{itemize}

\noindent What this list makes clear is that simply building PCM devices that `work' is not enough; these devices need to be competitive with other technologies on the market. It is this competition that provides the quantitative performance metrics that commercial PCM needs to target. 
The criteria indicated above (in \textbf{bold}) hold for any emerging SCM technology, not just PCM. If PCM cannot compete with those other technologies, commercial success will be eluded once more. The ability for a PCM device to meet these criteria is contingent on the processing and properties of the materials within the device, as well as on the device design itself. It is the perspective of the author that PCM limitations appear to stem from the materials themselves; reviewing the literature, the picture that has emerged over the past decades is that design engineers in this space are remarkably ingenious and have -- for their part -- found creative workarounds to a variety of the complications that have arisen. Rather, much of the yet-to-be-resolved trouble of PCM comes from the demands being placed upon the materials themselves. Finding solutions for this is where the materials science comes in.

\begin{figure}[h!]
    \centering
    \includegraphics[width=1\linewidth]{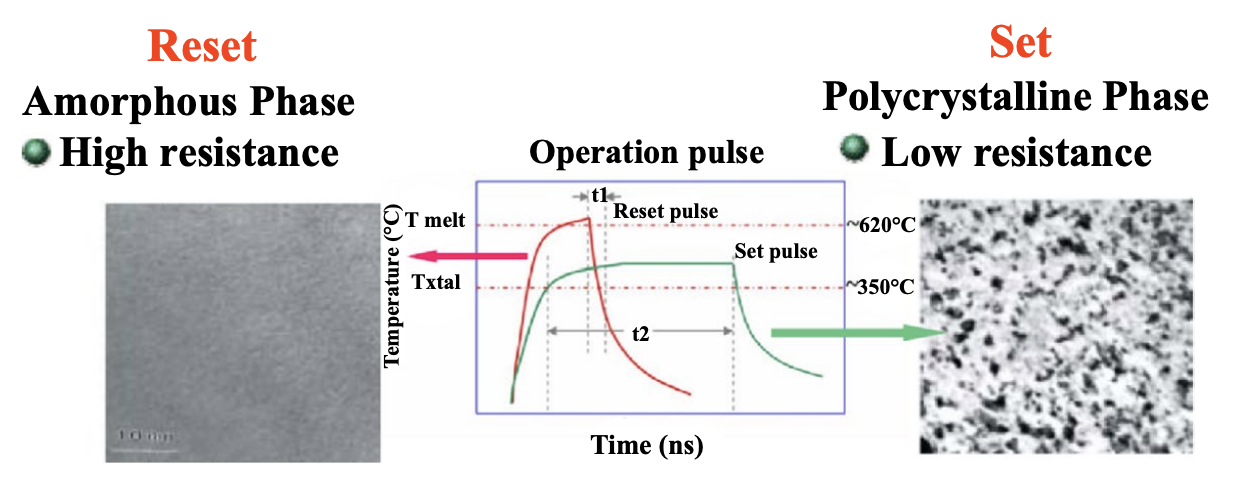}
    \caption{Phase change between high resistance and low resistance states, emphasizing the microstructure and temperature/time profile. \cite{Si_2021}}
    \label{fig:PCMphasechange}
\end{figure}

\begin{figure}[h!]
    \centering
    \includegraphics[width=1\linewidth]{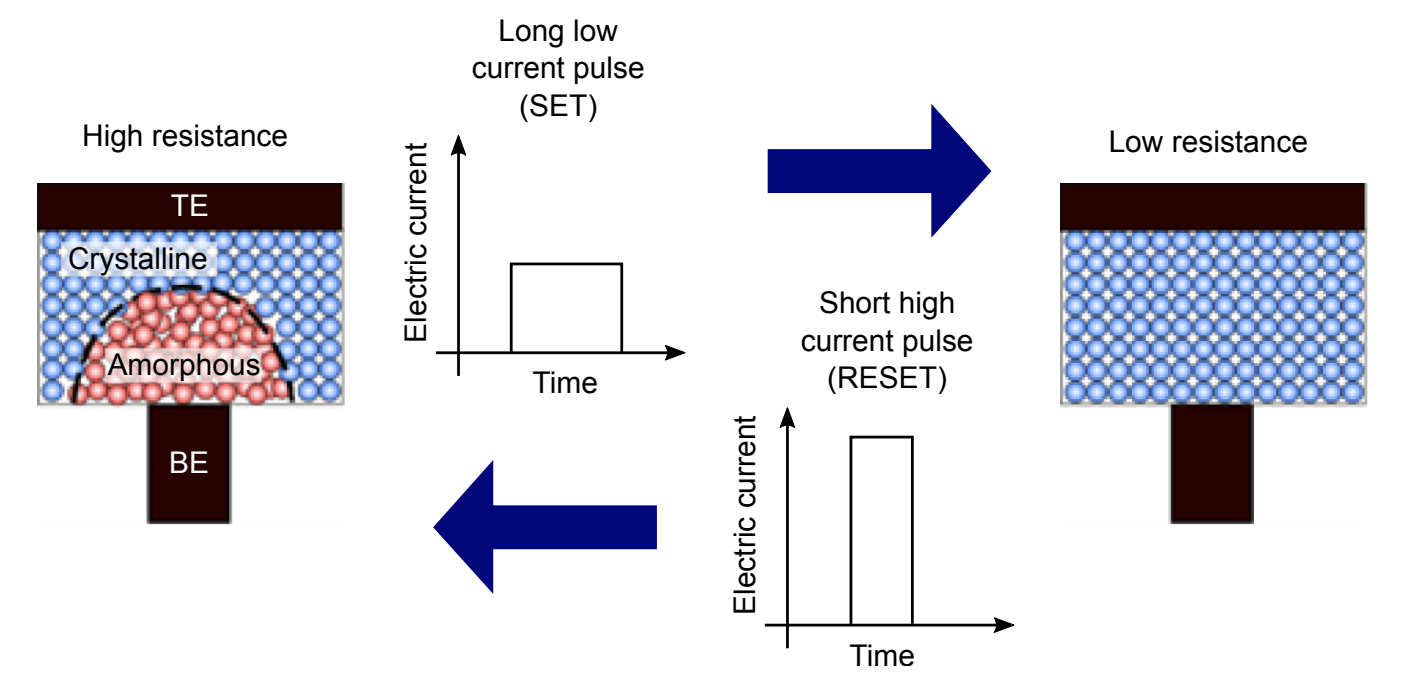}
    \caption{Phase change between high resistance and low resistance states, emphasizing the memory cell design and the amplitude of the SET/RESET current pulses. Designs vary, but all PCM devices essentially consist of a layer of phase-change material, sandwiched between two metal electrodes.\cite{Gallo_2020}}
    \label{fig:PCMoverview}
\end{figure}

As described previously on page 10, the conceptual framework of phase-change materials is simple: the crystalline phase has high electrical conductivity, the amorphous phase has low electrical conductivity, and a bit of information (0 or 1) is measured via probing the resistivity differences between the two states. The conversion between these two states is thermally achieved and controlled via electrical pulses to a heating element adjacent to the phase-change material. An overview of this conversion is shown in \textbf{Figure \ref{fig:PCMphasechange}} and \textbf{Figure \ref{fig:PCMoverview}}. The transition from a Reset phase to a Set phase is accomplished with a current pulse that heats the amorphous material to above its recrystallization temperature. (For the material shown in \textbf{Figure \ref{fig:PCMphasechange}}, the recrystallization temperature is about 350 $^\circ$C.) For the transition from the Set phase to the Reset phase, a current pulse again heats the material. However, this time, the pulse induces a higher temperature (to above the melting temperature, which for the material in \textbf{Figure \ref{fig:PCMphasechange}} is about 620 $^\circ$C) and then is abruptly cut off. This rapid quenching limits the atomic mobility, essentially `freezing' the atoms in place, and resulting in the disordered, amorphous, glassy state. Since both transitions rely on heating, thermal management is a key aspect of PCM. Not only is the presence of a heat sink implicit, but the energy consumption required for thermally cycling the system to sufficiently high temperatures can be a nontrivial sum (effectively increasing the carbon footprint). The thermal demands of PCM are one of the reasons that it actually performs better at smaller scales; the smaller the material quantity within an individual memory element, the less energy input required to achieve the desired temperature ranges.

\begin{figure}[h!]
    \centering
    \includegraphics[width=1\linewidth]{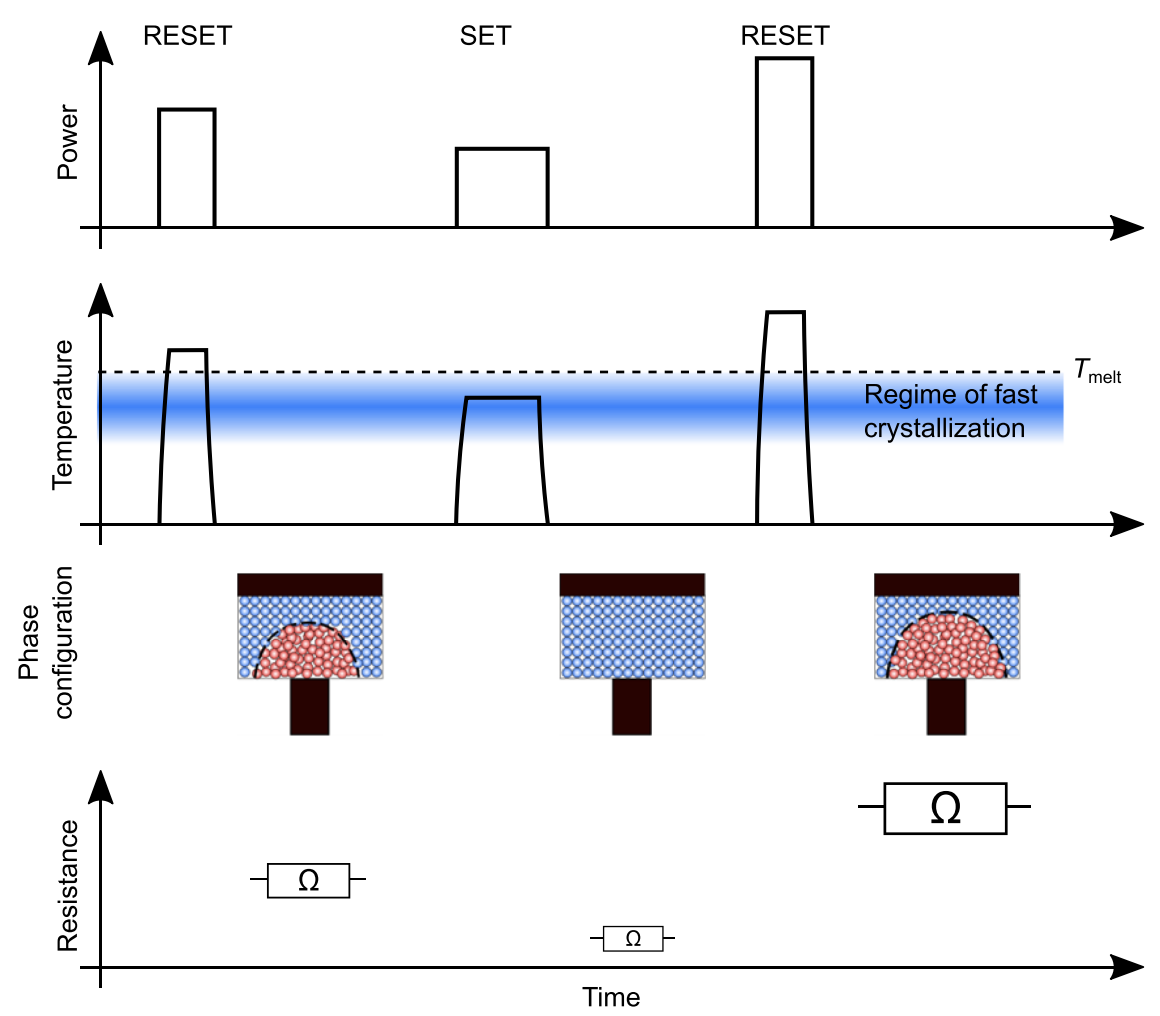}
    \caption{By adjusting the amplitude of the pulse power, the size of the amorphous region can be controlled and therefore the resistance as well. This is how intermediate resistance states can be introduced into PCM.\cite{Gallo_2020}}
    \label{fig:ControlAmorphousSize}
\end{figure}

\begin{figure}[b!]
    \centering
    \includegraphics[width=0.8\linewidth]{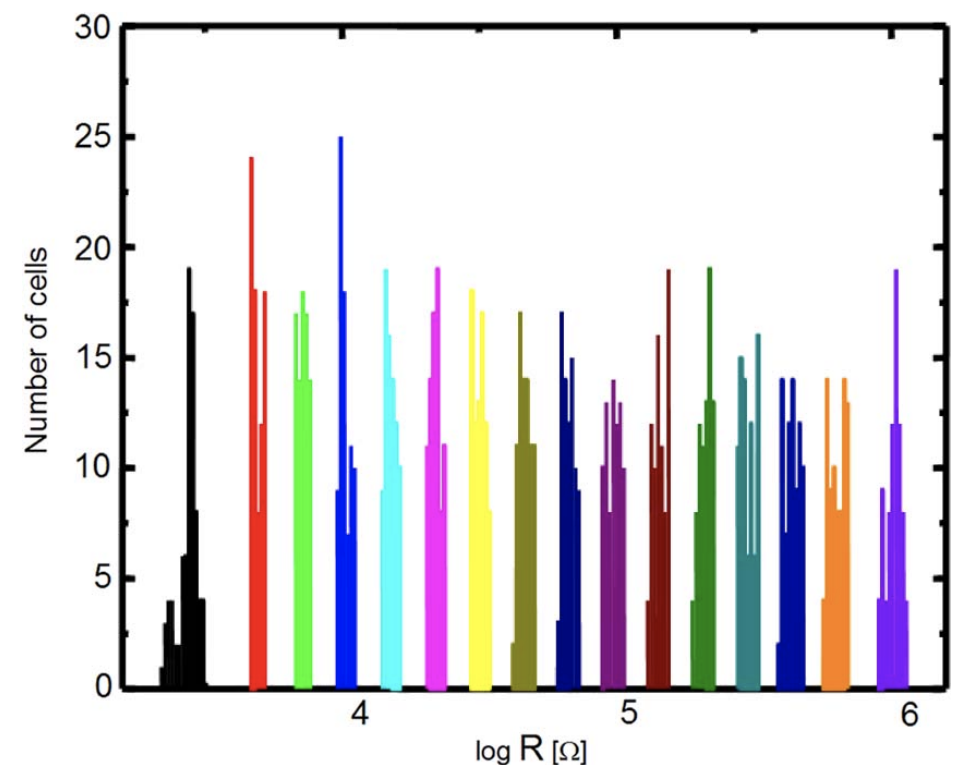}
    \caption{Demonstrated 4 bit/cell PCM architecture, obtained via 16-level programming. Strong signals with a narrow distribution and measurement sensitivity are particularly crucial for MLC use cases.\cite{Wong_2010}}
    \label{fig:fourbits}
\end{figure}

This may sound like a simple concept. In practice, there is a great deal of nuance, and even after 50 years of studying these materials, there still remain important unsolved questions.\cite{Fantini_2020} A good 2020 paper that summarizes what is known about the physics of phase-change materials is referenced.\cite{Gallo_2020} They also provide the following metrics:
``The key requirements for a PCM device to be used for electrical data storage is high endurance (typically $>$10$^8$ SET/RESET cycles before failure), low RESET current ($\le$200 µA highly desirable), fast SET speed ($\le$100 ns), high retention (typically 10 years at 85 $^\circ$C, but there are different requirements for embedded memories), good scalability ($<$45 nm node) and low intra- and inter-cell variability. \textit{While a single PCM device can be designed to easily meet \textbf{one} of the above constraints, the challenge is to build an array of devices that meets \textbf{all} of the above requirements.} Individual PCM devices have demonstrated $>$10$^{12}$ endurance cycles, $<$10 µA RESET current, $\sim$25 ns SET speed, projected 10 years retention at 210 $^\circ$C and sub-20 nm node scalability.''

While the initial description of PCM may sound binary (amorphous phase and crystalline phase), the reality is that they exist in a continuous spectrum of states, based on the degree of amorphization. The likelihood that the phase-change material is fully crystalline or fully amorphous is extremely low. On one hand, this can be a very useful feature; it enables the existence of multiple resistance levels. \textbf{Figure \ref{fig:ControlAmorphousSize}} shows how variations to the pulse power results in different observed resistivity. (In this case, the `mushroom' shape results from the device configuration, where the region near the bottom electrode experiences the greatest heating.) Since the information is encoded in the resistivity levels, introducing additional levels enables more information to be stored in the same physical volume of material. This is the physical phenomenon that MLC architecture leverages. The number of levels scales exponentially with the bits; a 1-bit design needs 2 stable levels, a 2-bit design needs 4 stable levels, a 3-bit design needs 8 stable levels, a 4-bit design needs 16 stable levels, and so forth. An example of one such 4-bit architecture is shown in \textbf{Figure \ref{fig:fourbits}}. On the other hand, the primary challenge issue of MLC is that the same physical basis that provides a continuum of resistivity states also introduces instabilities. Storing information through the extent of crystallization within the material is well and good, as long as these the atoms truly behave as `frozen' in place. 
However, PCM devices display a phenomenon known as `resistance drift'; the resistance of the cells tends to gradually increase over time. A variety of theories have been proposed to explain this, but the most commonly accepted explanation in the literature is that it orginates from the instability of the amorphous regions. The amorphous regions are non-equilibrium metastable states, so there exists a thermodynamic driving force for the structure to rearrange towards more energetically favorable configurations. According to the 2020 paper:\cite{Gallo_2020}
``Resistance drift in PCM devices has been mostly explained as a consequence of spontaneous structural relaxation of the amorphous phase-change material... When the molten phase-change material is quenched rapidly, the atomic configurations are frozen into a highly stressed glass state. Over time, the atomic configuration of this state will relax towards an energetically more favorable ‘ideal glass’ configuration.'' Note that the spontaneous structural relaxation does not mean increased crystallization; if that were the case, the resistance would presumably decrease over time instead, because the crystalline state is lower resistance. Resistance drift presents a significant problem for MLC architectures, as the ability to reliably distinguish between states may deteriorate as drift progresses. However, this does not pose a significant problem for two-tier PCM (SLC architecture). Drift primarily affects the amorphous states, not the crystalline states, and since the amorphous state is the high-resistance state in SLC, it is not of significant practical concern that the high-resistance state is increasing its resistance over time. In this case, drift is actually somewhat beneficial for the device longevity.

\begin{figure}[h!]
    \centering
    \includegraphics[width=1\linewidth]{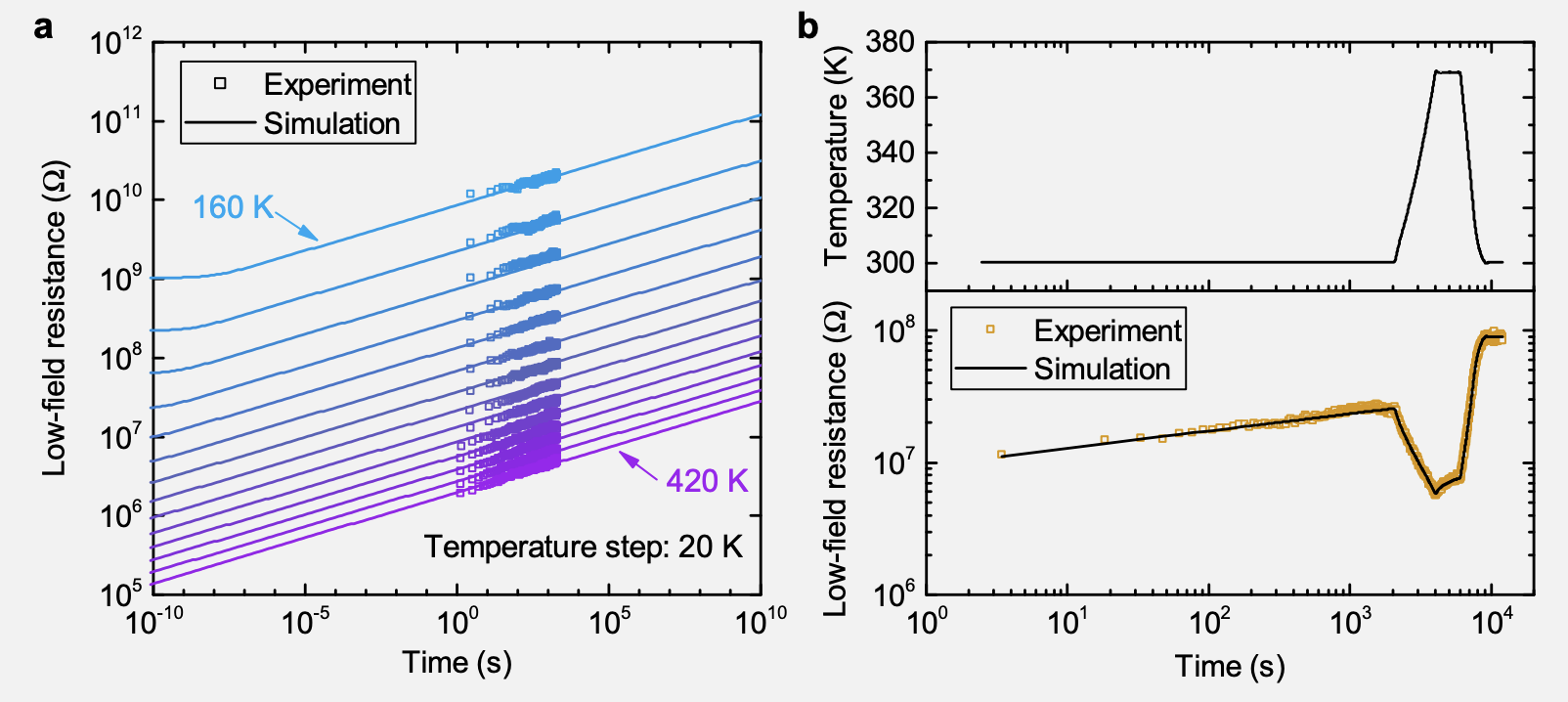}
    \caption{Measurements of resistance drift of 90-nm doped Ge$_2$Sb$_2$Te$_5$ mushroom-type PCM cells obtained under low-field (applied constant voltage of 0.2 V to the cell), (a) across a variety of temperatures, (b) upon the application of a time-varying temperature profile, shown on top.}
    \label{fig:experimentvssimulation}
\end{figure}

Yet, if the goal is to build devices for neuromorphic computing applications, the MLC architecture is relevant and drift ought to be avoided at all costs. Over the years, a variety of explanations for drift of the amorphous phase have been proposed, including stress release, decrease in defect density, shift of the Fermi level, and increased band gap.\cite{Wong_2010}
Unfortunately for materials selection purposes, the precise mechanisms are still not fully understood. 
For instance, the Gibbs relaxation model describes drift as the relaxation of structural defects; the defects have a distribution of activation energies associated with their removal, and defects with lower activation energies are removed first. The implication here is that the defects increase the electrical conductivity, and by removing them, the resistance increases.\footnote{\scriptsize{Note that, in general, defects can both increase or decrease the conductivity; defects that act as `shallow traps' can provide donor/acceptor levels to semiconductors, which increases the conductivity, while defects that act as `deep traps' can decrease the conductivity by both pinning the Fermi level to the middle of the bandgap and by acting as carrier recombination centers.}}
Another model, the `collective relaxation' model, describes drift as the entire amorphous system collectively relaxing towards the energetically favorable `ideal glass' state via transitions between neighboring `unrelaxed' amorphous states. Both models account for the temperature dependence and logarithmic behavior that drift exhibits; to quote the 2020 paper:\cite{Gallo_2020} ``As of now, it is not possible to discriminate between the two models from existing experimental resistance drift data.'' Consider the close fit between experimental data and simulation data shown in \textbf{Figure \ref{fig:experimentvssimulation}}; the simulation in question leveraged the collective relaxation model, but as the 2020 paper states: ``The dependencies of the resistance on time and temperature from experimental measurements reported in [\textbf{Figure \ref{fig:experimentvssimulation}}] can be equally well captured by both the collective relaxation model and the two-state model for relaxation based on Gibbs approach.'' 
In fairness, this is not always a bottleneck for practical design purposes; all models introduce assumptions and approximations, and if a given model fits the data well enough for a certain regime, it may be sufficiently accurate regardless. Rather, such uncertainties make materials screening more difficult.

In addition to the myriad of questions pertaining to resistance drift (even whether the bandgap stays constant or increases during drift is still an open topic), other aspects of phase-change materials with unresolved questions include:\footnote{\scriptsize{To avoid distraction, an in-depth discussion of each of these is not provided here.}}

\begin{itemize}
    \item Origin of the peculiar I-V curve of phase-change materials (i.e. origin of the threshold switching mechanism). This is perhaps the most prominent of all the open questions, since the threshold switching behavior is the key distinguishing factor of phase-change materials.
    
    \item The Poole-Frenkel conduction model is often used to describe the variation of conductivity with applied voltage (for small voltages, it predicts a linear current, and for high voltages it predicts exponential current), but the physical origin of the parameter $\delta$ used in the Poole-Frenkel conduction model is not well-understood, nor is the influence of different types of defect states on transport.
    \item Whether the crystallization kinetics of PCM at elevated temperatures are nucleation-driven (a stochastic process) or growth-driven (a deterministic process).
    \item The effect that the thermal cycling and the inhomogeneous temperature distributions have on the nucleation and growth kinetics.
    \item Whether the Arrhenius-type temperature dependence of the growth velocity occurs in the glass or supercooled liquid state remains an open question. This is difficult to study because precisely measuring the glass transition temperature of phase-change materials is difficult.
    \item The emergence of physical properties (like magnetoelectric behavior) from the superlattice multilayer structure.
    \item Origin of $\frac{1}{f}$ noise in phase-change materials; noise was been one of the least studied topics of PCM device physics.
    \item How to obtain desirable properties without the use of elements like arsenic or tellurium, which are particularly unpleasant to work with.
    \item How physical properties of PCM materials change in the sub-15 nm range (i.e. which mechanisms remain valid in ultra-scaled devices).
    
\end{itemize}

\noindent With such looming knowledge holes, it is evidently clear that the process of designing ``better'' phase-change materials is not just an engineering question; it is intimately intertwined with fundamental materials science research as well. As the former is pursued, so too does the latter advance. There have been plenty of papers published which attempt to address these questions.  However, such explanations do not appear to be unifying; conflicting data often arises. A recurring theme that arises in PCM research is that different material systems may have different mechanisms that contribute to the overall observed behavior. Thus, tuning the features of one system may need to be accomplished through different routes than that for another system. 

While lively debate continues to circulate regarding the theories that govern the behavior of PCM materials, significant experimental data has been gathered regardless. Focusing on what is known, we can overlay a physical picture atop the desired engineering criteria as follows:

\begin{itemize}
    \item \textit{Low latencies:} The operating speed of PCM is limited by the speed of crystallization in the material, specifically by the `SET’ operation; typical figures for the SET pulse and RESET pulse are about 150 ns and 120 ns, respectively.\cite{Qiu_2015} This is because — as shown in \textbf{Figure \ref{fig:PCMphasechange}} and \textbf{Figure \ref{fig:PCMoverview}} — a short high pulse leads to the amorphous (RESET) state, while a longer pulse leads to the crystalline (SET) state, i.e. the write speed is constrained by how long it takes to crystallize the amorphous region. Many of the commonly studied fast switching phase-change materials have a simple cubic or rocksalt structure, so switching between amorphous and crystalline states requires little atomic movement. A physical quantity that counteracts the driving force, limiting crystallization, is the viscosity. Fast switching phase-change materials tend to exhibit low ionicity and low tendency towards hybridization, and resonance bonding is believed to play an important role.\cite{Wong_2010} Furthermore, the SLC architecture has faster write operations than MLC does, because MLC uses an iterative program-and-verify procedure.\cite{Qiu_2015} Scaling appears to have a positive effect; for ultrascaled devices, switching speeds as low as 1 ns have been demonstrated.\cite{Wong_2010}
        
    \item \textit{High memory density:} MLC architecture increases the memory density, but it is not the only way to do so. The other primary approach that has been explored is three-dimensional device design. The `crosspoint' architecture is particularly popular in this regard. To quote a 2020 paper:\cite{Kim_2020_evolution} ``The X-point structure has been thought of as the most ideal structure for memory devices. It has an ultimate cell areal density of 4F$^2$ and requires only two photo-mask steps for patterning, which makes it possible to minimize the overall number of process steps. In addition, it is stackable, which enables 3-D structures. Based on these advantages, intensive efforts have been made to achieve this structure.'' An integral aspect of this structure is the selector (in \textbf{Figure \ref{fig:XpointConferenceSlide}}, the selector is the `diode,' labeled underneath the `memory element'). The selector is largely responsible for the device size, since it occupies the most layout area in the memory cell. Aside from the diode, other selectors include:\cite{Kim_2020_evolution} ``the stackable amorphous selectors, such as the ovonic threshold switch \textbf{(OTS)} and mixed ion–electron conductor \textbf{(MIEC)}. After intensive study and research, Intel-Micron announced the successful development and mass production for stackable 3D X-point memory using an amorphous selector. SK-Hynix also announced the successful development of stackable X-point PCM using a novel two-terminal selector.’’ \textbf{Table \ref{tab:selectorproscons}} compares the pros and cons of three different selectors. While scaling of the memory cell improves memory density, it is noted that -- particularly below the 10 nm range -- many properties of phase-change materials are size-dependent, such as crystallization temperatures and times, activation energies for crystallization, melting temperatures, resistances, and optical and thermal properties.\cite{Wong_2010} These nuances have not been fully explored, but in a convenient twist of fate, it appears that the phase change properties remain even at very small scales. \textbf{Figure \ref{fig:tinyTEM}} shows TEM images of two different sizes of GeTe nanoparticles; the paper states:\cite{Wong_2010} ``These nanoparticles are as small as about two to three times the lattice constant, so this will be close to the ultimate scaling limit of phase-change technology as far as the phase-change materials themselves are concerned.''

\begin{table}[t!]
    \scriptsize{

    \begin{tabular}{|>{\centering\arraybackslash}p{1.8cm}|>{\centering\arraybackslash}p{1.7cm}|>{\centering\arraybackslash}p{1.7cm}|>{\centering\arraybackslash}p{1.7cm}|}
        \hline
        & \textbf{OTS} & \textbf{MIEC} & \textbf{Doped $\alpha$-Si} \\ \hline
        \textbf{Material} & AsTeGeSiN & Cu-based & As-SiO$_2$ \\ \hline
        \textbf{$I_{\text{off}}$} & 10nA@0.5 $V_{\text{th}}$ & 100pA@0.5 $V_{\text{th}}$ & 20nA@0.7 $V_{\text{th}}$ \\ \hline
        \textbf{$I_{\text{on}}$} & 100$\mu$A (CC) & 10nA (w/o CC) & $>$100$\mu$A \\ \hline
        \textbf{Instability} & RTN \& \emph{drift} & Small RTN & Huge RTN \\ \hline
        \textbf{$V_{\text{th}}$ adjustability} & Good & No & Limited \\ \hline
        \textbf{Endurance} & $>10^8$ & $>10^5$ & $>10^5$ \\ \hline
    \end{tabular} \\ \\
    }
    \caption{Pros and cons of three popular PCM selectors.\cite{Kim_2020_evolution}}.
    \label{tab:selectorproscons}
\end{table}

\begin{figure}[b!]
    \centering
    \includegraphics[width=1\linewidth]{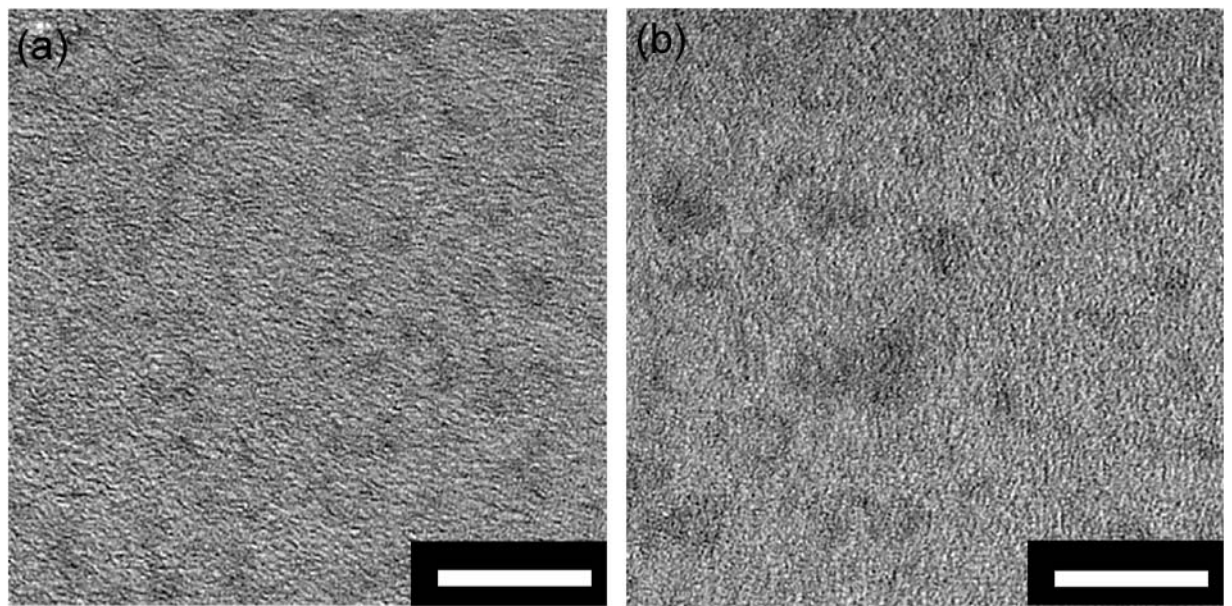}
    \caption{TEM images demonstrating the phase-change effect is still present in nanoparticles of size (a) 1.8 $\pm$ 0.44 nm, and (6) 2.6 $\pm$ 0.39 nm. The scale bar in the images is 10 nm. \cite{Wong_2010}}
    \label{fig:tinyTEM}
\end{figure}

    \item \textit{High device longevity:} One contributor to the high endurance generally displayed by PCM is that -- unlike with DRAM -- reads with PCM are non-destructive. Rather, the primary wear mechanism in PCM is the writes (i.e. the thermal cycling between the SET and RESET states). It is believed that the thermal expansion and contraction that occurs during writes degrades the electrode-storage contact.\cite{Wong_2010} The high cycling endurance of common PCM materials is generally attributed to the crystal structure. For instance, the rocksalt structure of the GST alloy (Ge$_2$Sb$_2$Te$_5$) is shown in \textbf{Figure \ref{fig:rocksalt}}. To quote a review article:\cite{Lacaita_2006} ``The chemical short-range order of the ternary GST system is almost the same in the two phases. It means that in the amorphous state, interactions between different Ge$_2$Sb$_2$Te$_5$ building blocks are weakened. The structure is therefore allowed to relax, but neither strong covalent bonds are broken nor atoms drastically change their position in the lattice. The Te-sublattice is partially preserved as well as the conservation of the local structure around the Sb atoms. The original crystalline structure can be therefore quickly and reliably recovered.'' PCM materials do facilitate extensive cycling, but they are

    \noindent not invulnerable to failure. The three main failure modes are:\cite{Burr_2016_review}
    \begin{enumerate}
        \item `cell-open'
        \item `stuck-low' (RESET failure)
        \item `stuck-high' (SET failure).
    \end{enumerate}
    In cell-open failure, voids agglomerate and accumulate towards the bottom electrode; when the voids are large enough to completely block the current from reaching the bottom electrode, an `open' cell will result. The mechanical stress of thermal expansion/contraction helps drive void formation. In stuck-low failure (for the GST system), the repeated cycling promotes Sb enrichment in the active volume, which decreases the resistance and increases the RESET current; this failure mode is recognized by a `right-shift' of R-I characteristics.\cite{Burr_2016_review} The Sb enrichment (phase segregation) is driven by incongruent melting and recrystallization at the liquid-solid boundary. Cell designs that induce complete melting help prevent stuck-low failure by suppressing phase segregation. In stuck-high failure, a combination of void formation and phase segregation increases the resistance. In the stuck-high situation, the device stays in a high-resistance state and refuses to SET, while in the stuck-low situation, the device stays in a low-resistance state and refuses to RESET.\cite{Kim_2020_endurance} All of these failure modes ultimately result from undesired atomic migration. Atomic migration is -- to some degree -- inevitable, but doping, materials selection, and design architecture adjustments can yield orders of magnitude improvements.

\begin{figure}[t!]
    \centering
    \includegraphics[width=0.6\linewidth]{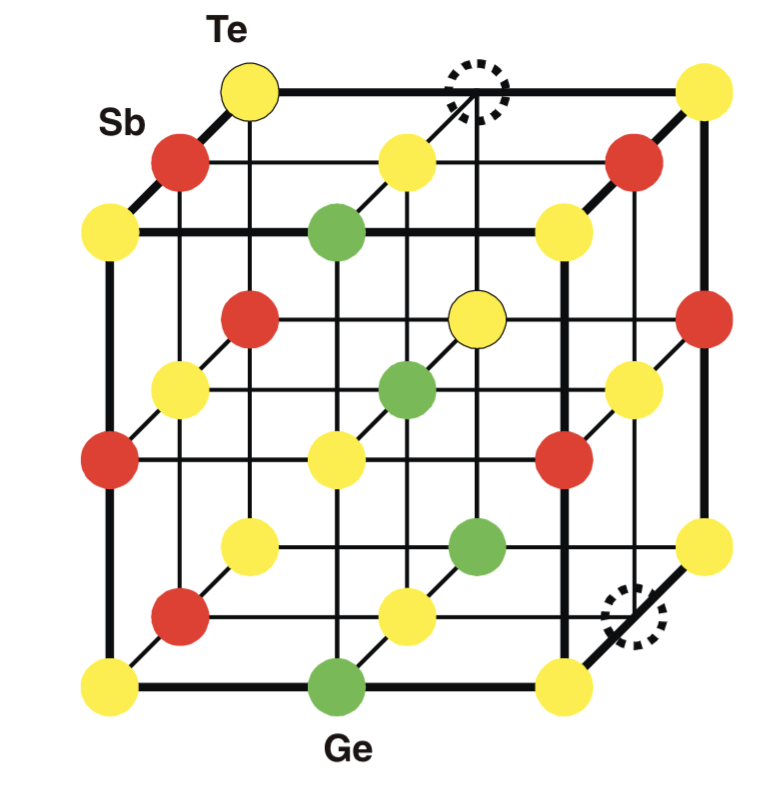}
    \caption{Rocksalt structure of the GST crystal (Ge$_2$Sb$_2$Te$_5$). One sub-lattice is comprised of tellurium atoms, while the other sub-lattice is comprised of antimony, germanium, and vacancies.\cite{Lacaita_2006}}
    \label{fig:rocksalt}
\end{figure}    

    \item \textit{Minimal power consumption:} The greatest power consumption is generally attributed to the RESET operation, since the current required to melt the material is higher than than the current required for recrystallization. As a 2016 review article strongly states:\cite{Burr_2016_review} ``The higher temperatures needed for melting (600 or higher) cause the RESET operation to dominate power considerations for PCM technology.'' Indeed, the commonly employed phase-change materials have melting temperatures in the range of 500-700 $^\circ$C.\cite{Wong_2010} Materials with lower cohesive energies (correlated to lower melting temperature) will lead to lower power consumption, with the caveat that the device stability may be lessened as well. Material characteristics that would significantly reduce power consumption for the RESET operation are:\cite{Burr_2016_review} low heat capacity, low heat of fusion, low melting point, low thermal conductivity, and high electrical resistance in the crystalline phase.
    The device design can also have a great impact on the required RESET current.\footnote{\scriptsize{Furthermore, creative device architectures like interfacial PCM \textbf{(iPCM)} appear to exhibit switching behaviors that depend on the short-range movement of atoms through Te-Te conduction channels, instead of on the order-disorder transition.\cite{Burr_2016_review} Being able to display threshold switching without melting would reduce the power consumption and improve endurance. For instance, an iPCM-inspired device comprised of Sn$_{10}$Te$_{90}$/Sb$_2$Te$_3$ superlattice displayed RESET current as small as 3 $\mu$A.\cite{Soeya_2013}}}
    According to the 2016 review article:\cite{Burr_2016_review} ``A thermally confined electrode (TaN/TiN/TaN) could drastically reduce the RESET current compared to a solid TiN electrode.'' While most of the efforts towards decreasing power consumption focus on the RESET current, a 2020 review article claims that -- depending on the architecture -- as the devices scale:\cite{Kim_2020_evolution} ``What actually dominates the overall power consumption is not the Reset write but the Set write. ... Below 2× nm technodes with a C-type PCM, greater importance should be put on the Set operation.''

    \item \textit{Multiple bits per cell:} The physical basis behind storing multiple bits per cell is described on page 17. To recap: by varying the ratio of amorphous/crystalline regions, intermediate resistance states can be introduced, and additional information can be stored by leveraging these additional tiers. The ratio can be adjusted in multiple ways. Applying a RESET pulse that dissipates more power will allow the melting temperature to be reached further away from the bottom electrode, resulting in a larger amorphous region. Also, varying the width of the SET pulse (or the length of its trailing edge) will affect the proportion of the cell volume that is able to recrystallize. Overall, successfully implementing this multi-level cell \textbf{(MLC)} architecture requires sufficiently large differences in resistance between the amorphous and crystalline states, tolerating or preventing resistance drift, and sufficiently sensitive reads that can distinguish between resistance levels. 

    \item \textit{Nonvolatile:} Nonvolatile memory retains its data even if the external power supply is removed. For PCM to behave as nonvolatile, the amorphous and crystalline microstructures (and, for MLC, the continuum of states in-between) need to be sufficiently stable, such that atomic mobility is minimized and the encoded resistance levels are preserved. Thermodynamic instability can be acceptable if the kinetics are sufficiently suppressed. The targets for data retention time are usually quoted for 10 years at some specified temperature (such as room temperature, 85$^\circ$C, 150$^\circ$C, 210$^\circ$C, etc.). It is noted that the desired difference in response time for crystallization between the actively-writing-data scenario (nanoseconds) and the dormant-storage scenario (years) spans 17 orders of magnitude.\cite{Wong_2010} While  clever ways to achieve both low latencies and long-term data retention do exist, optimizing for either scenario tends to come at the expense of the other.\footnote{\scriptsize{The gallium antimonide (GaSb) system is interesting, in that it offers both fast switching speed and good thermal stabilty, but it undergoes a decrease (about 5\%) in mass density upon crystallization (i.e. the amorphous phase is denser than the crystalline phase); this is in contrast to typical phase-change materials like GST, which tend to increase in mass density (about 5-6\%) upon crystallization.\cite{Burr_2016_review}}} One of the reasons the GST system has attracted so much focus is that it has provided a good compromise between speed of transformation and stability.\cite{Fantini_2020}
\end{itemize}

\noindent Two key takeaways from this overview are that (1) tailoring materials for application in PCM means navigating a complex, dynamic space in which the theories themselves which govern the material properties are still evolving, and that (2) implementing the desired engineering specifications for PCM devices often leads to contradictory requirements for the material properties. For instance, doping GST with extra nitrogen decreases the RESET current (good for energy consumption) and suppresses elemental segregation (good for endurance), but at the cost of a slower SET speed (bad for latencies).\cite{Burr_2016_review} To quote the 2010 review article:\cite{Wong_2010} ``It is clear that the search for the best phase-change material is a multiparameter optimization process with some seemingly contradictory requirements such as high stability of the amorphous phase at operating temperature, but very fast crystallization of the amorphous phase at switching temperature.'' Accordingly, computational tools and workflows can make the task of sifting through this complicated landscape of design consideration much more manageable. For the remainder of this case study, we consider a sample PCM materials informatics workflow and discuss what the computational infrastructure for implementing this workflow in accordance with the NASA 2040 Vision (\textbf{Table \ref{tab:NASA2040Vision}}) might look like.

\begin{figure}[t!]
    \centering
    \includegraphics[width=1.1\linewidth]{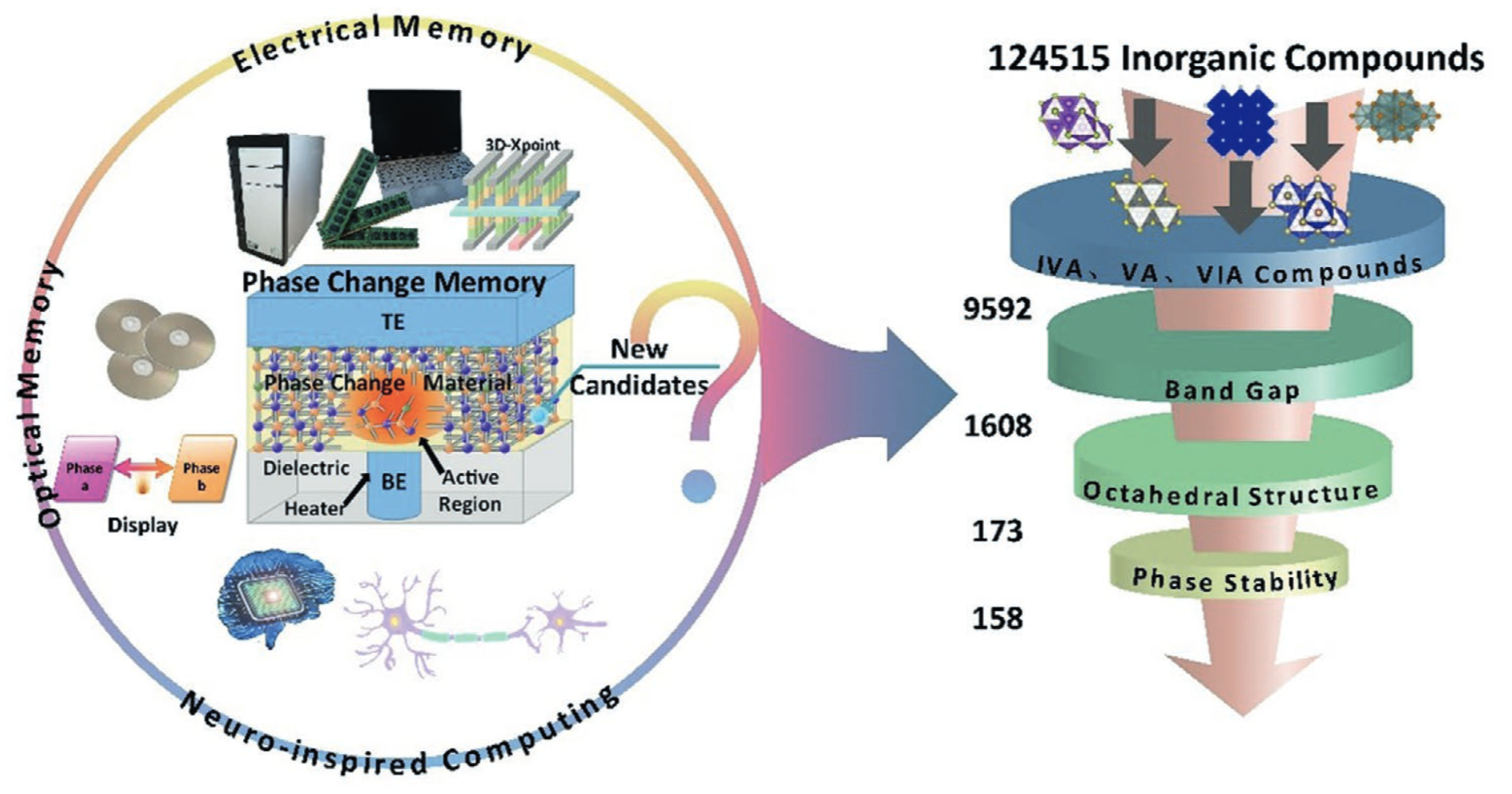}
    \caption{Computational screening approach taken by Liu et al. in their 2021 paper. 124515 material structures taken from the Materials Project database were filtered. \cite{Liu_2021_screening}}
    \label{fig:screening1}
\end{figure}

\begin{figure}[b!]
    \centering
    \includegraphics[width=1\linewidth]{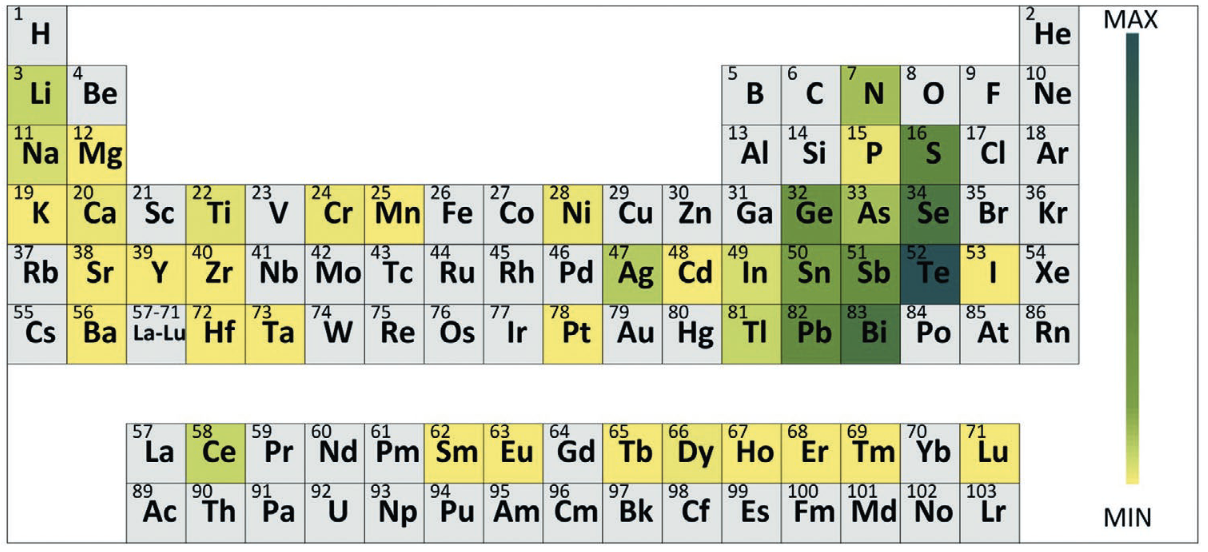}
    \caption{Distribution of elements present in the final, filtered 158 candidate structures. Elements that did not appear in any of the final structures are shown in grey.\cite{Liu_2021_screening}}
    \label{fig:screening2}
\end{figure}

A sample PCM materials informatics workflow is taken from the following 2021 paper: \textit{High-Throughput Screening for Phase-Change Memory Materials} by Liu et al.\cite{Liu_2021_screening} The researchers claim that ``the  work  offers  the  first  systematic  high-throughput  screening  of  PCM  materials  from  more  than  120,000 inorganic crystal structures,'' and they provide a step-by-step description of their workflow in the Experimental section. The objective was to accelerate discovery and development of suitable material systems by data-mining the Material Project database (described on page 4). They describe their motivation:\cite{Liu_2021_screening} ``Currently,  the  improvement  of  performances  by  searching  new  materials  is  mainly  based  on  experimental   `trial-and-error'  methods.   Such   a   systematic   searching is often limited by experimental conditions and often requires  very  long  development  duration  and  high  cost.  For  example, it has been more than 20 years to the discovery of current  mainstream  GST  materials since  PCM  technology  was  proposed. In this work, we employ 124,515 inorganic crystal structures in  the  Materials  Project  \textbf{(MP)}  database  for  high-throughput  screening  and  mining  PCM  related  materials  based  on  the  structural  and  electronic-property  descriptors.'' The four tiers of screening that they implemented are visualized in \textbf{Figure \ref{fig:screening1}} and described below; the elemental distribution of the 158 candidate structures that made it through all four tiers is shown in \textbf{Figure \ref{fig:screening2}}.

\begin{enumerate}
    \item Noting that PCM materials tend to possess chemical compounds containing elements in groups IVA, VA, and VIA (constructing a p-orbital bonding network, with bonding character between ionic and covalent), Liu et al. apply a preliminary screening of elemental compositions. They chose to exclude the oxides, due to their tendency to form stronger ionic bonds, destroying the p-bonding network.
    \item Noting that PCM materials usually possess narrow bandgaps (tellurides $\sim$0.2-0.9 eV, selenides $\sim$0.4-1.3 eV, sulfides $\sim$0.9-1.8 eV)\cite{Xu_2021_screening}, Liu et al. apply a secondary screening based on the bandgap data; specifically, only materials with calculated bandgap below 1 eV were considered. This cut-off metric is more qualitative than quantitative, because the bandgap values listed in the Materials Project database come from DFT calculations using the GGA and GGA+U functionals (in which bandgaps tend to be underestimated by $\sim$40\%, compared to experimentally measured values).\cite{MaterialsProjectBandgaps} However, through this 1 eV cut-off, materials with notably larger bandgaps are excluded.
    \item Recognizing that the octahedral (O:6) and trigonal non-coplanar (TY:3) structures facilitate the p-orbital bonding motif present in the commonly studied phase-change materials, Liu et al. apply a tertiary screening where materials that do not possess coordination environments similar to O:6 or TY:3 are filtered out.
    \item Choosing to leverage `energy above hull,' $\Delta$E$_{\text{hull}}$, as a convenient proxy for thermodynamic stability ($\Delta$E$_{\text{hull}}$ can be calculated via DFT and, accordingly, is included within the scope of the Materials Project data), Liu et al. apply a final screening where only materials with $\Delta$E$_{\text{hull}}<$100 meV atom$^{-1}$ are considered sufficiently stable. This value was selected to mimic the cut-off previously employed in a 2011 paper published by several of the researchers intimately tied to the Materials Project.\cite{Mueller_2011} In that context, the cut-off was selected due to its noted relevance to battery materials. Regardless, $\Delta$E$_{\text{hull}}$ is relatively well regarded in the computational materials community as a predictor of energetic stability, with the caveat that neither temperature nor pressure effects are accounted for in the DFT calculations.
\end{enumerate}

After applying the four tiers of screening, 158 compounds remained. Essentially, in lieu of a comprehensive theory that predicts the threshold switching property, this screening served to identify compounds that are likely to display similar bonding characteristics to the popular GST alloy and would therefore be likely to exhibit the threshold switching behavior as well. Next, the researchers calculated three metrics to narrow their search further: (1) Born effective charge, $Z$*, (2) cohesive energy, $E_C$, and (3) `degree of 90$^\circ$ bond angle deviation', $D_{\text{BAD}}$.

\begin{itemize}
    \item \textit{$Z$*:} The Born effective charge was taken as a proxy descriptor for the electronic polarizability, as the researchers noted that many PCM materials display high values of the latter. $Z$* was obtained via data from DFT calculations implemented in VASP.
    According to the researchers: ``The calculations of $Z$* for the selected 158 materials were performed using conventional unit cells. A unified scheme called \emph{automatic\_density} in the Python Materials Genomics (pymatgen) package was used to generate k-point grids for all of the 158 materials. The generated densities of the grids were 500 per atom for structural relaxation and 1000 per atom for the $Z$* calculations. the $Z$* for an atom was averaged by the diagonal elements of the $Z$* tensor in which the values of all the elements are transformed to their absolute values. Then, the final $Z$* for a material was averaged by all atoms in the calculation models.''
    \item \textit{$E_C$:} Melting temperature is a difficult value to obtain computationally, so the cohesive energy serves as a proxy descriptor for it. While the researchers did not clarify whether this value was obtained directly from the Materials Project database or obtained from their own DFT calculations, $E_C$ is generally a simple value to obtain. For $n$ atoms of type $i$ in the bulk structure (with $E_i$ as the energy of an isolated atom of type $i$),
\begin{equation*}
    E_C = E_{\text{bulk}} - \sum n_i E_i
\end{equation*}

    \item \textit{$D_{\text{BAD}}$:} The degree of 90$^\circ$ bond angle deviation is taken as a proxy descriptor for the operating speed (conversion between crystalline and amorphous phases). The premise is that 90$^\circ$ arises from pure p-orbital bonding, which would presumably promote fast crystallization. On the other hand, deviations from 90$^\circ$ would imply additional stability of the amorphous phase, presumably promoting fast amorphization. The researchers state: ``A suitable $D_{\text{BAD}}$ could display a compatibility of fast crystallization and fast amorphization in a PCM material. ... The [bond angle deviation was] obtained by [averaging of] 1000 transient structures that were intercepted from the trajectory of the 300 [Kelvin] MD simulations.'' $D_{\text{BAD}}$ is obtained from the following equation, where $X_i$ is the value of the bond angle with index $i$ and $n$ is the total number of bond angles:
\begin{equation*}
    D_{\text{BAD}} = \sqrt{\frac{1}{n-1} \sum_{i=1}^n (X_i -90)^2}
\end{equation*}

\end{itemize}

\begin{figure}[b!]
    \centering
    \includegraphics[width=0.6\linewidth]{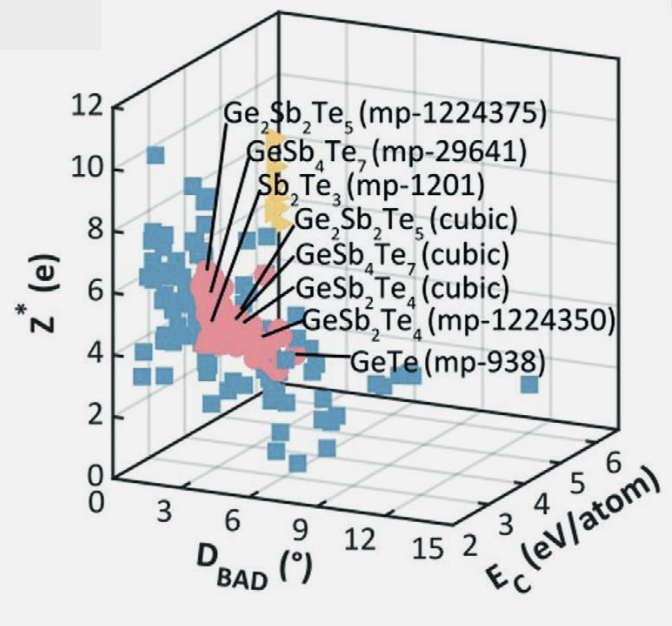}
    \caption{3D plot of results from PCM materials screening. Materials with values close to those of the GST system (5\% threshold) are in red. Materials (rare earth nitrides) of notably high cohesive energy ($>$6 eV atom$^{-1}$) are in yellow. The others are in blue. The `mp-*' label is the ID of the material in the Materials Project database.\cite{Liu_2021_screening}}
    \label{fig:screeninglast}
\end{figure}

\noindent The values that they obtained for $Z$*, $E_C$, and $D_{\text{BAD}}$ for all 158 structures are tabulated in the supplementary information for the paper. A graphical mapping of the results is shown in \textbf{Figure \ref{fig:screeninglast}}, with several of the common phase-change materials indicated.

\begin{figure}[t!]
    \centering
    \includegraphics[width=0.6\linewidth]{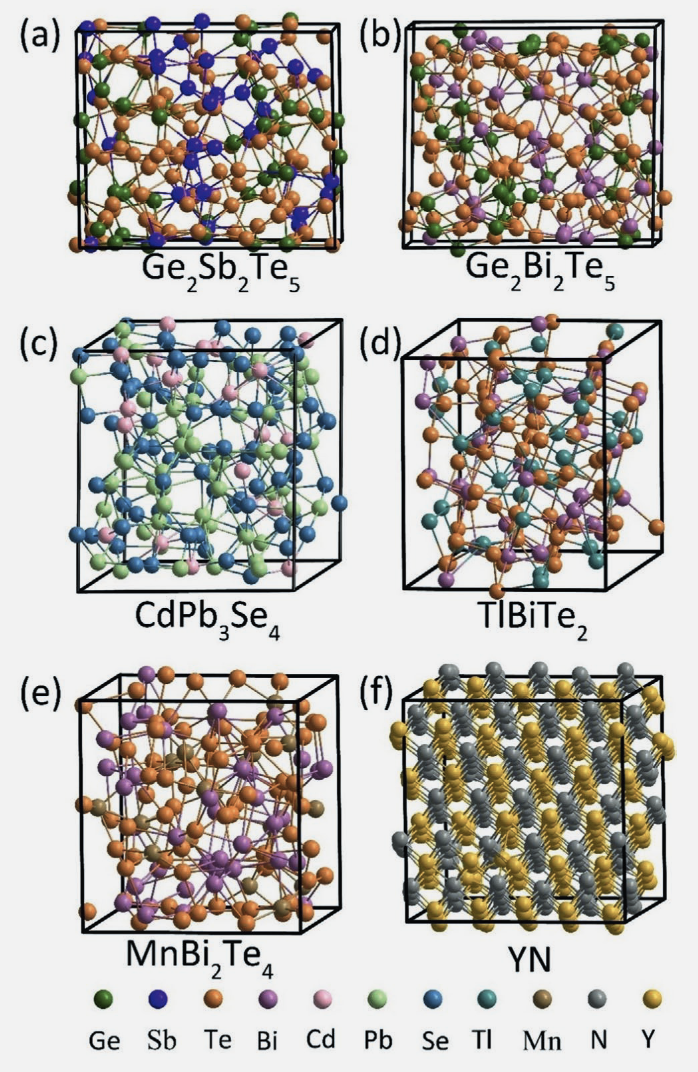}
    \caption{Amorphous structures obtained for (a) reference GST structure, (b) Ge$_2$Bi$_2$Te$_5$, (c) CdPb$_3$Se$_4$, (d) TlBiTe$_2$, (e) MnBi$_2$Te$_4$, (f) rare earth nitride structure for comparison.\cite{Liu_2021_screening}}
    \label{fig:happyamorphous}
\end{figure}

\begin{figure}[b!]
    \centering
    \includegraphics[width=0.9\linewidth]{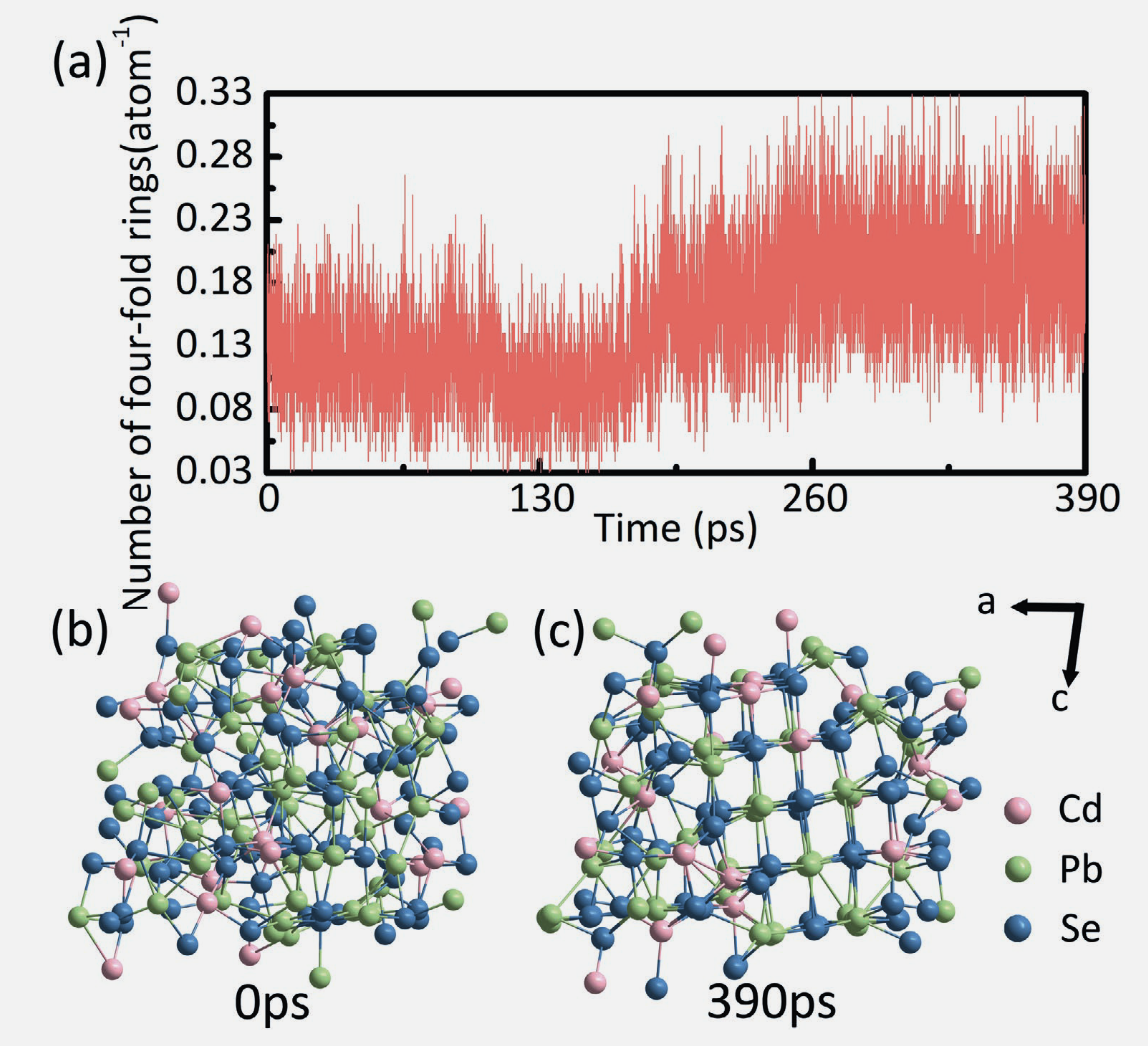}
    \caption{Nucleation and recrystallization AIMD simulation of CdPb$_3$Se$_4$ from 0 to 390 picoseconds. (a) Tracking the number of four-fold ring motifs present over time. (b) The structure at 0 ps. (c) The structure at 390 ps.\cite{Liu_2021_screening}}
    \label{fig:aimd}
\end{figure}

Speculating that the materials with similar Born effective charge, cohesive energy, and degree of 90$^\circ$ bond angle deviation to the GST family of PCM materials would be likely desirable candidates for PCM, the researchers proposed the 52 materials shown in red in \textbf{Figure \ref{fig:screeninglast}} as meriting further investigation. The vast majority are tellurium compounds, but not all of them. For instance, InBi$_3$Se$_6$, CdPb$_3$Se$_4$, AgBiSe$_2$, KBiSe$_2$, Pb$_2$Bi$_2$Se$_5$, PbBi$_2$Se$_4$, Bi$_2$Se$_3$, and Bi$_8$Se$_9$ are all selenium-based instead.The only material of the 52 that did not contain selenium or tellurium was Sn$_4$P$_3$. Lastly, acknowledging the key role of the amorphous phase in PCM, the researchers selected four candidate structures (Ge$_2$Bi$_2$Te$_5$, CdPb$_3$Se$_4$, TlBiTe$_2$, and MnBi$_2$Te$_4$), performed melt-quenching molecular dynamics \textbf{(MD)} on all four, and performed ab-initio MD \textbf{(AIMD)} on just one, CdPb$_3$Se$_4$. The melt-quenching MD provided the amorphous structures, and the AIMD simulated the recrystallization process from amorphous to crystalline phases. The reason for limiting these additional calculations to only a few materials was the computational cost involved. The amorphous structures obtained from melt-quenching MD are shown in \textbf{Figure \ref{fig:happyamorphous}}. Results from the AIMD simulation of CdPb$_3$Se$_4$ are shown in \textbf{Figure \ref{fig:aimd}}.

\begin{figure}[h!]
    \centering
    \includegraphics[width=1\linewidth]{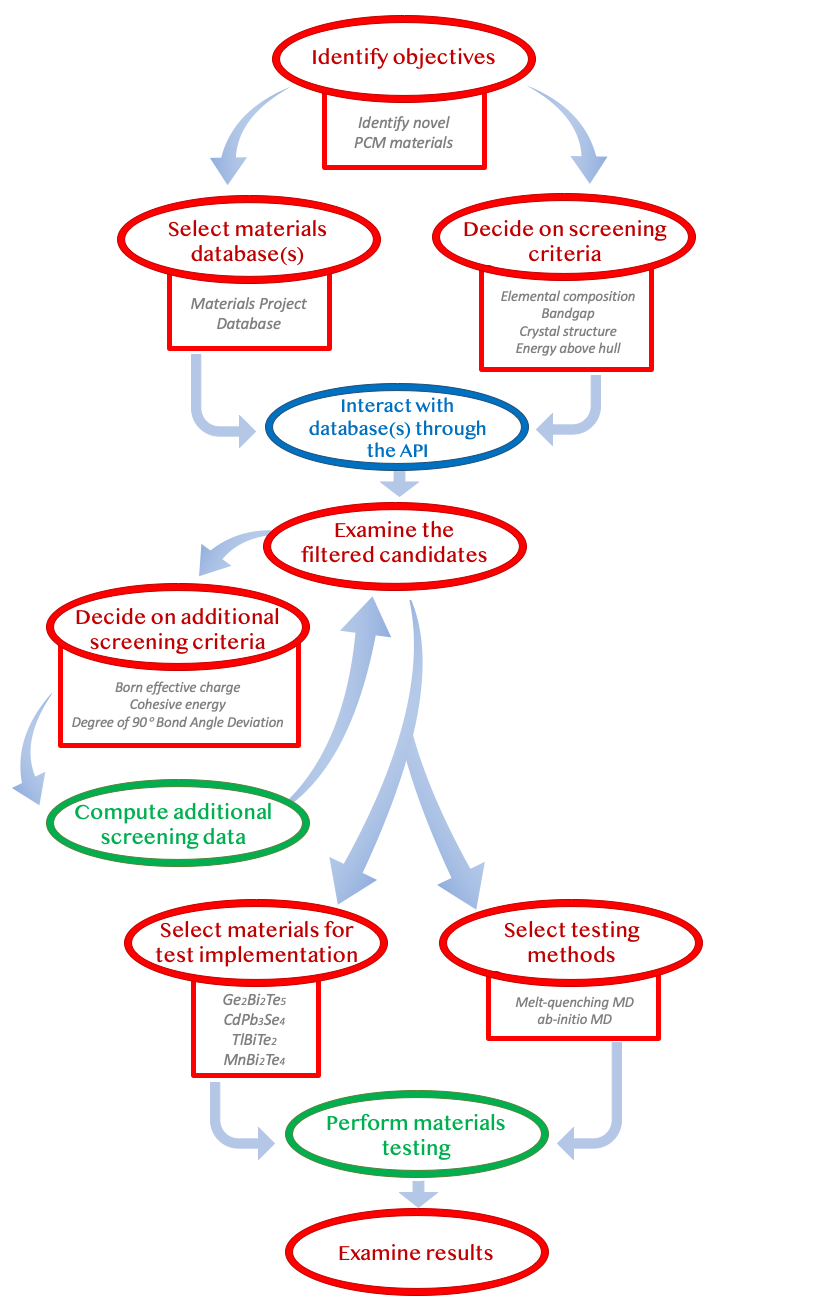}
    \caption{Computational workflow carried out by Liu et al., as described above. Tasks in red denote designer choices, tasks in green leverage compute resources, and tasks in blue leverage database/storage resources. (Branching of arrows implies that both tasks are performed simultaneously.)}
    \label{fig:muffin1}
\end{figure}

Overall, the computational materials exploration workflow performed by Liu et al. is summarized in \textbf{Figure \ref{fig:muffin1}}. 

\noindent While this workflow may have succeeded at identifying novel materials for potential PCM application, it does illustrate many of the current limitations of the materials data infrastructure \textbf{(MDI)}. In \textbf{Table \ref{tab:NASA2040Vision}}, the present-day description of the MDI is that the ``design of materials and systems is disconnected; stages of the product development lifecycle are segmented; tools, ontologies, and methodologies are domain-specific; materials properties are based on empiricism; product certification relies heavily on physical testing.'' This workflow matches that description.
In particular, the results obtained from this workflow do not satisfy the desired criteria described in \textbf{Figure \ref{fig:VennMaterials}}. The property predictions were heavily influenced by assumptions (assumptions that stemmed from mostly empirical observations of the properties of other PCM materials, rather than comprehensive theoretical understanding), verification was extremely minimal (only a few materials simulated in more detail, and no experimental testing), and an economics assessment (materials accessibility, cost, fabrication routes) was entirely absent. This workflow may be used to augment experimental discovery by providing recommendations of material systems to investigate, but ultimately, it remains significantly detached from the downstream implementation of materials in real-world PCM devices.

\begin{figure}[h!]
    \centering
    \includegraphics[width=1\linewidth]{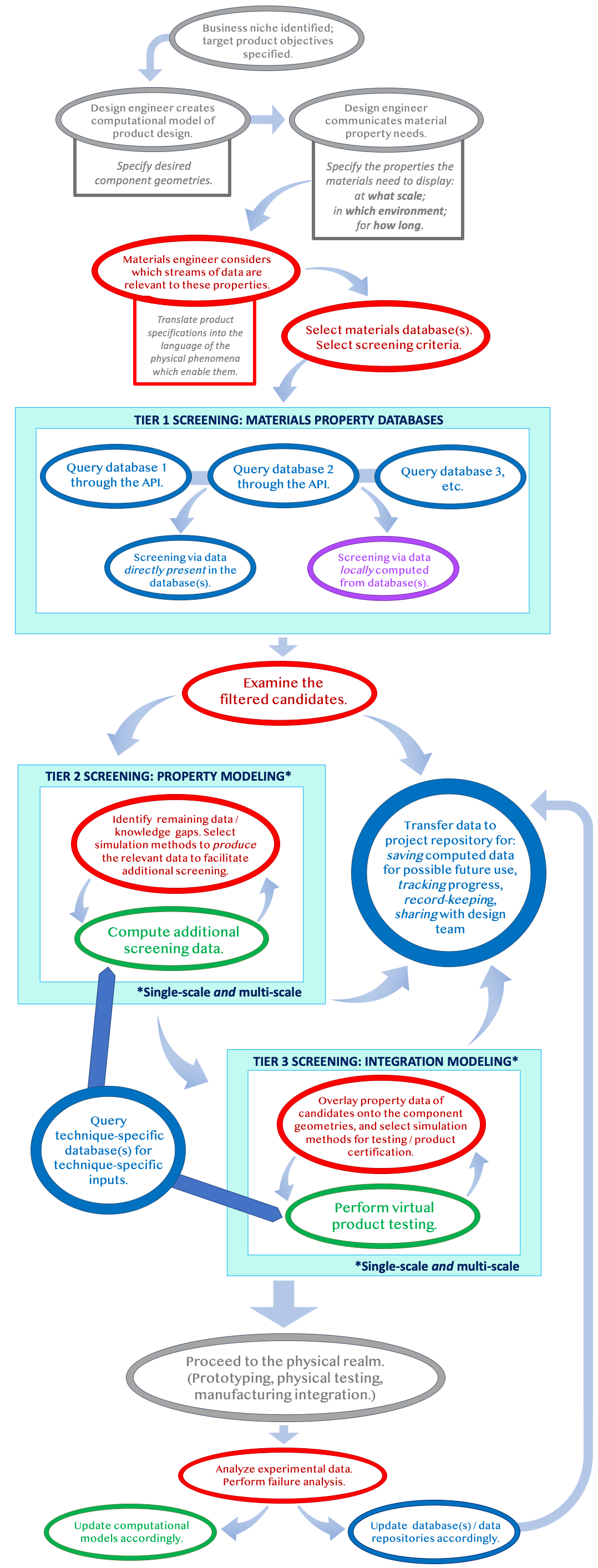}
    \caption{Generalized computational materials engineering workflow that fulfills the NASA 2040 Vision (\textbf{Table \ref{tab:NASA2040Vision}}). Tasks that require decision-making by the materials engineer are in red, tasks that leverage compute resources are in green, tasks that leverage database/storage resources are in blue, a task that would leverage non-Von-Neumann in-storage computing is in purple, and non-computational-materials tasks are in grey.}
    \label{fig:muffin2}
\end{figure}

In contrast, what might an accelerated, improved, future workflow look like? Consider a workflow that fulfills the NASA 2040 Vision, described in \textbf{Table \ref{tab:NASA2040Vision}}: the  ``design of materials and systems is integrated; stages of the product development lifecycle are seamlessly joined; tools, ontologies, and methodologies are usable across the community; materials properties are virtually determined; product certification relies heavily on simulation.'' An example of how this vision can be overlaid onto a materials engineering workflow is given in \textbf{Figure \ref{fig:muffin2}}.

At present, computational materials modeling is performed primarily in academic environments and/or for research purposes; materials modeling efforts in industry tend to be limited to a few niche areas. In contrast, the workflow shown in \textbf{Figure \ref{fig:muffin2}} describes a process where simulation of materials is integrated \emph{directly} into industrial product development. 
The value of this is multifaceted:
\begin{itemize}
    \item The process is highly adaptive and responsive to changes such as: \\ large or small design modifications, additional processing treatments, alternative manufacturing routes, novel emerging materials, advances in data-driven design optimization and analytics, fluctuations in the supply chain, etc.

    \item The process is well-suited for high-throughput screening. 
    \item Waste from trial-and-error prototyping is drastically reduced.
    \item Facilitates a high degree of design tailoring.
    \item Bottlenecks are simple to identify, quantify, and communicate.

    \item The process can be augmented with advanced tools and techniques.

    \item Risk management and uncertainty quantification are data-driven and straightforward. As the project progresses through the process stages, uncertainty is lessened and the degree to which the candidate materials overlap with the implementation sweet spot (\textbf{shown in Figure \ref{fig:GeneralVenn}}) increases significantly. 
    
    \item Transparency and record-keeping are inherent in the process. Clear explanations for why specific design features were implemented and brought to market are consolidated and accessible, enabling all members of the design team with authorized access to the project repository to perform professional due diligence (identifying issues and communicating knowledge gaps that arise).

    \item Capability to probe Process-Structure-Property \textbf{(PSP)} relationships (recall \textbf{Figure \ref{fig:OlsonDiagram}}) relevant to the candidate material systems is included.

\end{itemize}

The workflow can be scaled according to the project scope and needs; depending on the complexity and desired extent of materials exploration and optimization, the work can be undertaken by a single (well-trained) computational materials engineer or distributed across a larger team.

Lastly, let us consider the application of the workflow in \textbf{Figure \ref{fig:muffin2}} to the accelerated design and implementation of phase-change materials for PCM, as informed by the prior discussion. The first step is establishing the specific PCM niche to target. A great deal of scientific computing work requires solving systems of linear equations, however solving such system iteratively can be computationally expensive. In a 2019 paper, researchers from the University of California, Berkeley, proposed an innovative approach; they state:\cite{Sun_2019_berkeley} ``Linear algebra is involved in virtually all scientific and engineering disciplines, e.g., physics, statistics, machine learning, and signal processing. ... In-memory computing with analog resistive memories has shown high efficiencies of time and energy, through realizing matrix-vector multiplication in \textit{one step} with Ohm’s law and Kirchhoff’s law. However, solving matrix equations in a \textit{single operation} remains an open challenge. Here, we show that a feedback circuit with cross-point resistive memories can solve algebraic problems such as systems of linear equations, matrix eigenvectors, and differential equations in \textit{just one step}.'' 
The prototype demonstrated in this paper leveraged filamentary ReRAM, with a HfO$_2$ dielectric as the memory material.
Perhaps Company X is interested in commercializing this idea, in producing resistive computing devices that can solve a system of linear equations in a single step. However, Company X wishes to leverage PCM ReRAM instead of filamentary ReRAM.
With this target product objective specified, the design engineering team gets to work, creating preliminary computational models of possible cross-point array device architectures and product geometries. 
However, the engineers recognize that their designs rely on the material behaving in a very specific fashion, and furthermore, aspects of the design geometry -- such as size/placement of the heating elements -- will be governed by materials-specific features. They communicate their objectives to their materials engineer \textbf{(ME)}, describing which device metrics the material ought to facilitate (e.g. low latencies, low power consumption,  nonvolatility) and which material properties they believe to be absolutely crucial (e.g. a stable spectrum of distinguishable resistance states). They provide tentative estimates of the environment in which the material will dwell (e.g. 100--200$^\circ$C, with oxygen present), the scale of the part geometries (e.g. memory cell is under 5 nm), and their targets for the device longevity (e.g. 10 year lifetime). Furthermore, they note that they are interested in testing out a titanium top electrode and a graphitic carbon bottom electrode (since that configuration worked well in a previous product), although they would be interested to hear what other recommendations the ME may have for them if necessary. The ME mulls over this information, considers the behavior of phase-change materials, and recognizes that several of the device objectives lead to conflicting material property requirements. Furthermore, while the ME has worked with other phase-change chalcogenides in the past, the new proposed designs are sufficiently ambitious that a novel materials system may be necessary to implement. However, the sheer volume of data published on potential PCM materials is immense. Combing through the literature and hoping to stumble across a sufficiently studied materials system that will satisfy all the criteria simultaneously would be like searching for a needle in a haystack (and the papers the ME wants to read never seem to be open-access anyway).
Natural language processing tools might help, but Company X has not purchased a subscription to these services, and besides, the ME intends to take a more systematic approach. Fortunately, the year is 2040 and the materials data infrastructure is sufficiently evolved to enable the analysis that the ME is trained to perform. 

The ME happily begins the materials design process by identifying a set of preliminary screening criteria and selecting several databases to query. By 2040, a dazzling array of materials databases have emerged, each with their own unique motivations. Some databases are large, others are small. Some databases are open-source, others are proprietary. Some databases are well-maintained, others are not. Some databases are populated primarily with experimental data, in which case the measurement technique exerted significant influence over the recorded value. Other databases solely contain computed data, in which case the values are reflections of the computational techniques and input assumptions employed. 
The data itself appears in a wide array of formats, reflecting the wide array of use cases. The NASA 2040 Vision article predicted that:\cite{NASA_2040_Vision} ``Unlike current databases which are largely textual and tabulated, materials databases in the near future will also contain spatial and temporally resolved microstructure information (potentially from multiple imaging modalities), local (spatial) and global property data, thermodynamic data, and metadata detailing the history of the sample, etc. In addition, this is likely to be a blend of experimental characterization and testing as well as modeling and simulation;'' in the hypothetical future scenario described here, this prediction has indeed been realized.
Of course, not all the data is usable. Ever-present and lurking is the possibility that the databases contain outdated information, falsified information, or misleading information. To avoid such dangers, cross-validation is desired where possible. The ME understands how to navigate this heterogeneous landscape of databases; rather than expecting flawless, comprehensive property data that target the use case perfectly, the ME focuses on targeting a few specific properties that will guide the exploration, while curating both an open mind and a heavy dose of skepticism. 

For the preliminary screening, the ME decides on 7 main databases: (1) a large open-source (national-lab-curated) database containing computationally-generated chemical and electronic structure data; (2) a large government-curated database of experimental thermodynamic measurements; (3) a smaller proprietary database curated specifically for the semiconductor industry that the ME convinced Company X to purchase temporary access to; (4) a database of materials that Company X has fabricated or obtained from vendors in the past; (5) an extremely large (albeit much lower-quality) open-source database provided by a large tech company and populated by automated natural language processing and data-scraping of literature articles; (6) a ``materials marketplace'' database where many materials supply and fabrication companies listed their current product offerings to gain enhanced visibility to potential customers; (7) an internationally collaborative database between industry, government, and academia providing consolidated information on the environmental hazards, processing concerns, and externality costs of various materials. The data formats, reliability, relevance, data gathering methods, breadth, and degree of structuring all vary among these databases, but together, they form a large materials data pool that the ME can leverage.\footnote{\scriptsize{To expand the materials property offerings, several of these databases leverage in-storage computing capabilities for providing data that are downstream derivatives of other data within the database (similar to the concept of ``schema-on-read''). However, the in-storage compute functionality is not visible to the front-end database user.}} Fortunately, the ME is benefiting from previously established community-adopted standards; each of these databases supports common API interfaces, facilitating interoperability. As such, much of the effort of consolidating multiple data streams has been offloaded to the infrastructure, allowing the ME to focus their energy primarily on the queries themselves. 

ME dutifully creates a `wish list' of properties, as well as `deal-breakers.' The search begins.\footnote{\scriptsize{In addition to simply having the technical ability to interact with the tools of the materials data infrastructure and possessing a greater professional network, the key distinguishing factor between a senior-career ME and a junior-career ME in this hypothetical year-2040 scenario is the ability to formulate good questions (and good queries). Adaptable, knowledge-informed interrogation (and understanding the limits of a given information source) is a skill that can require significant practical experience.}}
First the ME eliminates any materials that have been previously flagged in their respective database as unverifiable, outdated, or falsified. 
Next, they filter results further by immediately rejecting any material that matches at least one of the dealbreakers. Many properties can be tuned later on through processing variations or even design modifications. Such compromises can be made later, but now is not the time. The ME's queries slash through the candidates, swiftly narrowing them down.
Earlier in their career, the ME would carefully and tentatively apply successfully targeted layers of criteria, not wanting to accidentally filter something that might be helpful.
Now, a more seasoned veteran in their field, the ME takes relish in a different approach: crafting variations upon extremely stringent criteria and seeing if anything makes it through. 
In addition to personally querying the databases, the ME deploys an artificial-intelligence tool to identify potential candidates algorithmically as well. The ME has found that this tool works best for exploring materials that are similar to or derivatives of other previously specified materials, so they provide it with a list of already well-known phase-change materials.
Although not strictly necessary, another artificial-intelligence tool that the ME is fond of takes the `wishlist' criteria as input and provides as output recommendations for other related properties that (while not previously specified) have a high likelihood of being linked to the properties necessary for the final application. Such as tool is helpful for identifying potential correlations as well as for assisting the ME in staying up-to-date on experimental and computational techniques; as the theory and characterization techniques evolve, so too do the properties that researchers invest effort in collecting.
Furthermore, the property measurements and computed values present within the databases are inherently -- at their core -- approximations correlated to the underlying physics at play. The ME is, in a sense, leveraging the properties as a communication bridge to the underlying, emergent complexity of the physics below. Having sufficiently filtered the databases with respect to electronic and thermodynamic properties, the ME examines the surviving candidates, considers cost estimates and accessibility, and transfers the list to the repository shared with the members of the design team. After a meeting with the designers, the ME moves onto the next step: property simulation, where properties deemed sufficiently important for implementation are investigated computationally in greater depth. 

At this stage of the workflow, the ME formulates questions for the materials simulations to address. The role of simulation is to provide targeted insights on pre-specified features, not to fully replicate the material behaviors in its entirety; complex models are not necessarily better, and in many cases they can obscure the physical relationships present within the system. Rather, the ME assesses the extent of the remaining knowledge gaps and selects computational techniques that are well-suited for addressing these gaps. Conveniently, thanks to further evolution of the materials data infrastructure, a government-funded database has emerged to consolidate the modeling techniques that can be employed at different length- and time-scales. Listings are vetted by the community, metrics and uses cases are described, and sample implementations are included (similar to the `AI-model gardens' repository concept pioneered by researchers from the University of Chicago and Argonne National Laboratory).\footnote{\scriptsize{One of the researchers describes the `AI-model garden' concept thusly:\cite{AI_model_gardens} ``The Garden is a place where a community can be built around AI in science... We see Model Gardens as places where models in similar domains are kept, tended, and shared, and where researchers can discover a validated model that fits their needs and immediately run it with four lines of code, turning what used to require months of effort into minutes or seconds of effort.''}} Resources like this have helped tools, ontologies, and methodologies to be usable across the community (as the NASA 2040 Vision foretold). Leveraging these techniques does still require training and nuance -- naturally, there remains a vibrant and flourishing community of specialist practitioners -- but much of the manual effort involved with setting up and deploying a new simulation has been alleviated. Documentation is plentiful and calibration tests are readily accessible. Furthermore, simulation inputs can be stores and shared, enabling easy verification of results by others. Conveniences like these save the ME a great deal of time, helping them to focus their attention primarily on probing the materials properties.
Several of the well-established computational techniques have established databases specifically for them (for example, the OpenKIM repository for DFT potentials).\cite{OpenKIM} The property simulations do require compute resources, for which ME has been allocated a budget.
The ME considers the factors in \textbf{Figure \ref{fig:VennSimulation}} and makes their selections accordingly.
Once the property simulation and screening is complete, the ME sends the data to the shared repository location for the design team to access. After another discussion with the design team, the virtual integration testing begins. The property data of the remaining material candidates is overlaid atop the design geometries, and tests probing the property behavior at individual-feature-level and device-level are conducted. Just as with the prior modeling, compute resources are necessary. Materials that satisfy threshold criteria via virtual testing (using community specifications, such as those curated by ASTM)\cite{ASTM}
are certified for the next stage: prototyping, physical testing, and manufacturing integration. Accordingly, the ME is involved in the collection, processing, and interpretation of the physical data.  These stages of the design cycle are reminiscent of older workflows, prior to the shift towards virtual materials design. 
The key distinction between those older workflows and the 2040 workflow employed by the ME is that the materials data infrastructure has evolved such that in the 2040 workflow, the cost incurred by integrating targeted, quality materials informatics and modeling efforts into the early design stages is much more feasible.

\section{\textbf{\textcolor{blue}{Conclusion}}}

This report covered a wide array of topics. It started by introducing the historical development of ICME and material data infrastructure in the United States, highlighting the unique needs of this field, the growth of the community, present-day obstacles, and key aspects that remain in development. Then, a brief overview of data storage architectures (file, block, object) was provided. The physical hardware of storage systems was touched on, describing the `hierarchy' of computer storage/memory in the process. Emergent storage class memory (SCM) was discussed in more depth, with a specific focus on Intel Optane persistent memory devices and the PCM technology that they use. The discussion of Optane transitioned into the introduction of DAOS, a next-generation object storage system that can leverage SCM. The experience of using object storage was described, including a brief overview of APIs for this context. Several comments were included on setting up and interacting with DAOS, as well.
The literature on novel computing and storage paradigms that SCM may enable is immense, but a small sampling of examples was provided. Lastly, aspects of a sample materials engineering problem (leveraging present-day and future materials data infrastructure to accelerate the design of phase-change materials in PCM devices) were discussed. In conclusion, 
the challenges of the 21$^{\text{st}}$ century are nontrivial, but the tools that engineers can build for addressing them are rapidly improving, and a robust materials data infrastructure plays a crucial role in the development of these tools. 
Providing attention to the technological building blocks that comprise such infrastructure is a worthy endeavor.

\vskip3pt
\ack{This work was made possible through the author's participation in the 2022 GRIPS-Berlin summer program,\cite{GRIPS_Berlin2022} funded by the Institute of Pure and Applied Mathematics (IPAM) at the University of California, Los Angeles (UCLA), and hosted at the Zuse Institute of Berlin (ZIB), which is part of the MODAL research campus\cite{MODAL} and home to a Tier-2 German supercomputing facility\cite{Tier2}. GRIPS stands for `Graduate-level Research in Industrial Projects for Students.' This is an 8-week summer program that provides participants with the experience of working abroad on projects that exist at the intersection of industry and academia. The 2022 program had three industrial sponsors, each contributing their own project opportunity. The project that the author worked on, originally titled `High performance computing (HPC) for real-world simulations', was sponsored by Hewlett Packard Enterprise (HPE). In addition, the author was provided with mentorship, feedback, and support by members of the Algorithms for Innovative Architectures (A4IA) group at the ZIB. The author would like to express her gratitude to the individuals Steffen Christgau, Utz-Uwe Haus, and Thomas Steinke for their assistance on the project. Furthermore, the efforts of the GRIPS program organizers and support staff at the ZIB and IPAM towards facilitating such a rich experience are greatly appreciated.}
\vskip5pt
\noindent \textit{ Stephanie R. Taylor (University of California, Los Angeles, USA)}
\vskip3pt
\noindent E-mail: stephformankind@ucla.edu




\begin{thebibliography}{}

\scriptsize{

\bibitem{NASA_2040_Vision}
Liu, X.; Furrer, D.; Kosters, J.; Holmes, J. ``Vision 2040: A Roadmap for Integrated, Multiscale Modeling and Simulation of Materials and Systems,'' \textit{NASA}, 2018, \textit{CR—2018-219771}.

\bibitem{Cyberinfrastructure_2004}
Novotny, M.A.; Ceperley, D.; Jayanthi, C.S.; Martin, R.M. ``Materials Research Cyberscience Enabled by Cyberinfrastructure,'' \textit{National Science Foundation}, 2004.

\bibitem{ICMEReport_2008}
``Integrated Computational Materials Engineering: A Transformational Discipline for Improved Competitiveness and National Security,'' \textit{National Research Council}, 2008. \textit{[doi.org/10.17226/12199]}.

\bibitem{Ford_2006}
Allison, J.; Li, M.; Wolverton, C.; Su, X. ``Virtual aluminum castings: An industrial application of ICME,'' \textit{JOM}, 2006, \textbf{58}, pg. 28-35.

\bibitem{MGI_2011}
``Materials Genome Initiative for Global Competitiveness,'' \textit{National Science and Technology Council (NSTC)}, 2011. 

\bibitem{MGI_2014}
``Materials Genome Initiative Strategic Plan (2014),'' \textit{National Science and Technology Council (NSTC)}, 2014.

\bibitem{BuildingMDI_2017}
``Building a Materials Data Infrastructure: Opening New Pathways to Discovery and Innovation in Science and Engineering,'' \textit{The Minerals, Metals \& Materials Society (TMS)}, 2017.

\bibitem{MGI_2021}
``Materials Genome Initiative Strategic Plan (2021),'' \textit{National Science and Technology Council (NSTC)}, 2021.

\bibitem{MaterialsProject}
``The Materials Project,'' \textit{Lawrence Berkeley National Laboratory}, 2022. \textit{[materialsproject.org]}.

\bibitem{Donegan_2020}
Donegan, S.P.; Groeber, M.A. ``Data Structures and Workflows for ICME,'' (2020). In: Ghosh, S.; Woodward, C.; Przybyla, C. (eds) ``Integrated Computational Materials Engineering (ICME)'', \textit{Springer}, 2020. 


\bibitem{PIF_Citrine_2016}
Michel, K.; Meredig, B. ``Beyond bulk single crystals: A data format for all materials structure–property–processing relationships,'' \textit{MRS Bulletin}, 2016, \textbf{41}, 617-623.




\bibitem{Zwart_2020}
Zwart, S.P. ``The ecological impact of high-performance computing in astrophysics,'' \textit{Nature Astronomy}, 2020, \textbf{4}, pg. 819-822.

\bibitem{Redhat_Graphic}
``File storage, block storage, or object storage?'' \textit{Redhat.com}, 2018. \textit{[www.redhat.com/en/topics/data-storage/file-block-object-storage]}.

\bibitem{AFileIsNotAFile}
Harter, T.; Dragga, C.; Vaughn, M.; Arpaci-Dusseau, A.C.; Arpaci-Dusseau, R.H. ``A File is Not a File: Understanding the I/O Behavior of Apple Desktop Applications,'' \textit{ACM Transactions on Computer Systems}, 2012, \textbf{30}(3).

\bibitem{IEEE_POSIX1}
``IEEE Standard Portable Operating System Interface for Computer Environments,'' \textit{IEEE}, 1988. \textit{[standards.ieee.org/ieee/1003.1/1388]}.

\bibitem{IEEE_POSIX2}
``IEEE Std 1003.1, 2017 Edition,'' \textit{The Open Group}, 2018.\\ \textit{[unix.org/version4/ieee\_std.html]}.

\bibitem{Alluxio_Graphic}
Fan, B.; Zhao, B. ``Speeding Up I/O for Machine Learning: Apple Case Study Using Tensorflow and Alluxio,'' \textit{Alluxio Online Meetup}, 2020. \textit{[www.slideshare.net/Alluxio/speeding-up-io-for-machine-learning-ft-apple-case-study-using-tensorflow-nfs-dc-os-alluxio]}.

\bibitem{POSIX_outdated_2016}
Atlidakis, V.; Andrus, J.; Geambasu, R.; Mitropoulos, D.; Nieh, J. ``POSIX Has Become Outdated," \textit{login Usenix Mag.}, 2016, \textbf{41}(3).

\bibitem{POSIX_IO_bad_2017}
Lockwood, G. ``What's so bad about POSIX I/O?'' \textit{The Next Platform}, 2017. \textit{[www.nextplatform.com/2017/09/11/whats-bad-posix-io]}.

\bibitem{Liu_2018}
J. Liu et al. ``Evaluation of HPC Application I/O on Object Storage Systems,'' \textit{2018 IEEE/ACM 3rd International Workshop on Parallel Data Storage \& Data Intensive Scalable Computing Systems (PDSW-DISCS)}, 2018, pg. 24-34.


\bibitem{Lustre}
``About the Lustre File System,'' \textit{Lustre}, 2022. \textit{[www.lustre.org/about]}.

\bibitem{HDFS}
``HDFS Architecture,'' \textit{Apache Hadoop}, 2022.\\ \textit{[hadoop.apache.org/docs/stable/hadoop-project-dist/hadoop-hdfs/HdfsDesign.html]}.

\bibitem{SpectrumScale}
``IBM Spectrum Scale,'' \textit{IBM}, 2022. \textit{[www.ibm.com/products/spectrum-scale]}.

\bibitem{BeeGFS}
``BeeGFS - The Leading Parallel File System,'' \textit{ThinkParQ}, 2022.\\ \textit{[thinkparq.com/products/beegfs]}.

\bibitem{IBM_FileBlockObject}
``Object vs. File vs. Block Storage: What’s the Difference?'' \textit{IBM Cloud Education}, 2021. \textit{[www.ibm.com/cloud/blog/object-vs-file-vs-block-storage]}.

\bibitem{Netflix}
``Migrating to Cloud - Lessons from Netflix, Brought Up to Date,'' \textit{Amazon Web Services Youtube Channel}, 2018. \textit{[www.youtube.com/watch?v=XrWII4ewrXA]}.

\bibitem{Amazon_CloudStorage}
``Cloud Storage on AWS,'' \textit{Amazon Web Services}, 2022.\\ \textit{[aws.amazon.com/products/storage]}

\bibitem{Amazon_S3}
``Amazon S3 Features,'' \textit{Amazon Web Services}, 2022. \textit{[aws.amazon.com/s3/features]}.

\bibitem{MicrosoftAzure}
``Introduction to Azure Blob storage,'' \textit{Microsoft Azure}, 2022.\\ \textit{[docs.microsoft.com/en-us/azure/storage/blobs/storage-blobs-introduction]}.

\bibitem{GoogleCloud}
``Cloud storage,'' \textit{Google Cloud}, 2022. \textit{[cloud.google.com/storage]}.

\bibitem{MinIO}
``Multi-cloud object storage,'' \textit{MinIO}, 2022. \textit{[min.io]}.

\bibitem{Ceph}
``Ceph object storage,'' \textit{Ceph}, 2022. \textit{[ceph.io/en/discover/technology]}.

\bibitem{Swift}
``Swift's documentation,'' \textit{Openstack}, 2022. \textit{[docs.openstack.org/swift/latest]}.

\bibitem{StorNext}
``StorNext Shared Storage,'' \textit{Quantum}, 2022. \textit{[www.quantum.com/en/products/file-system/stornext]}.


\bibitem{CephSoftwareStack}
``Ceph Architecture,'' \textit{Ceph}, 2022. \textit{[docs.ceph.com/en/latest/architecture]}.


\bibitem{DAOS}
``DAOS Storage Revolutionizes High-Performance Storage,'' \textit{Intel}, 2022. \textit{[www.intel.com/content/www/us/en/high-performance-computing/daos.html]}.

\bibitem{DAOS_2021_YoutubeUpdate}
Prantis, K. ``Intel DAOS 2.0: Storage Software Stack and Ecosystem Update,'' \textit{Tech Field Day Youtube Channel}, 2021. \textit{[www.youtube.com/watch?v=HELtx1GRnto]}.

\bibitem{IO500}
``IO 500,'' \textit{IO500 Foundation}, 2022. 
\textit{[io500.org/about]}.

\bibitem{DAOS_paper_2020}
Liang, Z.; Lombardi, J.; Chaarawi, M.; Hennecke, M. ``DAOS: A Scale-Out High Performance Storage Stack for Storage Class Memory.'' In: Panda, D. (eds) ``Supercomputing Frontiers'' \textit{Lecture Notes in Computer Science, Springer}, 2020, 12082.

\bibitem{BriefOverview_2019}
Mellor, C. ``A brief overview of Intel DAOS high performance storage,'' \textit{Blocks \& Files }, 2019. \textit{[blocksandfiles.com/2019/11/28/intel-daos-high-performance-storage-file-system-explainer]}.


\bibitem{MemoryHierarchyPic}
Leng, R.J. ``Computer Memory Hierarchy,'' \textit{bit-tech}, 2007. \textit{[www.bit-tech.net/reviews/tech/memory/the\_secrets\_of\_pc\_memory\_part\_1/3]}.

\bibitem{CachePic}
Jenkov, J. ``Modern Hardware,'' \textit{jenkov.com}, 2015. \textit{[jenkov.com/tutorials/java-performance/modern-hardware.html]}.


\bibitem{SRAMvsDRAMpic}
Pal, S.K. ``Different Types of RAM (Random Access Memory ),'' \textit{GeeksForGeeks}, 2021. \textit{[www.geeksforgeeks.org/different-types-ram-random-access-memory]}.

\bibitem{SSDpic}
Persaud, C. ``Everything you need to know about solid-state drives (SSD),'' \textit{University of Waterloo}, 2018. \textit{[uwaterloo.ca/arts-computing-newsletter/winter-2018/feature/everything-you-need-know-about-solid-state-drives-ssd]}.

\bibitem{FloatingGatePic}
Aravindan, A. ``Flash 101: Types of NAND Flash," \textit{Embedded}, 2018. \textit{[www.embedded.com/flash-101-types-of-nand-flash]}.

\bibitem{NAND_AgingMarker_2021}
Kim, M. et al. ``RealWear: Improving performance and lifetime of SSDs using a NAND aging marker,'' \textit{Performance Evaluation}, 2021, \textbf{145}, pg. 102153.

\bibitem{SSD_LifeExpectancy1}
``Understanding Life Expectancy of Flash Storage for LabVIEW Real-Time Systems,'' \textit{National Instruments Corporation}, 2022. \textit{[www.ni.com/en-us/support/documentation/supplemental/12/understanding-life-expectancy-of-flash-storage.html]}.

\bibitem{SSD_LifeExpectancy2}
Holland, T. ``SSD Lifespan: How Long do SSDs Really Last?'' \textit{The Ontrack Data Recovery Blog}, 2020. \textit{[www.ontrack.com/en-us/blog/how-long-do-ssds-really-last]}.

\bibitem{SSD_LifeExpectancy3}
Crider, M. ``How Long Do Solid State Drives Really Last?'' \textit{How-To Geek}, 2017. \textit{[www.howtogeek.com/322856/how-long-do-solid-state-drives-really-last]}.


\bibitem{SSD_7days}
Hruska, J. ``SSDs can lose data in as little as 7 days without power,'' \textit{ExtremeTech}, 2015. \textit{[www.extremetech.com/computing/205382-ssds-can-lose-data-in-as-little-as-7-days-without-power]}.



\bibitem{SSD_Humidity}
Maruf, A.; Brahmakshatriya, S.; Li, B.; Tiwari, D.; Quan, G.; Bhimani, J. ``Do Temperature and Humidity Exposures Hurt or Benefit Your SSDs?'' \textit{2022 Design, Automation \& Test in Europe Conference \& Exhibition (DATE)}, 2022, pg. 352-357

\bibitem{IBM_HDD}
``Hard Disk Drive (HDD) vs. Solid State Drive (SSD): What’s the Difference?'' \textit{IBM Cloud Education}, 2022. \textit{[www.ibm.com/cloud/blog/hard-disk-drive-vs-solid-state-drive]}.

\bibitem{OptaneRelease}
Alan. ``What is 3D XPoint?'' \textit{Utmel Electronic}, 2021.\\ \textit{[www.utmel.com/blog/categories/memory\%20chip/what-is-3d-xpoint]}.

\bibitem{HotColdOptanePic}
Marko, K. ``In the quest for faster storage, Intel hopes that Optane's time has come,'' \textit{diginomica}, 2020. \textit{[diginomica.com/quest-faster-storage-intel-hopes-optanes-time-has-come]}.



\bibitem{Optane3Modes}
Berrocal, E. ``Operating Modes of Intel® Optane™ DC Persistent Memory,'' \textit{Intel.} \textit{[www.intel.com/content/www/us/en/developer/videos/operating-modes-of-intel-optane-dc-persistent-memory.html]}

\bibitem{Optane3Modes2}
Kennedy, P. ``The Glorious Complexity of Intel Optane DIMMs and Why Micron Quit,'' \textit{ServeTheHome Youtube Channel}, 2021.\\ \textit{[www.youtube.com/watch?v=dOV3gGncGU8]}.


\bibitem{OptaneGenericPic}
Mellor, C. ``Just ONE THOUSAND times BETTER than FLASH! Intel, Micron's amazing claim,'' \textit{TheRegister}, 2015.\\ 
\textit{[www.theregister.com/2015/07/28/intel\_micron\_3d\_xpoint]}.

\bibitem{OptaneConferencePic}
Mellor, C. ``Peering closer at 3D XPoint memory: What are Intel, Micron up to?'' \textit{TheRegister}, 2015.\\ \textit{[www.theregister.com/2015/07/29/having\_a\_looks\_at\_imtfs\_crosspoint]}.

\bibitem{OptaneDescription}
Hruska, J. ``Intel, Micron reveal Xpoint, a new memory architecture that could outclass DDR4 and NAND,'' \textit{ExtremeTech}, 2015.\\ \textit{[www.extremetech.com/extreme/211087-intel-micron-reveal-xpoint-a-new-memory-architecture-that-claims-to-outclass-both-ddr4-and-nand]}.


\bibitem{XpointSEMpic2}
Potoroaca, A. ``Intel reveals Optane DIMMs for workstations, 665p SSD for consumers, future roadmap,'' \textit{TechSpot}, 2019. \textit{[www.techspot.com/news/82088-intel-reveals-optane-dimms-workstations-665p-ssd-consumers.html]}.

\bibitem{XpointSEMpic}
Tallis, B. ``Intel Shares New Optane And 3D NAND Roadmap - Barlow Pass DIMMs \& 144L QLC NAND in 2020,'' \textit{AnandTech}, 2019.\\ \textit{[www.anandtech.com/show/14903/intel-shares-new-optane-and-3d-nand-roadmap]}.

\bibitem{Ovskinsky_1968}
Ovshinsky, S.R. ``Reversible Electrical Switching Phenomena in Disordered Structures,'' \textit{Phys. Rev. Lett.}, 1968, \textbf{21}, 1450.

\bibitem{Neale_1973}
Neale, R.G.; Aseltine, J.A. ``The Application of Amorphous Materials
to Computer Memories,'' \textit{IEEE Transactions on Electron Devices}, 1973, \textbf{20}(2), 195.

\bibitem{Wuttig_2007}
Wuttig, M.; Yamada, N. ``Phase-change materials for rewriteable data storage,'' \textit{Nature Materials}, 2007, \textbf{6}, pg 824-832.

\bibitem{Gallo_2020}
Gallo, M.L.; Sebastian, A. ``An overview of phase-change memory device physics,'' \textit{J. Phys. D: Appl. Phys.}, 2020, \textbf{53}, 213002.

\bibitem{Moneta_2010}
Caulfield, A.M.; De, A.; Coburn, J.; Mollow, T.I.; Gupta, R.K.; Swanson, S. ``Moneta: A High-Performance Storage Array Architecture for Next-Generation, Non-volatile Memories,'' \textit{2010 43rd Annual IEEE/ACM International Symposium on Microarchitecture}, 2010, pp. 385-395.

\bibitem{SwansonQuote_2011}
``Phase Change Memory-Based ‘Moneta’system Points To The Future Of Computer Storage,'' \textit{UC San Diego, Jacobs School of Engineering, News Release}, 2011. \textit{[jacobsschool.ucsd.edu/news/release/1078]}.

\bibitem{SwansonGroup}
Swanson, S. ``Building software for persistent memory,'' \textit{Steven Swanson Group Website}, 2022. \textit{[swanson.ucsd.edu]}.

\bibitem{SwansonOptaneWhitepaper}
Swanson et al. ``Basic Performance Measurements of the Intel Optane DC Persistent Memory Module,'' \textit{arXiv}, 2019. \textit{[arXiv:1903.05714]}.




\bibitem{Pohm_1970}
Pohm, A.V.; Sie, C.H.; Uttecht, R.R.; Kao, V.; Agrawal, O. ``Chalcogenide glass bistable resistivity (Ovonic) memories,'' \textit{IEEE Transactions on Magnetics}, 1970, \textbf{6}(3), 592.




\bibitem{IntelOptaneDrop2022}
Coughlin, T. ``Intel Winding Down Its Optane Memory Business,'' \textit{Forbes}, 2022. \textit{[www.forbes.com/sites/tomcoughlin/2022/07/28/intel-winding-down-its-optane-memory-business]}.

\bibitem{CXL1}
Bowman, K.; Sharma, D.D. ``CXL Consortium Webinar: An introduction to compute express LinkTM (CXL) technology,'' \textit{CXL Webinars Youtube Channel}, 2020. \textit{[www.youtube.com/watch?v=RpAshNmpqLQ]}.

\bibitem{CXL2}
``Compute Express Link: The Breakthrough CPU-to-Device Interconnect,'' \textit{Compute Express Link}, 2022. \textit{[www.computeexpresslink.org]}.

\bibitem{CXL3}
Mann, T. ``Will CXL spell the end for boutique composable infrastructure vendors?'' \textit{The Register}, 2022.\\ \textit{[www.theregister.com/2022/07/10/cxl\_composable\_infrastructure]}.

\bibitem{CXL4}
Bai, E.L.; Raut, S.A. ``An Analysis on Compute Express Link with Rich Protocols and Use Cases for Data Centers,'' \textit{ICIPCN 2022: Third International Conference on Image Processing and Capsule Networks}, 2022, pg 787–800.


\bibitem{MicronLeaves2021}
Coughlin, T. ``Micron Ends 3D XPoint Memory,'' \textit{Forbes}, 2019.\\ \textit{[www.forbes.com/sites/tomcoughlin/2021/03/16/micron-ends-3d-xpoint-memory]}.

\bibitem{IntelOptaneDropSpeculation2022}
Coughlin, T. ``Is Intel Going To Drop Optane?'' \textit{Forbes}, 2022.\\ \textit{[www.forbes.com/sites/tomcoughlin/2022/02/28/is-intel-going-to-drop-optane]}.

\bibitem{AuroraPressRelease2019}
``U.S. Department of Energy and Intel to Deliver First Exascale Supercomputer,'' \textit{Intel Press Releases}, 2019. \textit{[www.intc.com/news-events/press-releases/detail/86/u-s-department-of-energy-and-intel-to-deliver-first]}.

\bibitem{OptaneSupply}
Mellor, C. ``Intel has Optane chip hoard with no plans to develop tech,'' \textit{Blocks \& Files}, 2022. \textit{[blocksandfiles.com/2022/05/02/intel-optane-chip-inventory]}.

\bibitem{KioxiaSCM}
``Kioxia Launches Second Generation of High-Performance, Cost-Effective XL-FLASH Storage Class Memory Solution,'' \textit{Business Wire}, 2022. \textit{[www.businesswire.com/news/home/20220801005862/en/Kioxia-Launches-Second-Generation-of-High-Performance-Cost-Effective-XL-FLASH\%E2\%84\%A2-Storage-Class-Memory-Solution]}.

\bibitem{KioxiaSCM2}
``XL-FLASH Storage Class Memory (SCM),'' \textit{KIOXIA America Inc.}, 2022. \textit{[americas.kioxia.com/en-us/business/memory/xlflash.html]}.

\bibitem{EverspinSCM}
``Everspin Announces New STT-MRAM EM128LX xSPI Memory,'' \textit{Business Wire}, 2022. \textit{[www.businesswire.com/news/home/20220801005856/en/Everspin-Announces-New-STT-MRAM-EM128LX-xSPI-Memory]}.

\bibitem{EverspinSCM2}
``Spin-transfer Torque MRAM Technology,'' \textit{Everspin Technologies}, 2022. \textit{[www.everspin.com/spin-transfer-torque-mram-technology]}.

\bibitem{SKHynixSCM}
Clinton. ``SK Hynix works on 3D Vertical XP memory as an alternative to 3D XPoint,'' \textit{Gadgetonus}, 2022. \textit{[gadgetonus.com/tech/5845.html]}.

\bibitem{SKHynixSCM2}
``Tomorrow Initiative: SK hynix’s New Roadmap for
Tomorrow’s Tech Ecosystem,'' \textit{SK Hynix}, 2022. \textit{[product.skhynix.com/support/forest.go]}.

\bibitem{DAOSEcosystemPic}
`ISC'22 DAOS Intro Video,' \textit{DAOS Youtube Channel}, 2022.\\ \textit{[www.youtube.com/watch?v=q4iYRwW5Uhw]}.

\bibitem{DAOSfutureDeveloperComments}
Nabarro, T.; Lombardi, J. ``Question about 3D Xpoint DIMM,'' \textit{DAOS group}, 2022. \textit{[daos.groups.io/g/daos/topic/question\_about\_3d\_xpoint\_dimm/92723028]}.

\bibitem{APIComicPic}
Kothalawala, A. ``What is an API? How does it work?'' \textit{Medium}, 2018.\\ \textit{[medium.com/@ama.thanu/what-is-an-api-how-does-it-work-f4ea552d741f]}.

\bibitem{Lamothe_2021}
Lamothe, M.; Gueheneuc, Y.G.; Shang, W. ``A Systematic Review of API Evolution Literature,'' \textit{ACM Computing Surveys}, 2021, \textbf{54}(8), 171.

\bibitem{MaterialsProjectAPI}
Ong S.P.; Cholia, S.; Jain, A.; Brafman, M.; Gunter, D.; Ceder, G.; Persson, K.A. ``The Materials Application Programming Interface (API): A simple, flexible and efficient API for materials data based on REpresentational State Transfer (REST) principles,'' \textit{Comput. Mater. Sci.}, 2015, \textbf{91}, pg 209-215.

\bibitem{MaterialsProjectAPI2}
Huck, P.; Jain, A.; Gunter, D.; Winston, D.; Persson, K. ``A Community Contribution Framework for Sharing Materials Data with Materials Project,'' \textit{2015 IEEE 11th International Conference on e-Science}, 2015, pg 535-541.

\bibitem{AFLOWAPI}
Taylor, R.H.; Rose, F.; Toher, C.; Levy, O.; Yang, K.; Nardelli, M.B.; Curtarolo, S. ``A RESTful API for exchanging materials data in the AFLOWLIB.org consortium,'' \textit{Comput. Mater. Sci.}, 2014, \textbf{93}, pg 178-192.

\bibitem{AFLOWAPI2}
Gossett et al. ``AFLOW-ML: A RESTful API for machine-learning predictions of materials properties,'' \textit{Comput. Mater. Sci.}, 2018, \textbf{152}, 134-145.

\bibitem{MaterialsCloudAPI}
Talirz et al. ``Materials Cloud, a platform for open computational science,'' \textit{Scientific Data}, 2020, \textbf{7}, 299.


\bibitem{AiiDA_API}
Huber et al. ``AiiDA 1.0, a scalable computational infrastructure for automated reproducible workflows and data provenance,'' \textit{Scientific Data}, 2020, \textbf{7}, 300.

\bibitem{NOMAD_API}
``NOMAD Repository and Archive: Introduction,'' \textit{NOvel Materials Discorvery (NOMAD)}, 2014. \textit{[nomad-lab.eu/prod/rae/docs/introduction.html]}.

\bibitem{OpenQuantumMaterialsDatabaseAPI}
``OQMD RESTful API,'' \textit{The OQMD development team}, 2019.\\ \textit{[static.oqmd.org/static/docs/restful.html]}.

\bibitem{OpenQuantumMaterialsDatabase}
Saal, J.E.; Kirklin, S.; Aykol, M.; Meredig, B.; Wolverton, C. ``Materials Design and Discovery with High-Throughput Density Functional Theory: The Open Quantum Materials Database (OQMD),'' \textit{JOM}, 2013, \textbf{65}(11), 1501.


\bibitem{CatalysisHubAPI}
``CatApp Database API,'' \textit{Catalysis-Hub}, 2022.\\ \textit{[www.catalysis-hub.org/graphQLConsole]}.

\bibitem{CatalysisHub}
Winther, K.T.; Hoffmann, M.J.; Boes, J.R.; Mamun, O.; Bajdich, M.; Bligaard, T. ``Catalysis-Hub.org, an open electronic structure database for surface reactions,'' \textit{Scientific Data}, 2019, \textbf{6}(75).

\bibitem{OpenMaterialsDatabaseAPI}
``OPTIMADE provider `Open Materials Database','' \textit{Materials Consortia \\OPTIMADE group}, 2022.\\ \textit{[www.optimade.org/providers-dashboard/providers/omdb.html]}.

\bibitem{OpenMaterialsDatabaseAPI2}
Armiento, R. ``Full httk API documentation,'' 2018.\\
\textit{[docs.httk.org/en/latest/httk\_base.html]}.

\bibitem{TCOD_API}
``How to query the COD database,'' \textit{Crystallography Open Database Wiki}, 2022. \textit{[wiki.crystallography.net/howtoquerycod]}.


\bibitem{CitrinationAPI}
O'Mara, J.; Meredig, B.; Michel, K. ``Materials Data Infrastructure: A Case Study of the Citrination Platform to Examine Data Import, Storage, and Access,'' \textit{JOM}, 2016, \textbf{68}, 2031-2034.

\bibitem{JARVIS_API}
Choudhary et al. ``The joint automated repository for various integrated simulations (JARVIS) for data-driven materials design,'' \textit{npj Computational Materials}, 2020, \textbf{6}(173).


\bibitem{InorganicCrystalStructureDatabaseAPI}
``ICSD Products: ICSD API Service,'' \textit{Inorganic Crystal Structure Database}, 2022. \textit{[icsd.products.fiz-karlsruhe.de/en/products/icsd-products]}.

\bibitem{InorganicCrystalStructureDatabase}
``NIST Inorganic Crystal Structure Database (ICSD) SRD 3,'' \textit{National Institute of Standards and Technology}, 2022. \textit{[www.nist.gov/srd/nist-standard-reference-database-3]}.

\bibitem{PaulingFile}
Villars, P.; Cenzual, K.; Gladyshevskii, R.; Iwata, S. ``PAULING FILE - towards a holistic view,'' \textit{Chem. Met. Alloys}, 2018, \textbf{11}, pg 43-76.

\bibitem{PaulingFile2}
``Materials Platform for Data Science (MPDS) API,'' \textit{Materials Phase Data System} and \textit{Tilde Materials Informatics}, 2020. \textit{[mpds.io/developer]}.

\bibitem{PubChemAPI}
``Programmatic Access,'' \textit{PubChem}, 2022.\\ \textit{[pubchemdocs.ncbi.nlm.nih.gov/programmatic-access]}.


\bibitem{OrganicMaterialsDatabase}
Borysov, S.S.; Geilhufe, R.M.; Balatsky, A.V. ``Organic materials database: An open-access online database for data mining.'' \textit{Plos one}, 2017, \textbf{12}(2), 0171501.

\bibitem{NRELMatDB}
``NREL Materials Database,'' \textit{National Renewable Energy Laboratory}, 2022. \textit{[materials.nrel.gov]}.



\bibitem{OPTIMADE_2021}
Anderson et al. ``OPTIMADE, an API for exchanging materials data,'' \textit{Sci Data}, 2021, \textbf{8}, 217.

\bibitem{RESTfulOrigin2007}
Richardson, L.; Ruby, S. ``RESTful Web Services,'' \textit{O'Reilly Media, Inc.}, 2007.

\bibitem{RESTorigin2000}
Fielding, R.T. ``Architectural Styles and the Design of Network-based Software Architectures,'' \textit{University of California, Irvine; PhD dissertation}, 2000. \textit{[www.ics.uci.edu/~fielding/pubs/dissertation/top.htm]}.

\bibitem{RESTorigin2002}
Fielding, R.T.; Taylor, R.N. ``Principled design of the modern Web architecture,'' \textit{ACM Transactions on Internet Technology}, 2002, \textbf{2}(2), pg 115-150.

\bibitem{RESTfulBook2019}
Subramanian, H.; Raj, P. ``Hands-On RESTful API Design Patterns and Best Practices,'' \textit{Packt Publishing}, 2019.


\bibitem{HPCFilesystemsVsObjectStorage}
``The NIH HPC Systems object store,'' \textit{Biowulf: High Performance Computing at the NIH}, 2022. \textit{[hpc.nih.gov/storage/object.html]}.

\bibitem{TestScript1}
``daos/src/tests/simple\_obj.c'', \textit{Github}, 2022.\\ \textit{[github.com/daos-stack/daos/blob/master/src/tests/simple\_obj.c]}

\bibitem{TestScript2}
``daos/src/tests/simple\_array.c'', \textit{Github}, 2022.\\ \textit{[github.com/daos-stack/daos/blob/master/src/tests/simple\_array.c]}.

\bibitem{TestScript3}
Lombardi, J. ``Re: DAOS starting example in C++,'' \textit{DAOS Groups}, 2020. \textit{[daos.groups.io/g/daos/message/1182]}.

\bibitem{NativeObjectInterface}
``Native Object Interface,'' \textit{DAOS v.2 website}, 2022.\\ \textit{[docs.daos.io/v2.0/user/interface]}.

\bibitem{keyvalueobjectDAOS}
``DAOS API Functions'' \textit{DAOS API Documentation}, 2022.\\ \textit{[docs.daos.io/v2.0/doxygen/html/globals\_func.html]}.

\bibitem{Soumagne_2022}
Soumagne et al. ``Accelerating HDF5 I/O for Exascale Using DAOS,'' \textit{IEEE Transactions on Parallel and Distributed Systems}, 2022, \textbf{33}(4), pg 903-914.

\bibitem{doxygen}
``DAOS API Documentation (Version 2),'' \textit{DAOS Developers}, 2022. \textit{[docs.daos.io/v2.0/doxygen/html]}.

\bibitem{DAOSgroup}
``daos@daos.groups.io: Topics,'' \textit{groups.io}, 2022. \textit{[daos.groups.io/g/daos/topics]}.

\bibitem{Kestor_2014}
Kestor, G.; Gioiosa, R.; Kerbyson, D.; Hoisie, A. ``Data movement in scientific applications,'' \textit{Pacific Northwest National Laboratory; Workshop on Modeling \& Simulation of Systems and Applications}, 2014.

\bibitem{Cicotti_2016}
Cicotti et al. ``Data movement in data-intensive high performance computing,'' \textit{In: Conquering Big Data with High Performance Computing}, 2016, pg 31-59.

\bibitem{Si_2021}
Si, M.; Cheng, H.; Ando, T.; Hu, G.; Ye, P. ``Overview and outlook of emerging non‑volatile memories,'' \textit{MRS Bulletin}, 2021, \textbf{46}, 946.


\bibitem{Cheng_2019_3D}
Cheng et al. ``3D cross-point phase-change memory for storage-class memory,'' \textit{J. Phys. D.: Appl. Phys.}, 2019, \textbf{52}, 473002.

\bibitem{Fantini_2020}
Fantini, P. ``Phase change memory applications: the history, the present and the future,'' \textit{J. Phys. D: Appl. Phys.}, 2020, \textbf{53}, 283002.

\bibitem{vonNeumannPicture}
``Von Neumann architecture,'' \textit{Wikipedia}, 2022.\\ \textit{[en.wikipedia.org/wiki/Von\_Neumann\_architecture]}.

\bibitem{Rosenburg_2017}
Rosenburg, J. ``Security in embedded systems,'' \textit{Chapter 6 in: Rugged Embedded Systems, Computing in Harsh Environments}, 2017, pg 149--205.

\bibitem{Burr_2016}
Burr et al. ``Neuromorphic computing using non-volatile memory,'' \textit{Advances in Physics: X}, 2017, \textbf{2}(1),  pg 89--124.

\bibitem{Shi_2021_book}
Shi, Z. ``Brain-like intelligence,'' \textit{Chapter 14, In: Intelligence Science, Leading the Age of Intelligence}, 2021, pg 537-593.

\bibitem{Markovic_2020}
Markovic, D.; Mizrahi, A.; Querlioz, D.; Grollier, J. ``Physics for neuromorphic computing,'' \textit{Nat. Rev. Phys.}, 2020, \textbf{2}, pg 499-510.

\bibitem{Zhu_2020_review}
Zhu, J.; Zhang, T.; Yang, Y.; Huang, R. ``A comprehensive review on emerging artificial neuromorphic devices,'' \textit{Applied Physics Reveiws}, 2020, \textbf{7}, 011312.

\bibitem{DAOSmarketingGraphic}
``DAOS: Revolutionizing High-Performance Storage with Intel Optane Technology,'' \textit{Intel}, 2022.\\ \textit{[www.intel.com/content/dam/www/public/us/en/documents/solution-briefs/high-performance-storage-brief.pdf]}.

\bibitem{Sebastian_2020}
Sebastian, A.; Gallo, M.; Khaddam-Aljameh, R.; Eleftheriou, E. ``Memory devices and applications for in-memory computing,'' \textit{Nature Nanotechnology}, 2020, \textbf{15}, 529-544.


\bibitem{MerrikhBayat_2018}
Merrikh-Bayat et al. ``High-performance mixed-signal neurocomputing with nanoscale floating-gate memory cell arrays,'' \textit{IEEE Transactions on Neural Networks and Learning Systems}, 2018, \textbf{29}(10), pg 4782-4790.

\bibitem{Xi_2021}
Xi et al. ``In-Memory Learning With Analog Resistive Switching Memory: A Review and Perspective,'' \textit{Proceedings of the IEEE}, 2021, \textbf{109}(1), 14.

\bibitem{Zou_2021_bottleneck}
Zou, X.; Xu, S.; Chen, X.; Yan, L.; Han, Y. ``Breaking the von Neumann bottleneck: Architecture-level processing-in-memory technology,'' \textit{Sci. China Inf. Sci.}, 2021, \textbf{64}(6), 160404.

\bibitem{Torabzadehkashi_2019}
Torabzadehkashi et al. ``Accelerating HPC applications using computational storage devices,'' \textit{IEEE 21st International Conference on High Performance Computing and Communications}, 2019, pg 1878-1885.

\bibitem{Lukken_2021}
Lukken, C.; Trivedi, A. ``Past, Present, and Future of Computational Storage: A Survey,'' \textit{arXiv}, 2021. \textit{[arxiv.org/pdf/2112.09691.pdf]}.

\bibitem{Aljameh_2022}
Khaddam-Aljameh et al. ``HERMES-Core -- A 1.59-TOPS/mm$^2$ PCM on 14-nm CMOS In-Memory Compute Core Using 300-ps/LSB Linearized CCO-Based ADCs,'' \textit{IEEE Journal of Solid-State Circuits}, 2022, \textbf{57}(4), 1027.

\bibitem{HeydariGorji_2022}
HeydariGorji et al. ``In-storage Processing of I/O Intensive Applications on Computational Storage Drives,'' \textit{23rd International Symposium on Quality Electronic Design}, 2022, pg 1-6.


\bibitem{Qiu_2015}
Qiu, M.; Ming, Z.; Li, J.; Gai, K.; Zong, Z. ``Phase-change memory optimization for green cloud with genetic algorithm,'' \textit{IEEE Transactions on Computers}, 2015, \textbf{64}(12), 3528.

\bibitem{Fridman_2021}
Fridman et al. ``Accessing the use cases of persistent memory in high-performance scientific computing,'' \textit{IEEE/ACM 11th Workshop on Fault Tolerance for HPC at eXtreme Scale (FTXS)}, 2021.

\bibitem{LopezGomez_2021}
Lopez-Gomez, G.; Blomer, J. ``Exploring object stores for high-energy physics data storage,'' \textit{EPJ Web of Conferences, CHEP 2021}, 2021, \textbf{251}, 02066.

\bibitem{Manubens_2022}
Manubens, N.; Quintino, T.; Smart, S.; Danovaro, E.; Jackson, A. ``DAOS as HPC storage, a view from numerical weather prediction,'' \textit{IEEE Transactions on Parallel and Distributed Systems}, 2022. \textit{Under review.} \textit{[arxiv.org/pdf/2208.06752.pdf]}.


\bibitem{Wong_2010}
Wong et al. ``Phase Change Memory,'' \textit{Proceedings of the IEEE}, 2010, \textbf{98}(12), 2201.

\bibitem{Kim_2020_evolution}
Kim, T.; Lee, S. ``Evolution of phase-change memory for the storage-class memory and beyond,'' \textit{IEEE Transactions on Electron Devices}, 2020, \textbf{67}(4), pg 1394-1406.

\bibitem{Lacaita_2006}
Lacaita, A.L. ``Phase change memories: State-of-the-art, challenges and perspectives,'' \textit{Solid-State Electronics}, 2006, \textbf{50}, pg 24-31.

\bibitem{Burr_2016_review}
Burr et al. ``Recent progress in phase-change memory technology,'' \textit{IEEE Journal on Emerging and Selected Topics in Circuits and Systems}, 2016, \textbf{6}(2), 146.

\bibitem{Kim_2020_endurance}
Kim, S.; Burr, G.W.; Kim, W.; Nam, S. ``Phase-change memory cycling endurance,'' \textit{MRS Bulletin}, 2019, \textbf{44}, pg 710-714.

\bibitem{Soeya_2013}
Soeya, S.; Shintani, T.; Odaka, T.; Kondou, R.; Tominaga, J. ``Ultra-low switching power, crystallographic analysis, and switching mechanism for Sn$_x$Te$_{100-x}$/Sb$_2$Te$_3$ diluted superlattice system,'' \textit{Appl. Phys. Lett.}, 2013, \textbf{103}, 053103.

\bibitem{Liu_2021_screening}
Liu, Y.; Li, X.; Zheng, H.; Chen, N.; Wang, X.; Zhang, X.; Sun, H.; Zhang, S. ``High-throughput screening for phase-change memory materials,'' \textit{Adv. Funct. Mater.}, 2021, \textbf{31}, 2009803.

\bibitem{Xu_2021_screening}
Xu et al. ``Materials screening for disorder-controlled chalcogenide crystals for phase-change memory applications,'' \textit{Adv. Mater.}, 2021, \textbf{33}, 2006221.

\bibitem{MaterialsProjectBandgaps}
``Electronic Structure Calculation Details,'' \textit{Materials Project Documentation}, 2022. \textit{[docs.materialsproject.org/methodology/materials-methodology/electronic-structure]}.

\bibitem{Mueller_2011}
Mueller, T.; Hautier, G.; Jain, A.; Ceder, G. ``Evaluation of tavorite-structured cathode materials for lithium-ion batteries using high-throughput computing,'' \textit{Chem. Mater.}, 2011, \textbf{23}(17), 3854-3862.








\bibitem{Sun_2019_berkeley}
Sun, Z.; Pedretti, G.; Ambrosi, E.; Bricalli, A.; Wang, W. ``Solving matrix equations in one step with cross-point resistive arrays,'' \textit{PNAS}, 2019, \textbf{116}(10), pg 4123-4128.

\bibitem{AI_model_gardens}
``UChicago/Argonne Researchers Will Cultivate AI Model `Gardens' With \$3.5M NSF Grant,'' \textit{UChicago CS News}, 2022. \textit{[cs.uchicago.edu/news/uchicago-argonne-researchers-will-cultivate-ai-model-gardens-with-3-5m-nsf-grant]}.

\bibitem{OpenKIM}
``OpenKIM: Interatomic Potentials and Analytics for Molecular Simulation,'' \textit{OpenKIM}, 2022. \textit{[openkim.org]}.

\bibitem{ASTM}
``Standards \& Publications,'' \textit{ASTM International}, 2022. \textit{[www.astm.org/products-services/standards-and-publications.html]}.


\bibitem{GRIPS_Berlin2022}
``Graduate-level Research in Industrial Projects for Students (GRIPS) – Berlin 2022
JUNE 20 - AUGUST 12, 2022,'' \textit{IPAM}, 2022.\\ \textit{[www.ipam.ucla.edu/programs/student-research-programs/graduate-level-research-in-industrial-projects-for-students-grips-berlin-2022]}.

\bibitem{MODAL}
``Research Campus MODAL,'' \textit{Zuse Institute Berlin}, 2022.\\
\textit{[www.zib.de/features/research-campus-modal]}.


\bibitem{Tier2}
``NHR Center at ZIB,'' \textit{Zuse Institute Berlin}, 2022. 
\textit{[https://www.zib.de/nhr]}.



}

\end{thebibliography}
\end{document}